%% file: apophis_arXiv.tex
\documentclass[12pt, twocolumn]{article}
\usepackage[top=1.2in,left=0.7in,footskip=0.5in,width=7.1in]{geometry}

\usepackage[utf8]{inputenc}

\usepackage{cite}

\usepackage{nameref,hyperref}
\hypersetup{
   colorlinks=true,             
   linkcolor=blue,              
   urlcolor=blue,               
   anchorcolor=blue,            
   citecolor=blue,              
}

\usepackage[right]{lineno}

\usepackage{microtype}
\DisableLigatures[f]{encoding = *, family = * }

\usepackage{changepage}

\usepackage[aboveskip=1pt,labelfont=bf,labelsep=period,singlelinecheck=off]{caption}

\usepackage{lastpage,fancyhdr}
\usepackage{epstopdf}

\usepackage{color}

\definecolor{Gray}{gray}{.25}

\usepackage{graphicx}

\usepackage{sidecap}

\usepackage{wrapfig}
\usepackage[pscoord]{eso-pic}
\usepackage[fulladjust]{marginnote}
\reversemarginpar

\usepackage{blindtext, multicol}
\usepackage{float}
\input{boxed1.sty}
\floatstyle{simplerule}
\restylefloat{figure}
\usepackage[table]{xcolor}
\definecolor{grey80}{rgb}{0.90,0.90,0.90}
\usepackage{amsmath}
\usepackage{natbib}
\begin{document}
   \onecolumn
   {\huge\textbf\newline{\bf First approximation for spacecraft motion relative to (99942) Apophis}}
   \newline
   \\
   S. Aljbaae\textsuperscript{1,*},
   D. M. Sanchez\textsuperscript{1},
   A. F. B. A. Prado\textsuperscript{1},
   J. Souchay\textsuperscript{2},
   M. O. Terra\textsuperscript{3},
   R. B. Negri\textsuperscript{1},
   L. O. Marchi\textsuperscript{1}\\
   \bigskip
   
   \noindent
   {\bf 1} Division of Space Mechanics and Control, INPE, C.P. 515, 12227-310 S\~ao Jos\'e dos Campos, SP, Brazil.\\
   {\bf 2} SYRTE, Observatoire de Paris, PSL Research University, CNRS, Sorbonne Universités, UPMC Univ. Paris 06, LNE, 61 avenue de l'Observatoire, 75014 Paris, France.\\
   {\bf 3} Instituto Tecnol\'ogico de Aeron\'autica, S\~ao Jos\'e Campos, SP, 12228-900, Brazil.
   \\
   \bigskip
   
   \noindent
   * \href{safwan.aljbaae@gmail.com}{safwan.aljbaae@gmail.com}

   \section*{Abstract}
      We aim at providing a preliminary approach on the dynamics of a spacecraft in orbit about the asteroid (99942) Apophis during its Earth close approach. The physical properties from the polyhedral shape of the target are derived assigning each tetrahedron to a point mass in its center. That considerably reduces the computation processing time compared to previous methods to evaluate the gravitational potential. The surfaces of section close to Apophis are build considering or not the gravitational perturbations of the Sun, the planets, and the SRP. The Earth is the one that most affects the invisticated region making the vast majority of the orbits to collide or escape from the system. Moreover, from numerical analysis of orbits started on March 1, 2029, the less perturbed region is characterized by the variation of the semimajor axis of 40-days orbits, which do not exceed 2 km very close to the central body ($a < 4$ km,  $e < 0.4$). However, no regions investigated could be a possible option for inserting a spacecraft into natural orbits around Apophis during the close approach with our planet. Finally, to solve the stabilization problem in the system, we apply a robust path following control law to control the orbital geometry of a spacecraft. At last, we present an example of successful operation of our orbit control with a total $\bigtriangleup v$ of 0.495 m/s for 60 days. All our results are gathered in the CPM-ASTEROID database, which will be regularly updated by considering other asteroids.\\
      {\bf Key words:} Celestial mechanics — minor planets, asteroids: individual (Apophis) — gravitation. \\
      \bigskip
   \twocolumn      
   \section{Introduction}\label{sec01_introduction}
      The asteroid (99942) Apophis was discovered 16 years ago, on June 13th. 2004 (\citet{smalley_2005}, cf. MPEC 2004-Y25). Soon after this discovery, its orbit was a particular subject of investigations. First simulations, badly constrained, lead to a hypothetic impact with the Earth in 2029, which although being quite improbable, could not be completely rejected. Then the hypothesis of impact with our planet was definitely rejected after more and more rigorous orbital simulations were carried out \citep{sansaturio_2008, bancelin_2012}. Nowadays, after numerous refinements on the determination of initial conditions and orbital simulations, it is a well-established fact that (99942) Apophis will pass at a distance of $\sim$ 38,000 km, roughly six planetary radii, from the Earth’s center, on April 13$^{\text{th}}$, 2029. Note that the asteroid will drive particular attention in the future, for other close encounters with our planet that are scheduled to occur in the XXI$^{\text{th.}}$ centuty. In addition to the dramatic orbital changes caused by the 2029 close encounter with the Earth, complementary studies were oriented towards two objectives: first, the study of the Yarkovsky effect, which has to be taken into account for post-2029 refined orbital models \citep{bottke_2006, giorgini_2008, chesley_2009}; second, the modeling of important changes of rotational parameters, as the rotation rate and the orientation of the axis of rotation \citep{scheeres_2005, souchay_2014, souchay_2018}. 
   
      In this paper we orientate our study towards an additional field of investigations: in a first step we determine the polyhedral shape and the gravity field of Apophis and, in a second step, we study the behavior of a test particle close to the asteroid, mainly perturbed by the gravitational action of the Earth and the Solar Radiation Pressure (referred as SRP hereafter) during the 2029 close encounter. In fact, Apophis appears to be in a state of non-principal axis rotation (tumbling). During the exceptional close approach, the tidal stresses and torques may cause resurfacing or reshaping of the body. However, this still a completely unknown interaction and very difficult to predict. As a preliminary step to propose a rendezvous mission to Apophis, we neglected the possible changes of the spin of the target and tried to approach a realistic analysis during the close approach. However, addressing this issue will be fundamental in future studies. In Sect. \ref{sec02_polyhedral_shape} we gather the necessary information to model the polyhedral shape of Apophis, from which we compute a 3D inertia tensor and calculate the spherical harmonic coefficients of the gravitational potential. Then, we construct a gravity model from which we can display the zero-gravity curves. In Sect. \ref{sec03_surfaces_section} we build the surfaces of section in a body-fixed frame. Sect. \ref{sec04_CE} concerns the specific case of the 2029 close encounter with the Earth, for which we model the modifications of the surfaces of section established in Sect.\ref{sec03_surfaces_section} and by taking into account the SRP, we deduce the equations of motion of a test particle surrounding the asteroid, during the close approach. In Sect. \ref{sec05_stability_research}, as an application of our inversions, we identify the less perturbed region around Apophis suitable to place a spacecraft around the asteroid. To compensate all the perturbations in the Apophis system, we apply orbital correction maneuvers in Sect. \ref{sec06_rbital_control}. Finally, a general presentation of our CPM-Asteroid (Close Proximity Motion relative to an Asteroid) database is presented in Sect. \ref{sec07_CPM-Asteroid}, which enables one to acquire insights into the orbital dynamics of a spacecraft near the asteroid (99942) Apophis.
   \section{Apophis shape model and Gravity field} \label{sec02_polyhedral_shape}
      
      To derive our model, we first describe some physical properties of Apophis, based in its polyhedral shape, considering an uniform density. We then consider a cloud of point masses (ideal spheres), which reproduces the total mass and moments of inertia of our target, as well as its gravitational field.
      
      \subsection{Physical Properties}
      
         \citet{muller_2014} observed Apophis with the Herschel Space Observatory Photodetector Array Camera and Spectrometer (PACS) instrument and classified our target as a Sq-class object most closely resembling LL ordinary chondrite meteorites. The author considered an Itokawa-like density ($\rho = 1.75 \pm 0.11$ g/cm$^3$ \citep{lowry_2014}) and estimated a mass ($M$) between $4.4$ and $6.2 \times 10^{10}$ kg. 
         
         Our research focused on the real Apophis polyhedral shape obtained from extensive photometric observations by \citet{pravec_2014} and available in the 3D Asteroid Catalogue\footnote{November 2019, \href{https://3d-asteroids.space/asteroids/99942-Apophis}{https://3d-asteroids.space/asteroids/99942-Apophis}} website. The authors showed that Apophis rotation is retrograde with a spin period of 30.4 h. According to \citet{durech_2010} and \citet{hanus_2017}, the photometry alone cannot provide information on asteroid sizes. Note that, concerning the asteroid (216) Kleopatra, \citet{descamps_2011} showed that, although its shape appears to be correct, the difference in the dimensions obtained from radar shape reconstruction and photometry can reach 20\%. Moreover, \citet{chanut_2015b} found that the behavior of the zero velocity curves and the dynamics differ substantially if one applies a scale-size of 1.15 relative to the original shape of (216) Kleopatra. For that reason, we started our work by checking if the dimension of the reconstructed shape of \citet{pravec_2014} corresponds to the observed diameter \citep{muller_2014}. We determine the coefficient ($\gamma$), which links the volume of the polyhedral shape ($V$) with the mass ($M_{P} = V\rho$) to be compatible with the mass ($M$), when considering the Itokawa-like density ($\rho$). In other words, we choose suitable coefficients that multiply the coordinates x, y and z of the shape to find the polyhedral volume compatible with the mass and density estimated from the observation. Our results are presented in Table \ref{table1_coefficient}. Recently, \citet{hanus_2017} presented shape models and volume for 40 asteroids from DAMIT\footnote{\href{http://astro.troja.mff.cuni.cz/projects/damit}{http://astro.troja.mff.cuni.cz/projects/damit}} shape modeling. The authors derived shape models and volumes for 41 asteroids using the ADAM algorithm from the inversion of their optical light curves, disk-integrated images from the Near-InfraRed Camera (Nirc2) at the Keck II telescope located at Maunakea in Hawaii and stellar occultation measurements. In this work, we just combined the volume with the mass estimation from the literature to derive the bulk density. We found a scale-size of $\gamma = 0.285 \pm 0.158$ relative to the polyhedral shape derived from \citet{pravec_2014}, that must be applied to obtain a mass of $5.31 \mp 0.9\times10^{10}$ kg and a diameter of 0.387 km. Recently, \citet{brozovic_2018} used the radar data to improve the shape of \citet{pravec_2014}. The authors obtained a shape with 2000 vertices and 3996 faces. We also found a scale-size of 1.152 should be applied to obtain the mass and density already mentioned.

\input{TABLES/table1_coefficient}

\begin{figure}[ht]
            The shape derived from \citet{pravec_2014}: 1014 vertices and 2024 faces\\
            \includegraphics[width=0.3\linewidth]{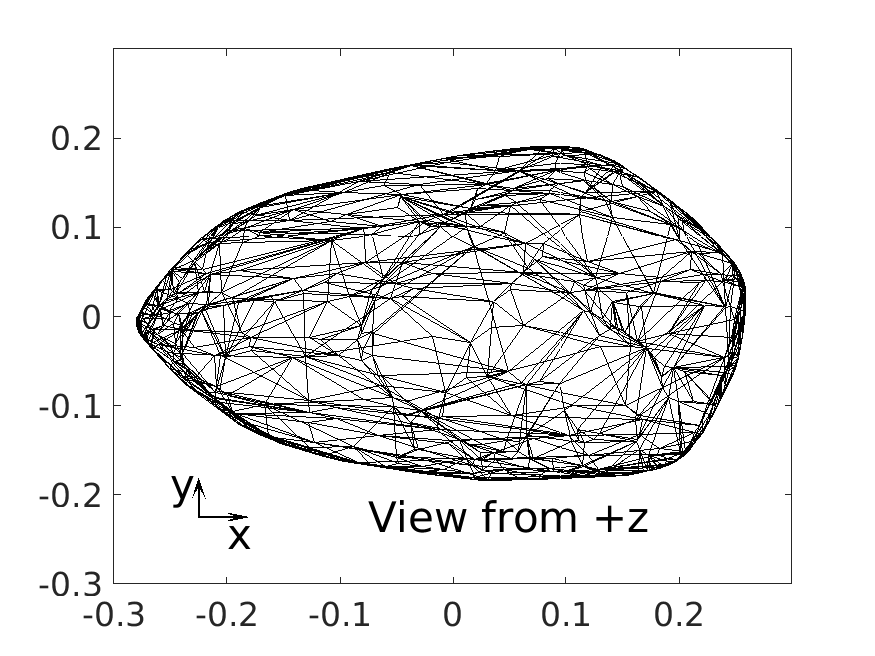}
            \includegraphics[width=0.3\linewidth]{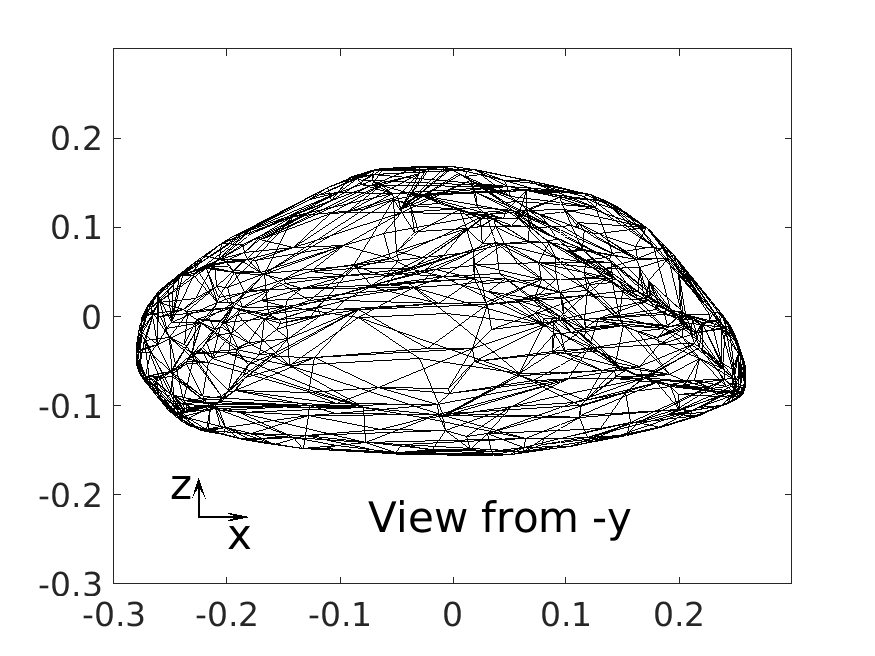}
            \includegraphics[width=0.3\linewidth]{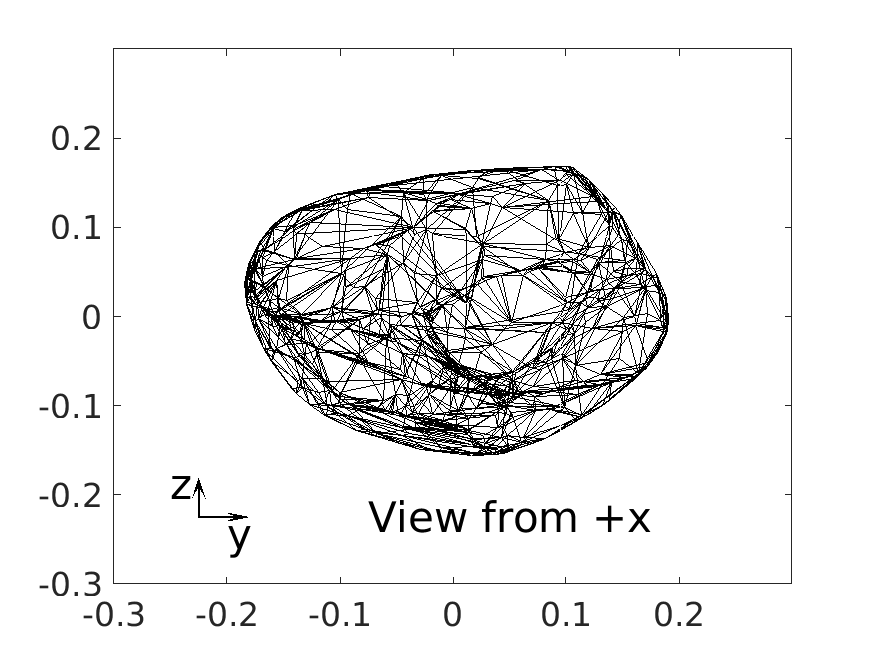}\\   
            \includegraphics[width=0.3\linewidth]{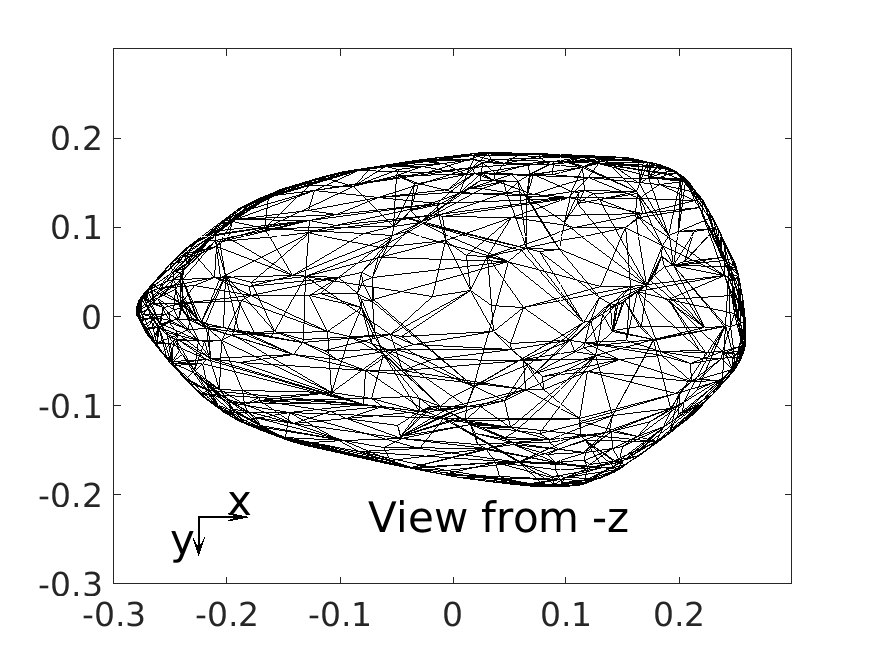}
            \includegraphics[width=0.3\linewidth]{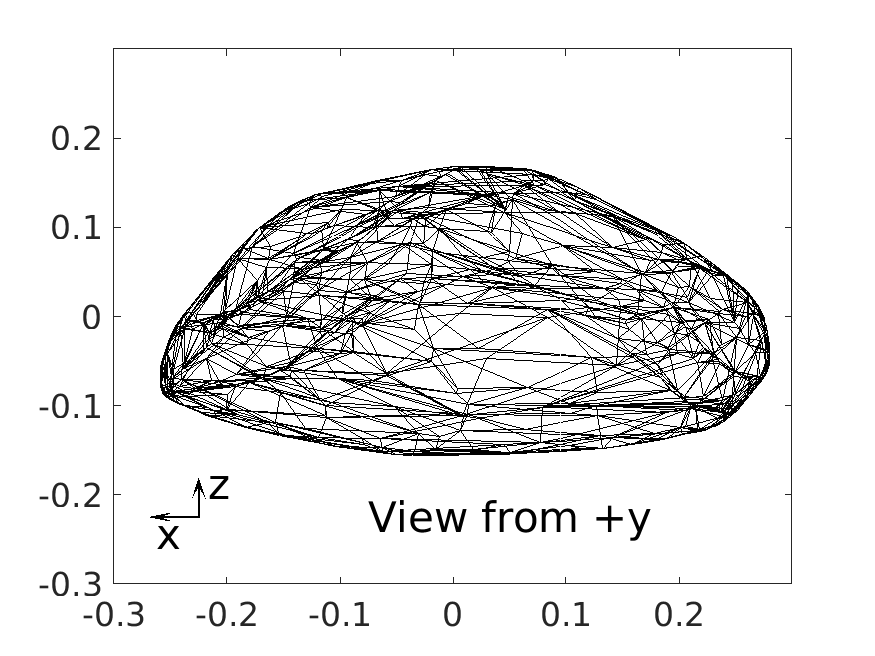}
            \includegraphics[width=0.3\linewidth]{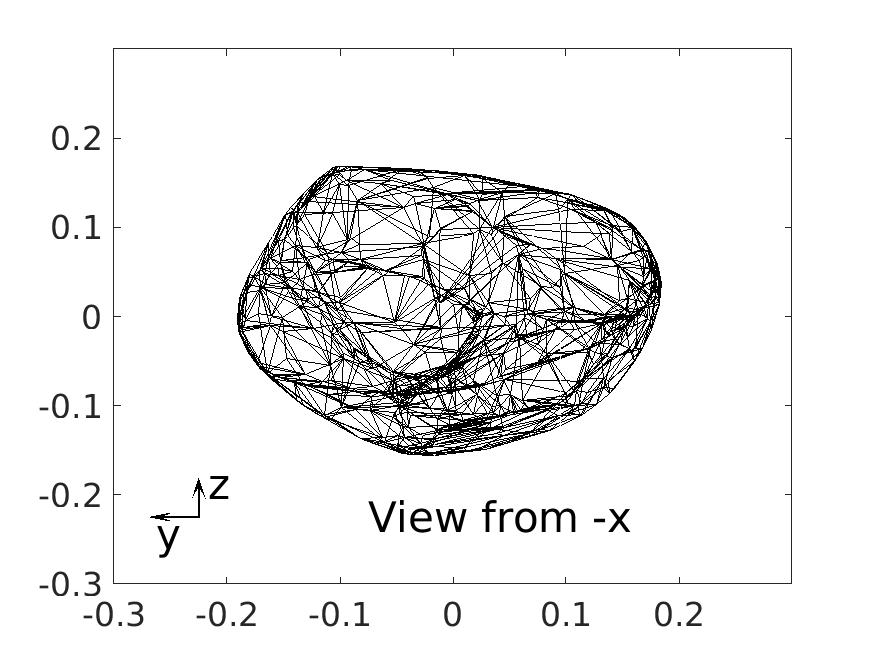}\\
            The shape derived from \citet{brozovic_2018}: 2000 vertices and 3996 faces\\
            \includegraphics[width=0.3\linewidth]{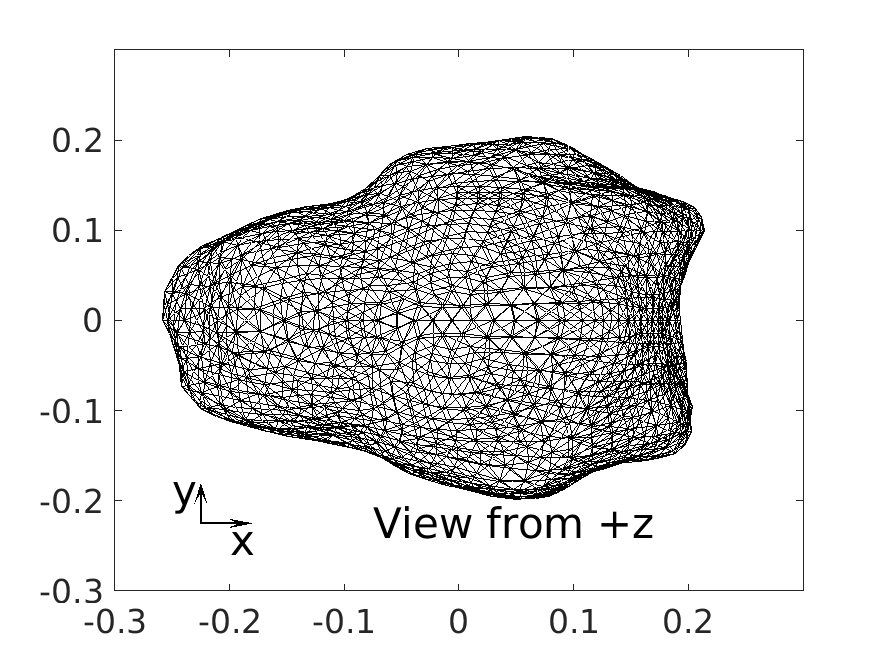}
            \includegraphics[width=0.3\linewidth]{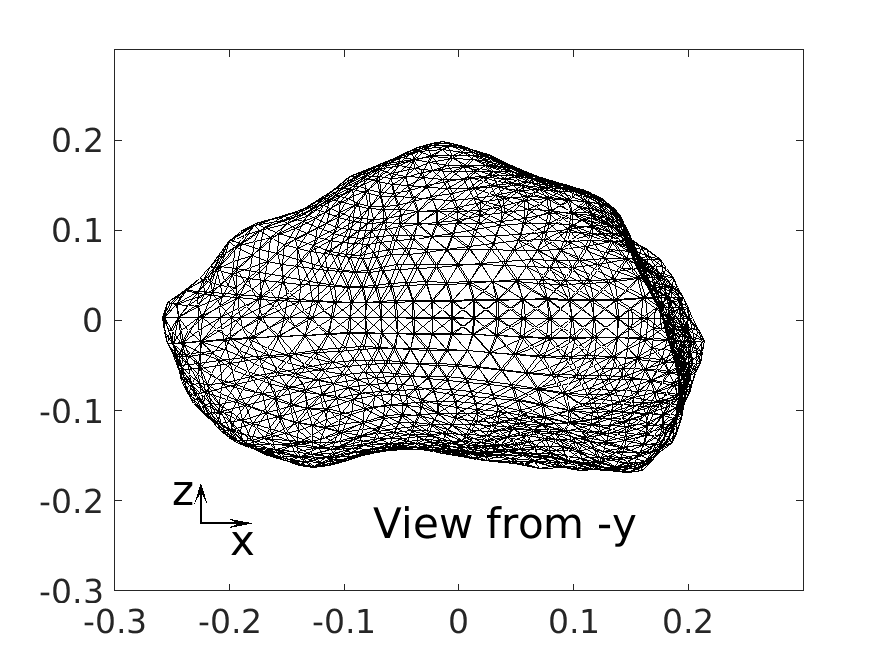}
            \includegraphics[width=0.3\linewidth]{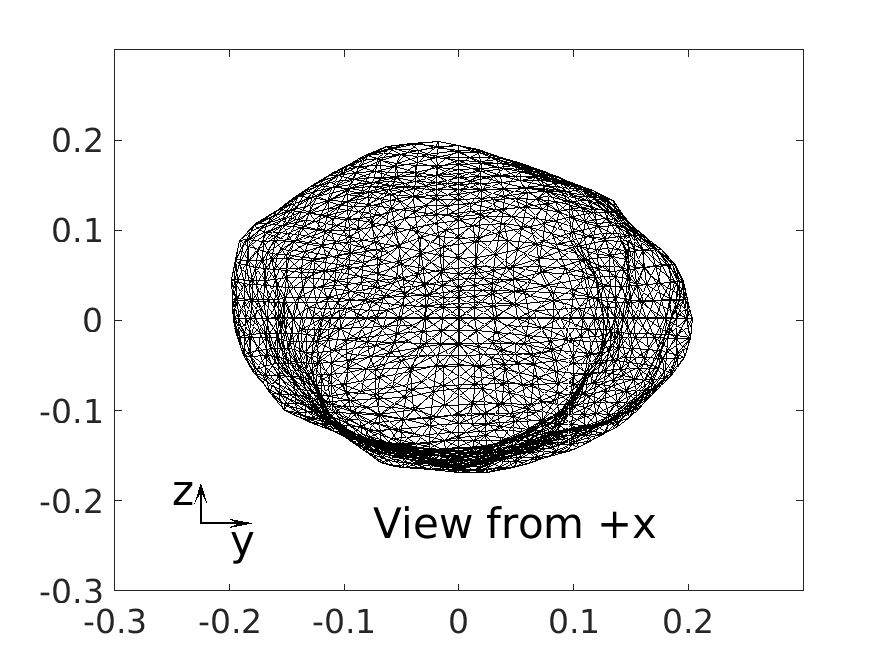}\\   
            \includegraphics[width=0.3\linewidth]{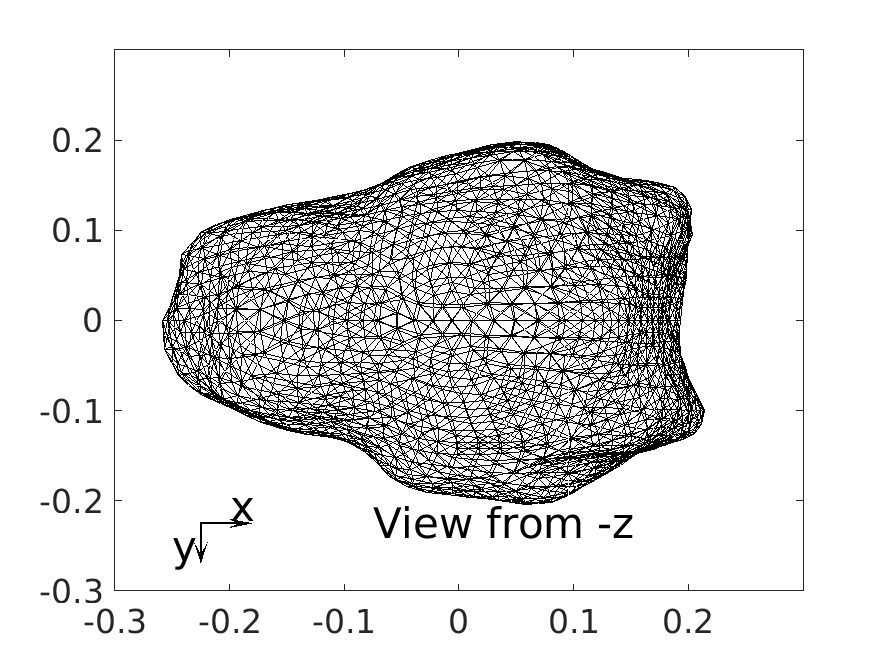}
            \includegraphics[width=0.3\linewidth]{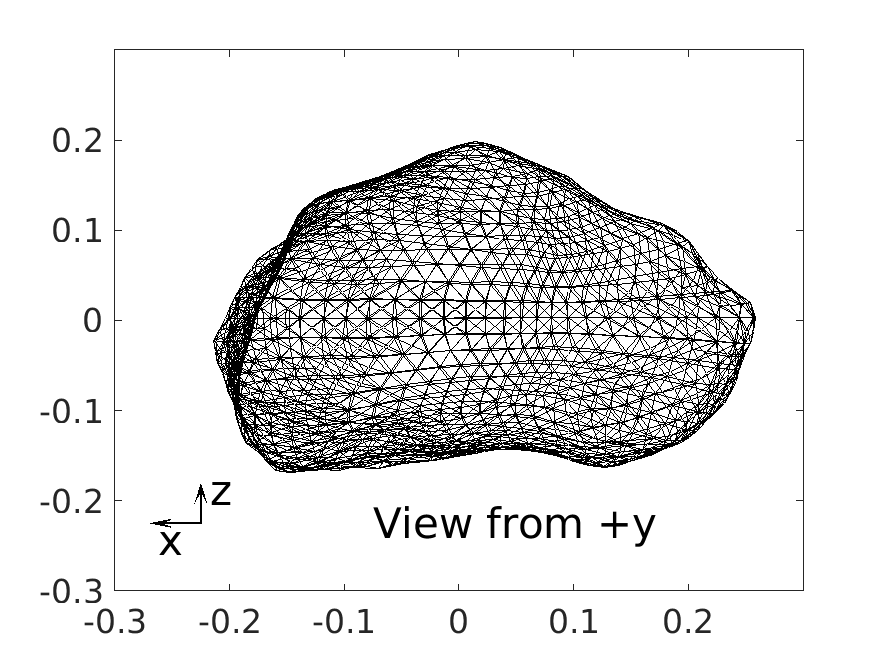}
            \includegraphics[width=0.3\linewidth]{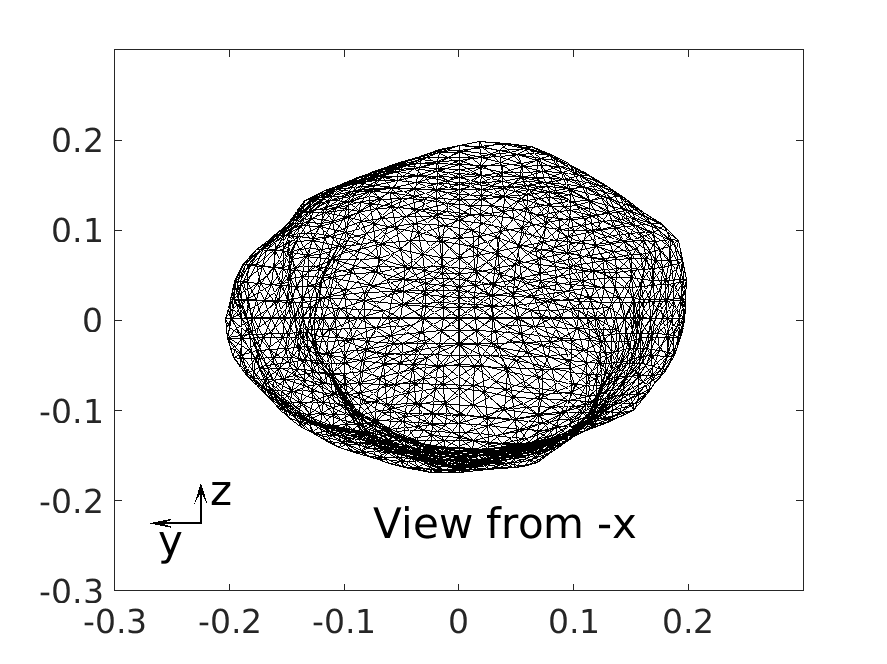}            
            \caption{The polyhedral shape of (99942) Apophis shown in 6 perspective views ($\pm$ x, $\pm$ y, and $\pm$ z) after rescaling the shape with the reported volume-equivalent.} \label{fig01_shape}
         \end{figure}

         Using the algorithm of \citet{mirtich_1996}, we computed the 3D inertia tensor derived from each polyhedral shape of (99942) Apophis with a uniform Itokawa-like density. We found that this tensor is diagonal. This means that the body is perfectly oriented along its principal axes of inertia. The shape of our target is presented in Fig. \ref{fig01_shape}. The overall dimensions of this shape are $(-0.280, 0.259)\times(-0.184, 0.191)\times(-0.156, 0.169)$ km in the x-, y-, and z-directions, respectively, and the polyhedral volume is $0.03034285$ km$^{3}$ (volume-equivalent diameter of 0.387 km).\\   
         
          We then applied the algorithm of \citet{werner_1997} to calculate the spherical harmonic coefficients $C_{n,m}$ and $S_{n,m}$ (up to degree 4, see Table \ref{table02_harmonics}). These coefficients are dimensionless, unnormalized parameters that describe the exterior gravitational potential $U$ of the body. They allow us to verify if the coordinate axes of the polyhedral shape are oriented with its principal axes of inertia. $C_{1,1}$ and $S_{1,1}$ should be zero if the expansion of the gravitational field is fixed around the center of mass; $C_{2,1}$, $S_{2,1}$ and $S_{2,2}$ should be zero if the axes are precisely oriented along the principal axes of inertia \citep{scheeres_2000}. However, we will not use these coefficients in our analyses. Our approach to calculating the exterior gravitational potential around the asteroid employs the polyhedral shape, that will be more accurate than the harmonic coefficients, even if these coefficients were measured up to a degree higher than four.

\input{TABLES/table02_harmonics}

     \subsection{The gravity model}       
      
\begin{figure}[!htp]
            \includegraphics[width=1\linewidth]{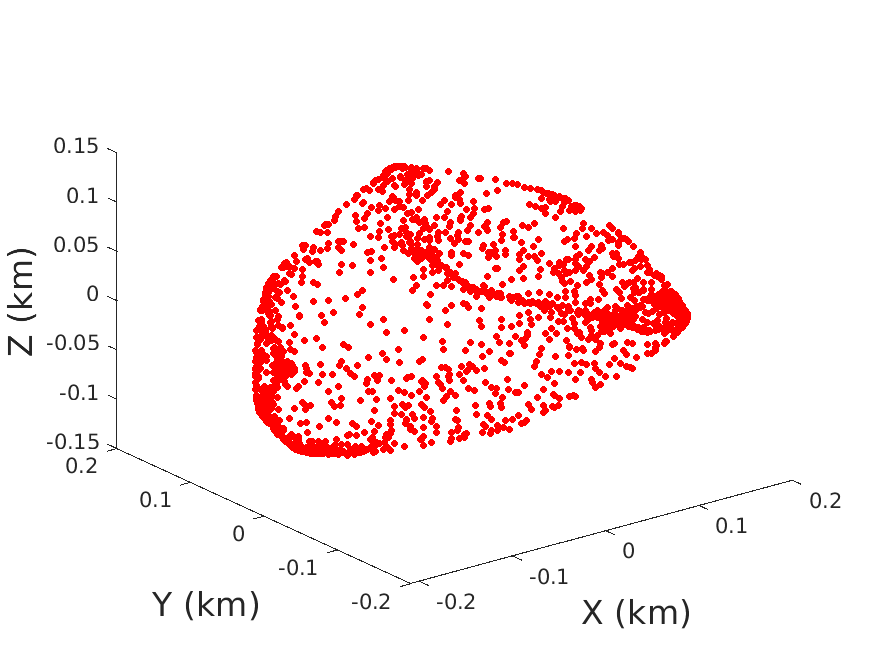}\\
            \caption{Representation of (99942) Apophis modeled by a cloud of 2024 point masses.} \label{fig02_center_tetrahedron}
         \end{figure}

         \citet{werner_1997} derived expressions to precisely evaluate the gravitational potential and acceleration components of a homogeneous polyhedron whose surface consists of a combination of planar triangles. \citet{tsoulis_2001} analyzed the singularities of the potential field caculated by the polyhedron. This method is considered as the best one to describe the gravitational field near or on the surface of a constant density polyhedron \citep{scheeres_1998, scheeres_2000}. Using the classical polyhedron method of \citet{tsoulis_2001}, we compared the polyhedral shape derived from \citet{pravec_2014} and \citet{brozovic_2018} in terms of the computational time and precision of orbit determination close to Apophis. A 60 days circular orbit with mechanical energy of $1.5\times 10^{-9}$ km$^{2}$s${-2}$ at a distance of 1 km from the center the target could be integrated with 463 minutes using the shape of \citet{pravec_2014} and in 884 minutes using the shape of \citet{brozovic_2018}. The total variation distance of this orbit from the central body is 90.30 m in the first case and 88.60 m in the second one. In fact, the difference between the 2 shapes is inversely proportional to the distance from the center. Based on these results, taking into consideration the fact that we did not investigate collisions with Apophis, we think that using the shape of \citet{pravec_2014} with 1014 vertices and 2024 faces is a reasonable approach for the suit of this preliminary study, noting the difference in the execution time. Yet, the main problem of the classical polyhedral approach is the large computational effort, depending on the number of triangular faces chosen. This issue has been reported in \citet{chanut_2015a} and \citet{aljbaae_2017} applying the mascon gravity framework using a shaped polyhedral source, dividing each tetrahedron into several parts \citep{venditti_2013}. Inspired by this last idea, we first calculated the mass of each tetrahedron of the Apophis shape and assigned it to a point mass in the center of the tetrahedron. Thus, we considered the asteroid as a sum of 2024 points that correspond to the number of faces in the shape (Fig. \ref{fig02_center_tetrahedron}).   
   
\begin{figure}[ht]
            \includegraphics[width=1\linewidth]{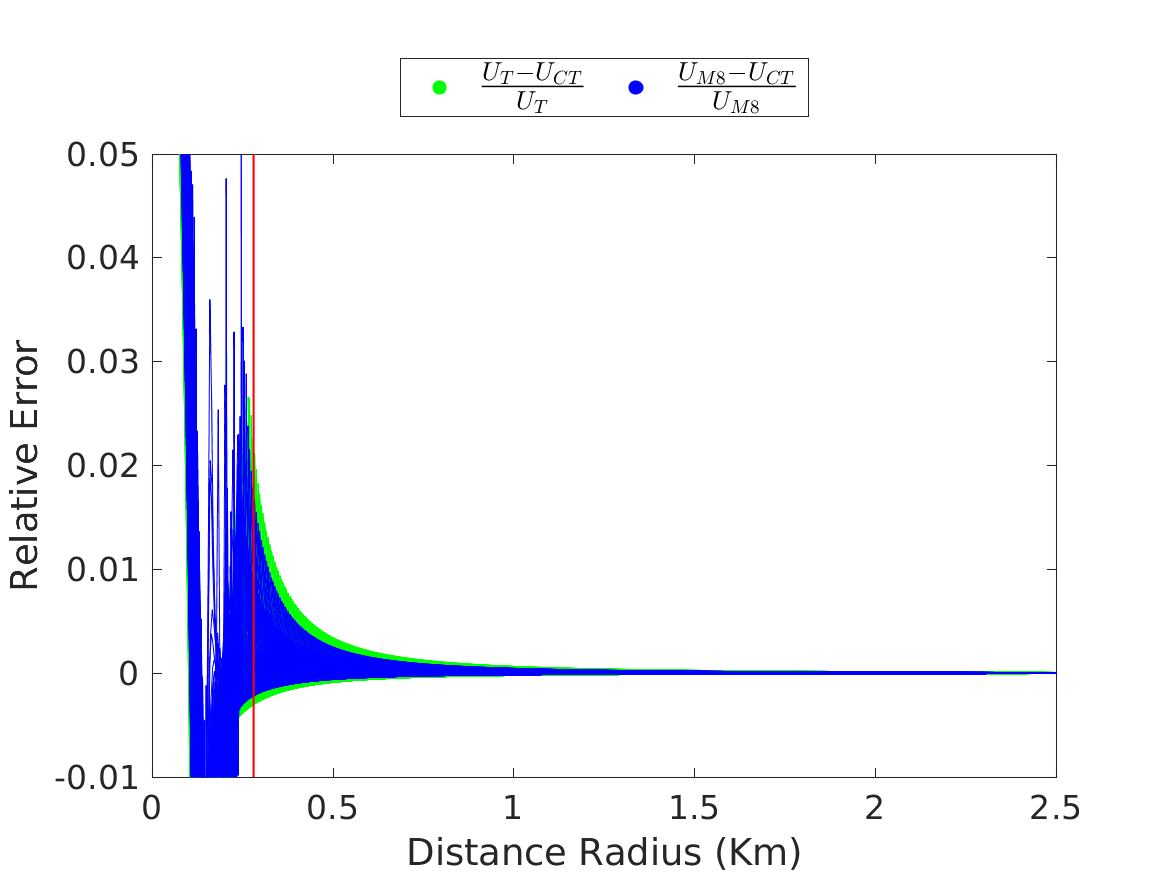}\\
            \includegraphics[width=1\linewidth]{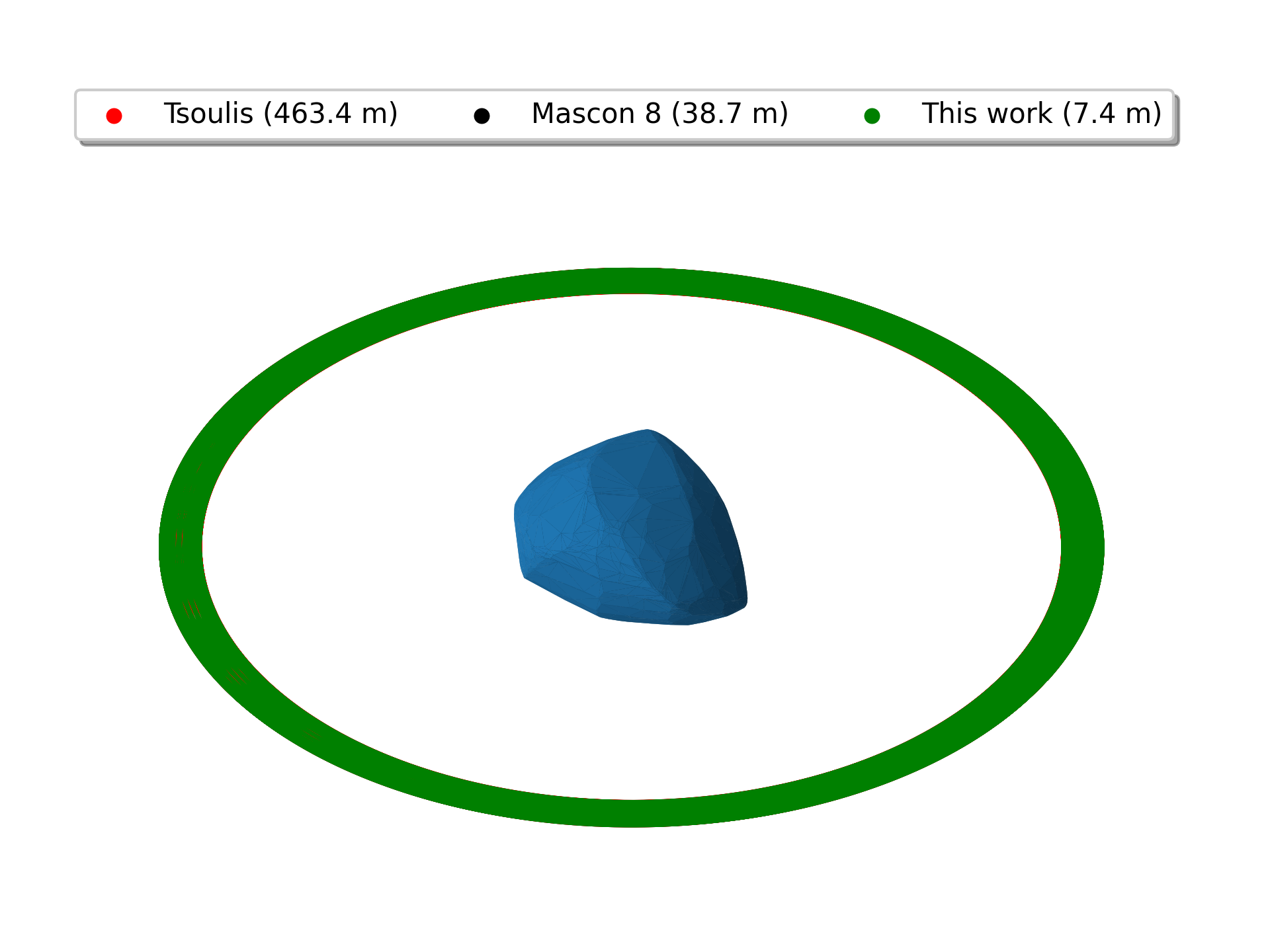}
            \caption{Top: Relative error of the gravitational potential considering the asteroid as a sum of 2024 points in the center of the tetrahedron ($U_{CT}$) with the classical polyhedron method ($U_{T}$) or the model Mascon 8 ($U_{M8}$). Bottom: Three circular orbits around Apophis with the same initial conditions integrated using three methods. The execution time of each orbit is shown in parentheses in the legend} \label{fig03_pot_relativ_error}
         \end{figure}

         To show the efficiency of the method presented here to calculate the gravitational potential, we performed a series of tests comparing the potential, $U_{CT}$, calculated by our method (considering the asteroid as a sum of 2024 points in the center of the tetrahedron) with the classical polyhedron method, $U_{T}$, \citet{tsoulis_2001} and the Mascon gravity approach, dividing the asteroid into 8 layers, $U_{M8}$, \citet{chanut_2017, aljbaae_2017}. In the top panel of Fig. \ref{fig03_pot_relativ_error}, we present the relative errors between $U_{CT}$ and $U_{T}$ or $U_{M8}$, which show that our results are in good agreement with these models outside the body (right side of the red line). In the bottom panel of Fig. \ref{fig03_pot_relativ_error}, we present three circular orbits around Apophis with the same initial conditions integrated using the three methods. The total variation distance from the central body, in each case, are presented in Table \ref{table05_processing_time}. In fact, we tried, in this work, to approach a realistic suite of simulations for motion about Apophis considering the real positions of the planets in our Solar System. To reach the minimum distance Earth-Apophis provided by the JPL's HORIZONS ephemerides, we considered a step-size of 30 seconds in our integration, that makes the use of the classical polyhedron method very heavy in terms of the execution time, as shown in Table \ref{table05_processing_time}. Even with this small step of integration our introduction of the gravitational potential modeling considerably reduced the processing time keeping the accuracy at satisfactory levels.         
         
\begin{figure}[!htp]
            \includegraphics[width=1\linewidth]{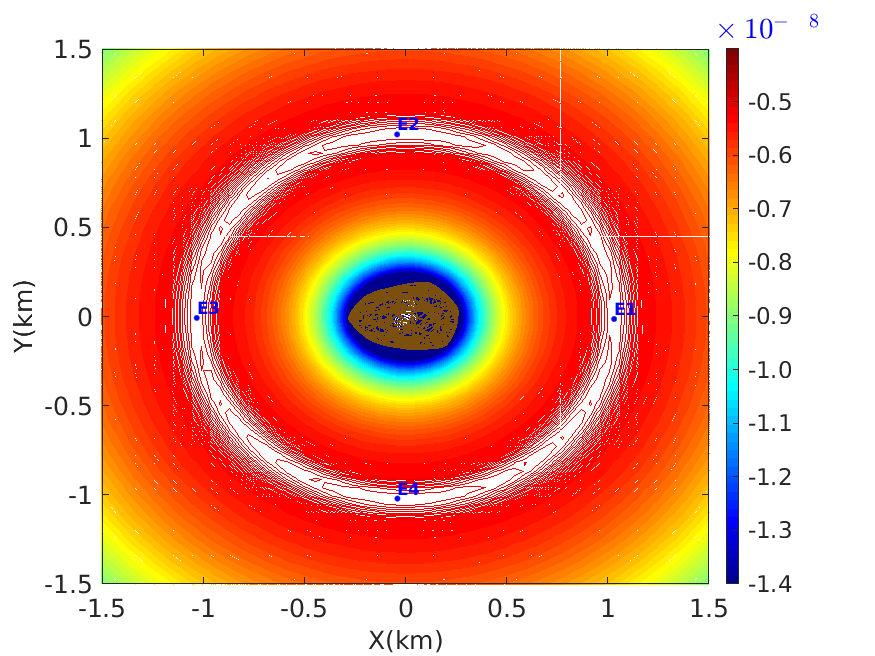}\\
            \caption{Zero-velocity surfaces in Apophis equatorial plane (z = 0) assuming a uniformly rotating 2024 points gravity field.} \label{fig04_zero_velocity}
         \end{figure}

\begin{figure}[!htp]
            \includegraphics[width=0.48\linewidth]{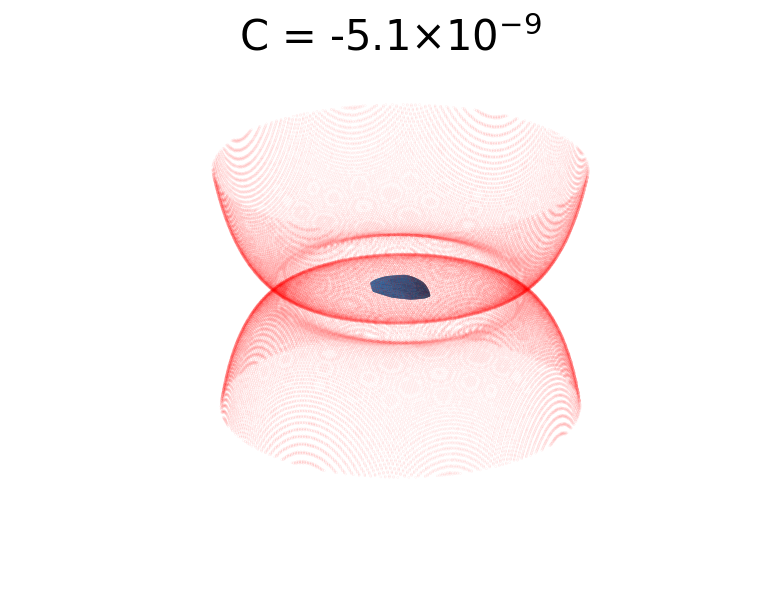}
            \includegraphics[width=0.48\linewidth]{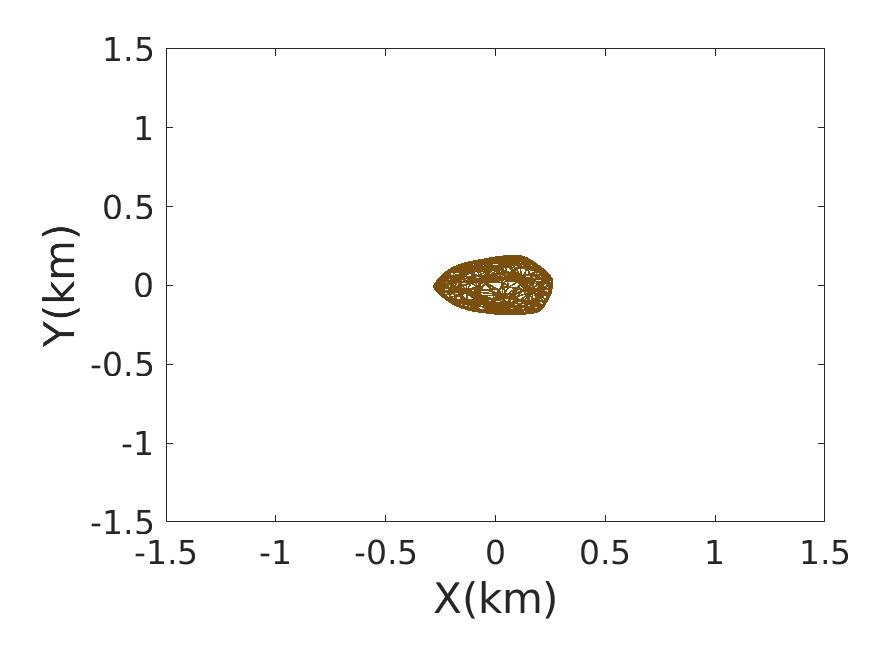}\\
            \includegraphics[width=0.48\linewidth]{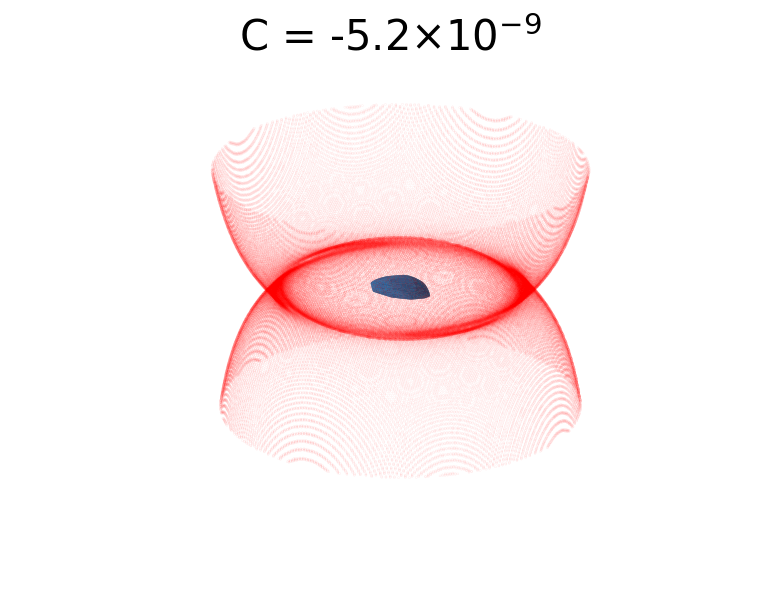}
            \includegraphics[width=0.48\linewidth]{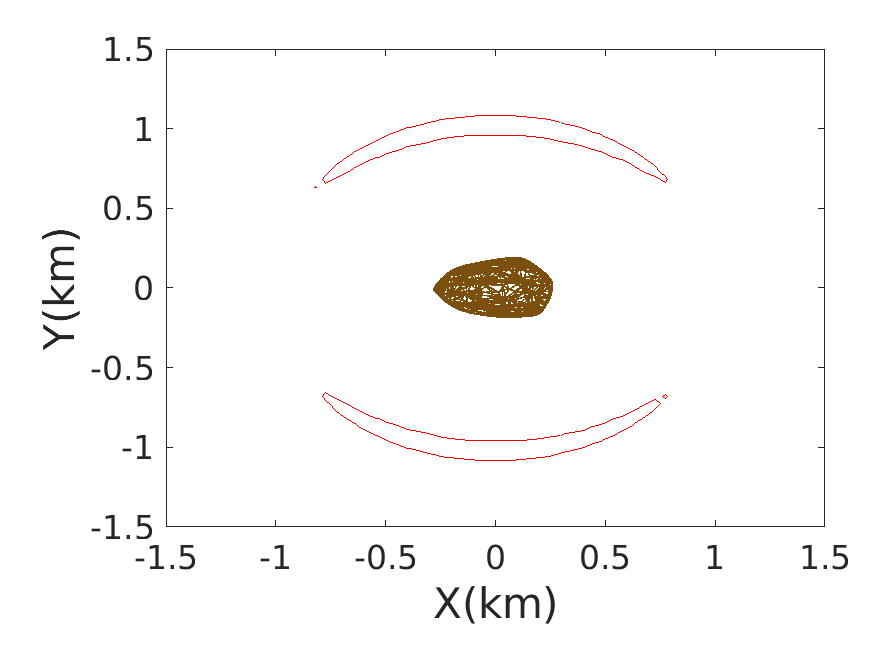}\\
            \includegraphics[width=0.48\linewidth]{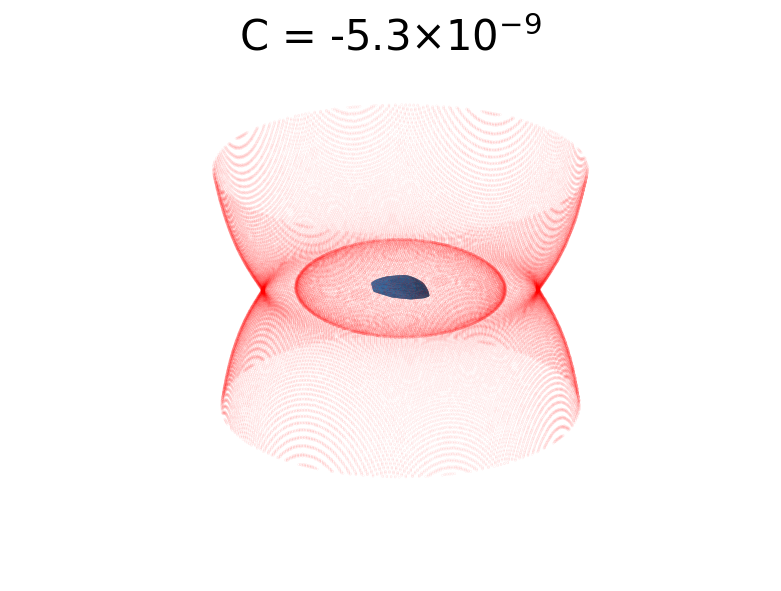}
            \includegraphics[width=0.48\linewidth]{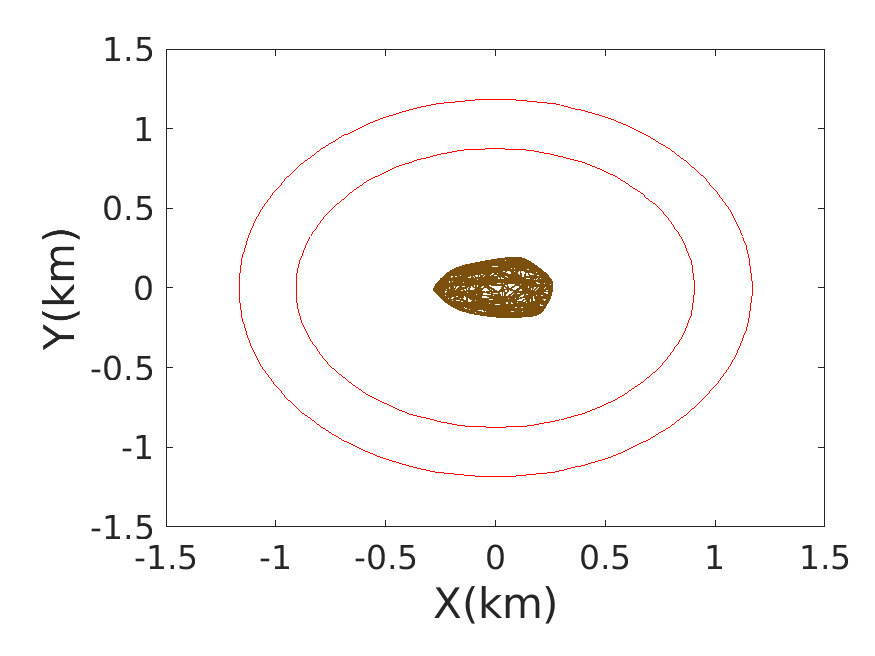}\\
            \includegraphics[width=0.48\linewidth]{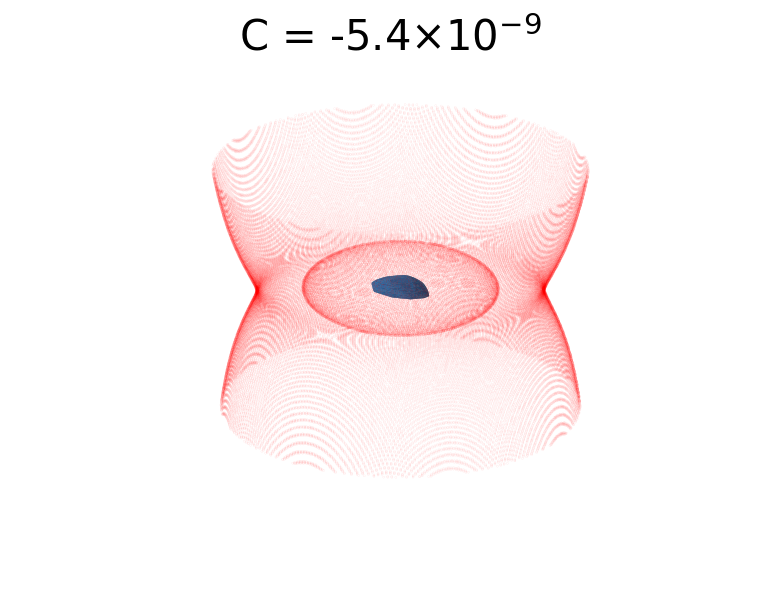}
            \includegraphics[width=0.48\linewidth]{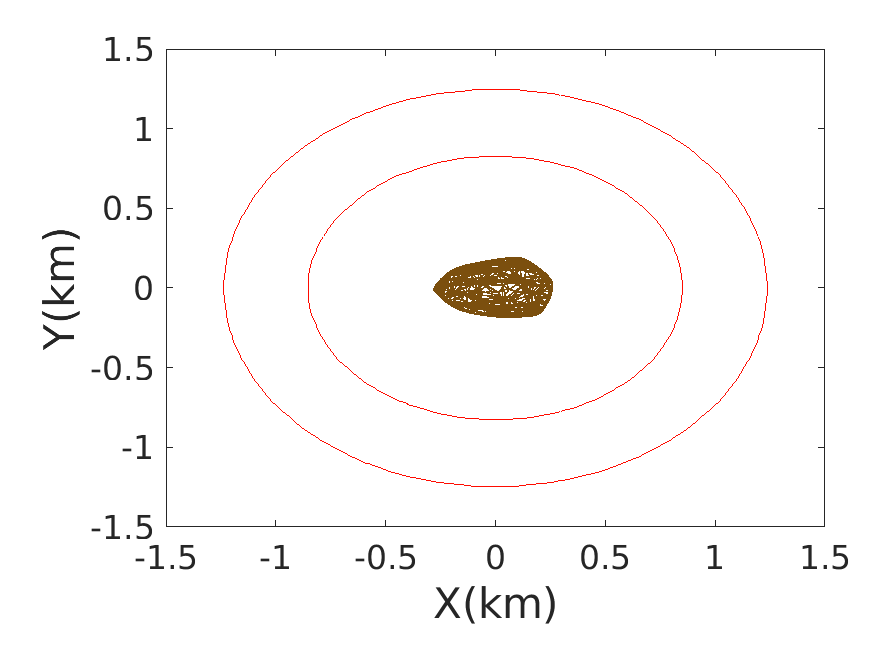}
            \caption{Zero-velocity surfaces around Apophis with different values of the Jacobi constant.} \label{fig04_zero_velocity_3d}
         \end{figure}

         As we already mentioned, Apophis is a tumbling asteroid, by consequence, there are no equilibrium points in the real problem. However, in order to ensure that our model has no deadlocks, we use the three already mentioned above methods to calculate the Zero-velocity curves, neglecting the tumbling state of the target and the existence of other celestial bodies. Generaly speaking, the geometry of these curves has important implications for the stability of the trajectories around the asteroid \citep{yu_2012}. In Figs. \ref{fig04_zero_velocity} and \ref{fig04_zero_velocity_3d} we present our results considering the asteroid as a sum of 2024 points (one point in the center of each tetrahedron). We identified four equilibrium points, two points along the x-axis and two points along the y-axis. The locations of the equilibrium points and the errors (\%) with respect to the classical polyhedron method \citep{tsoulis_2001} are listed in Table \ref{table03_eq_poins}.

\input{TABLES/table03_eq_poins}

         We also examined the stability of the equilibrium points by calculating the eigenvalues of the linearized system (Table \ref{table04_eigenvalues}). We found three pairs of purely imaginary eigenvalues for $E_2$ and $E_4$ points, showing that these solutions are linear stable with the linearized vector field exhibiting a behavior of the kind center x center x center. On the other hand, the linear analysis reveals that $E_1$ and $E_3$ points are unstable equilibria of the kind center x center x saddle, given that we obtained one pair of real eigenvalues with opposite signals and two pairs of purely imaginary eigenvalues for this case. Therefore, according to the classification proposed by \citet{jiang_2014} and \citet{wang_2014}, these solutions belong to Case 1 and 2, respectively. Moreover, our target can be classified as a type-I asteroid, according to the classification proposed by \citet{scheeres_1994}. \\

\input{TABLES/table04_eigenvalues}

         Finally, It is important to recall that, in terms of the execution time on computers Pentium 3.60GHz, our model reduced the computation processing time by more than 95\% compared to the classical polyhedron method, losing less than 2\% of the precision in the tested area (Table \ref{table05_processing_time}). However, this point will be much more important if we need integration for larger times. That motivated us to represent our target as a cloud of 2024 point masses for the rest of this work.
         
         \input{TABLES/table05_processing_time}

   \section{Surfaces of Section}\label{sec03_surfaces_section}

\begin{figure}[ht]
         \includegraphics[width=1\linewidth]{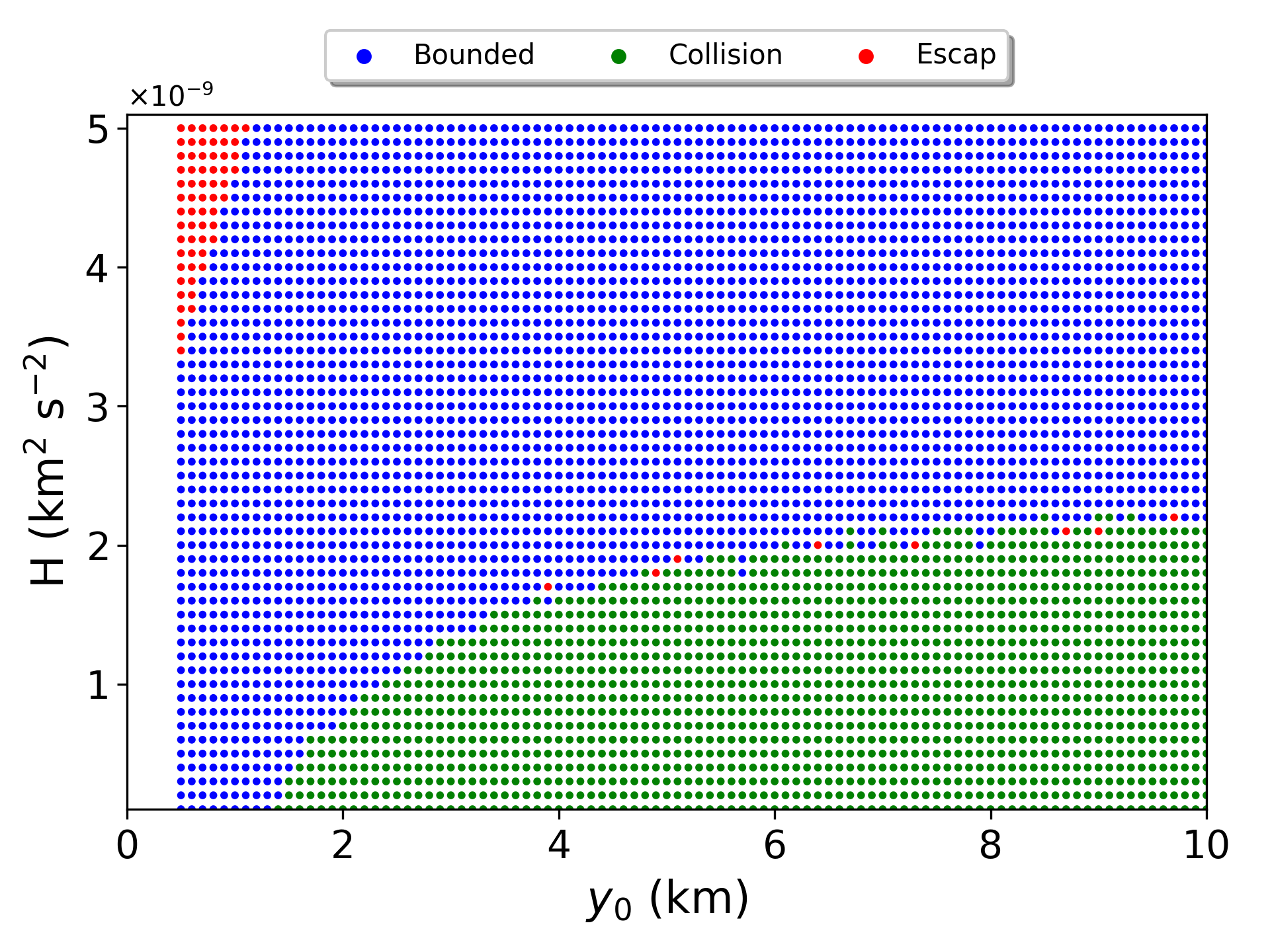}\\
         \caption{Type of orbits around the asteroid (99942) Apophis. Neglecting the perturbations of the planets in our Solar System and the SRP.} \label{fig06_type_orbits}
      \end{figure}

      In this section, we build the surfaces of section related to the potential of Apophis in the body-fixed frame. Our model is similar to that presented in \citet{borderes_2018, jiang_2016}. However, we use the mechanical energy of orbits around our target, as presented in \citet{scheeres_2000, aljbaae_2019}

      \begin{eqnarray*}\label{energies}
         H &=& \frac{1}{2}(\dot{x}^2+\dot{y}^2+\dot{z}^2) - \frac{1}{2} \omega^{2}(x^2 + y^2) -  U  \\
         U &=& +\sum_{i=1}^{2024}\frac{\mathcal{G}m_{i}}{r_{i}} \nonumber
      \end{eqnarray*}
      \textcolor{white}{.}\\
      where: $x,y,z$ and $\dot{x},\dot{y},\dot{z}$ are the location and velocity of the particle in the body-fixed frame of reference. $U$ is the gravitational potential of the asteroid, calculated using one point fixed in the center of each tetrahedron, as explained in the previous section. $\mathcal{G}m_{i}$ is the gravitational parameter of the $i$th. tetrahedron, with $\mathcal{G} = 6.6741 \times 10^{-20}$ km$^{3}$kg$^{-1}$s$^{-2}$. $r_{i}$ is the distance between the center of mass of the tetrahedron and the particle. The equatorial prograde motion of a massless particle around Apophis is determined numerically with the Runge-Kutta 7/8 integrator with variable step size, optimized for the accuracy of $10^{-12}$, covering a maximum of 200 years. We stop our integration after 3000 intersections between the trajectory and the plane $y = 0$. However, this does not necessarily ensure that the nature of all orbits remains unchanged in time, because some orbits may manifest a nonlinear behavior as time goes on. We distributed our initial conditions in the y-axis, with $x_{0} = z_{0} = \dot{y}_{0} = \dot{z}_{0} = 0$ and $\dot{x}_{0}$ was computed according to Eq. \ref{energies}. The values of $y_{0}$ are taken between 0.5 and 10 km from the asteroid center with an interval of 0.1 km. We first consider our target significantly far from any other celestial body, where the motion is dominated by the asteroid own gravitational field. In this section, we also neglected the SRP. The equation of motion used in this analysis is as follows:   
      \begin{eqnarray*}
         \ddot{\text{r}}&=&-2\Omega \times \dot{\text{r}} - \Omega \times ( \Omega \times \text{r})+U_{\text{r}}
         \label{Equations_motion3}
      \end{eqnarray*}
   
      where $\text{r}$ is the coordinate vector of the particle in the body-fixed frame, $\Omega$ is the rotation vector from the uniform rotation of (99942) Apophis, and $U_{\text{r}}$ is the gradient of the gravitational potential of the central body, calculated considering it as a sum of 2024 points (Sect. \ref{sec02_polyhedral_shape}). In Fig. \ref{fig06_type_orbits}, we present the initial conditions that generate orbits escaping from the system (red points), colliding with the central body (green points), and bounded orbits around our target (blue points). The orbit escapes from Apophis system when the distance from the central body becomes 10 times greater than the Apophis Hill sphere (34 km). We considered a relatively high distance to be sure that the orbits beyond this limit will certainly not return back. The collision with the central body occurs when the particle crosses the limit of the polyhedral shape of Apophis using the Computational Geometry Algorithms Library (CGAL\footnote{\href{https://www.cgal.org/}{https://www.cgal.org/}}). We notice that, no collision with the central body occurs for $H > 2.2 \times 10^{-9}$. The escapes from the system occur for an initial $H > 3.4 \times 10^{-9}$ with some particles escaping with $1.7 \times 10^{-9} \leq H \leq 2.2 \times 10^{-9}$. In agreement with \citet{aljbaae_2019, aljbaae_2020}, the particles escape from Apophis system when they are very close to the central body and have sufficient energy.
      
\begin{figure}[!htp]
         \centering
         \includegraphics[width=0.99\linewidth]{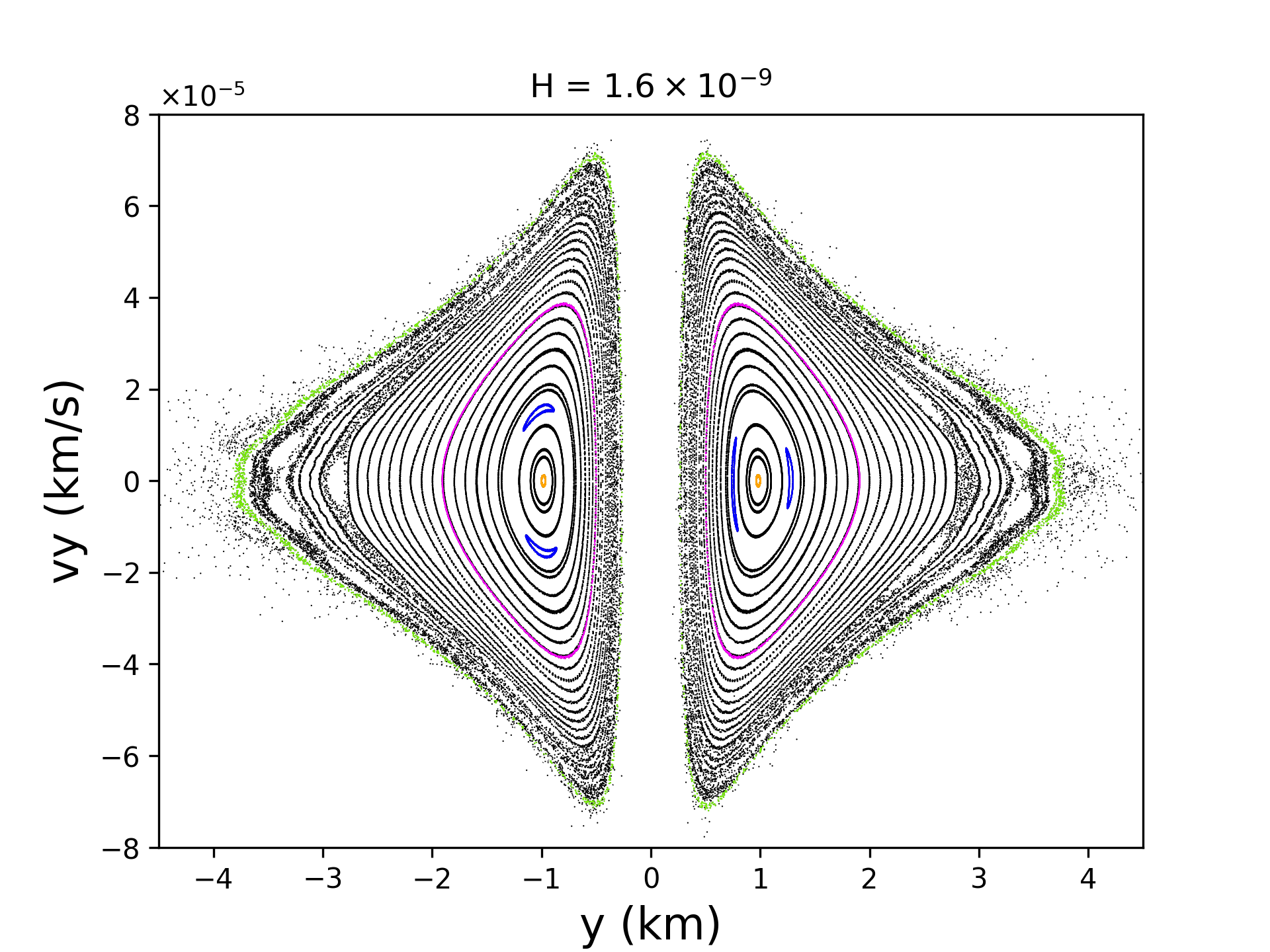}\\
         \includegraphics[width=0.50\linewidth]{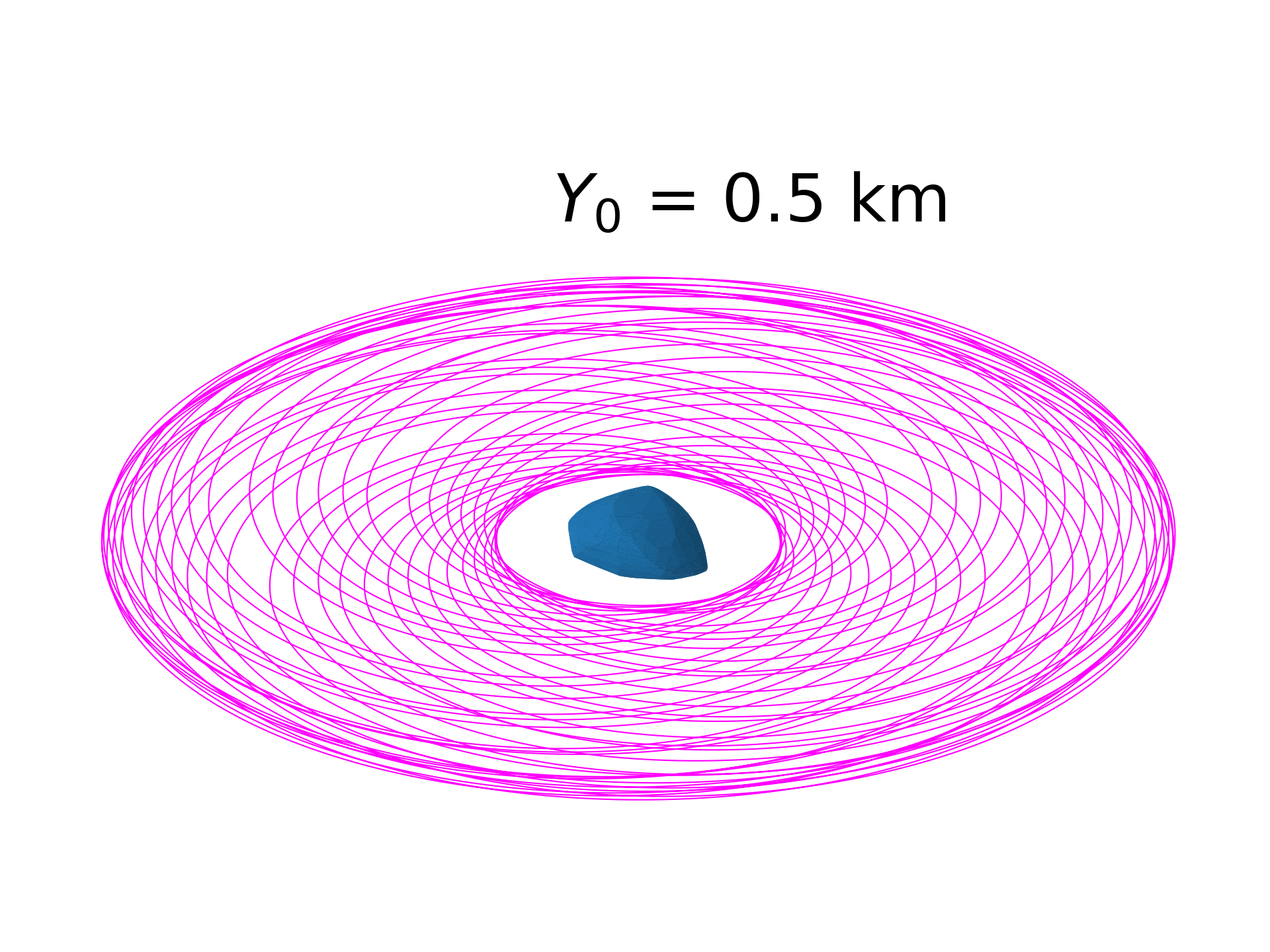}
         \includegraphics[width=0.40\linewidth]{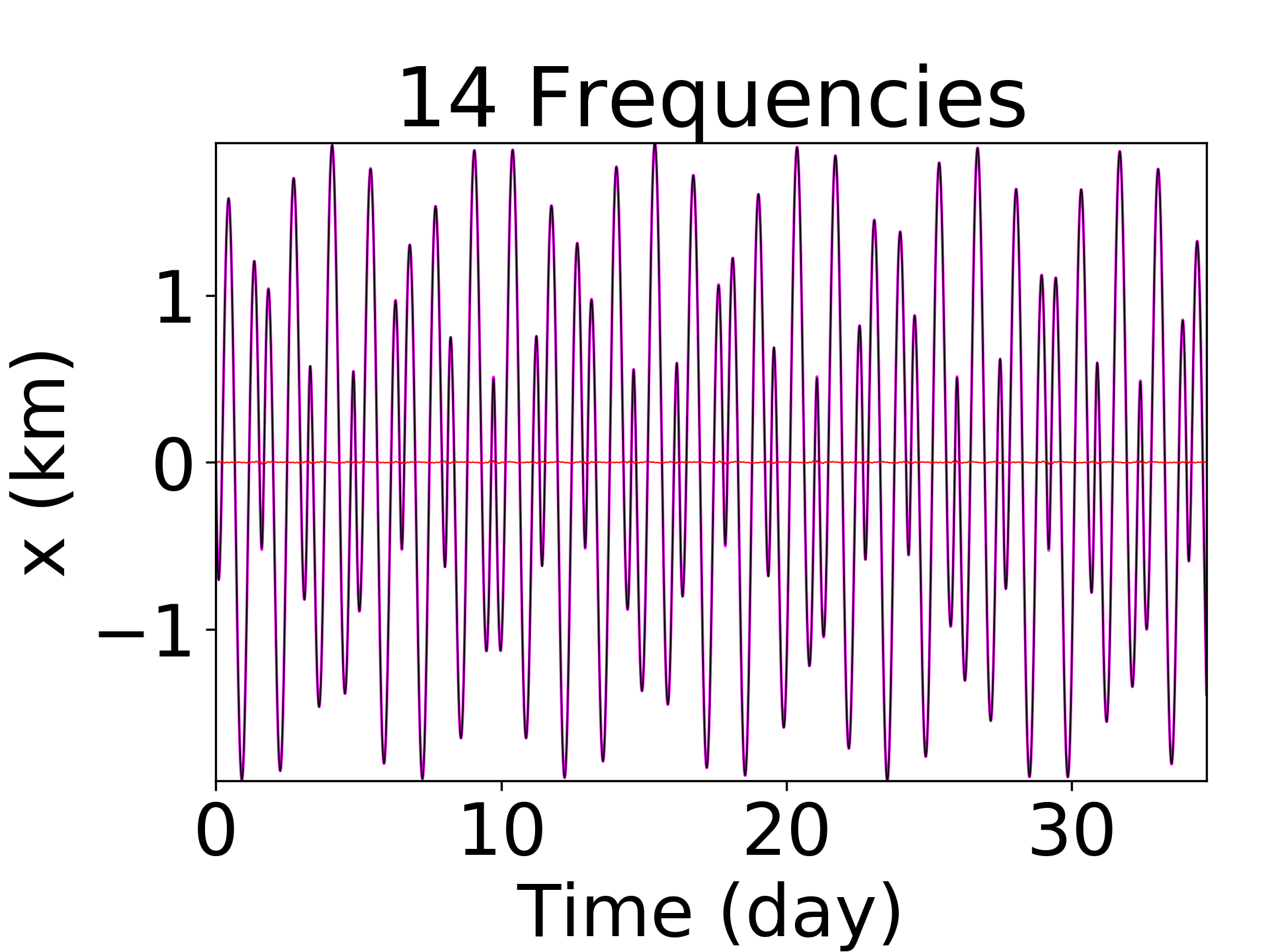}\\
         \includegraphics[width=0.50\linewidth]{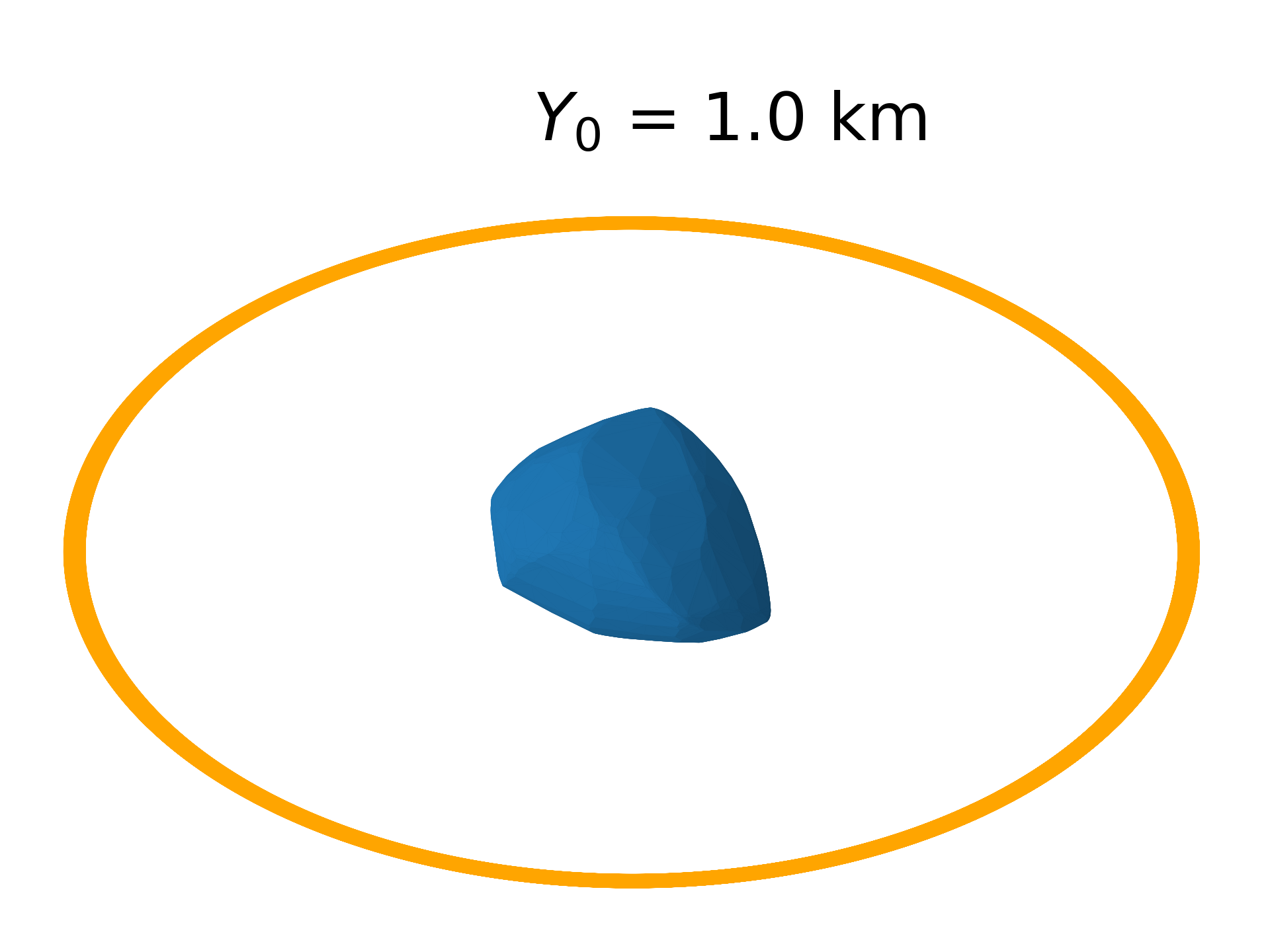}
         \includegraphics[width=0.40\linewidth]{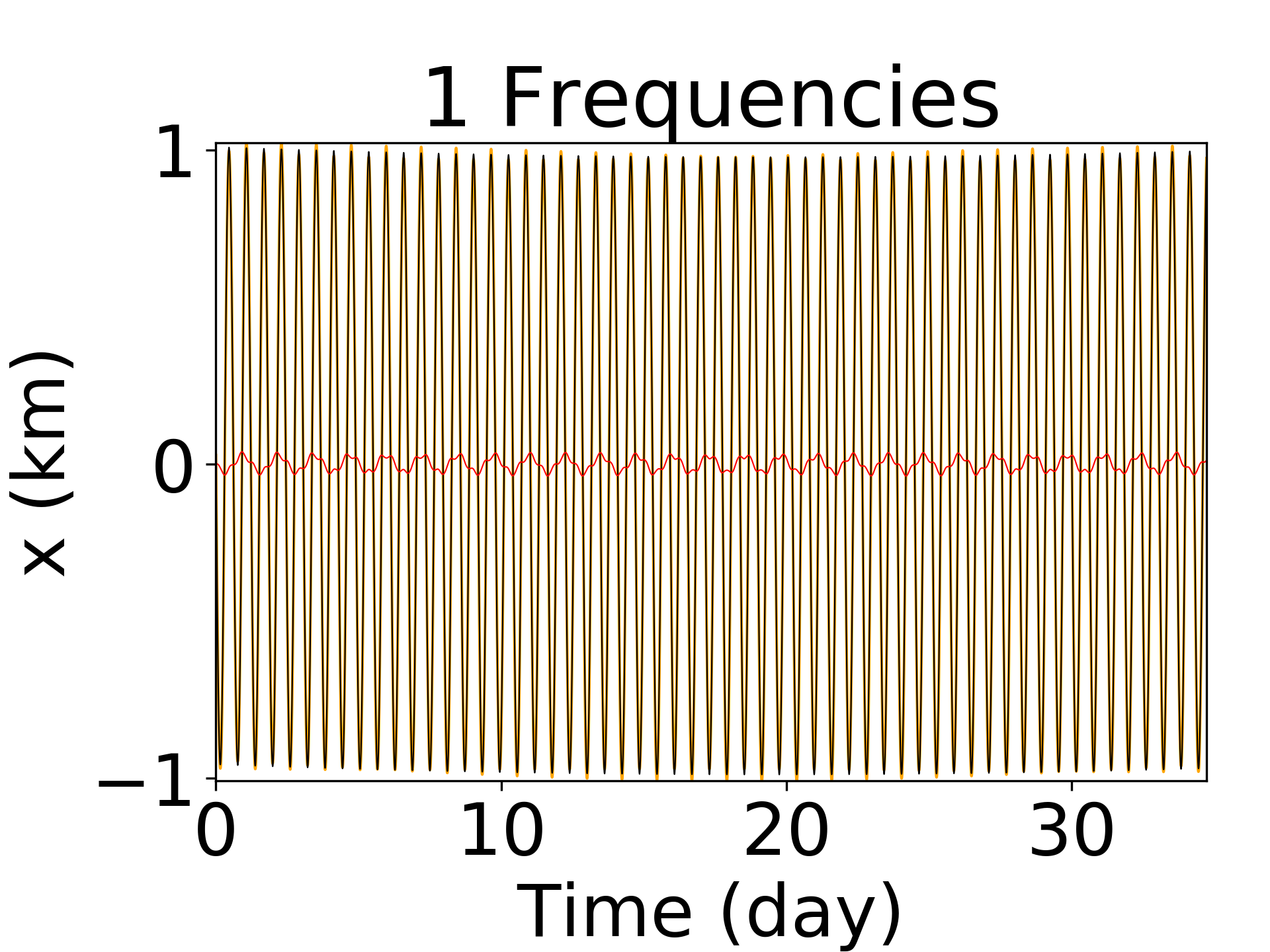}\\
         \includegraphics[width=0.50\linewidth]{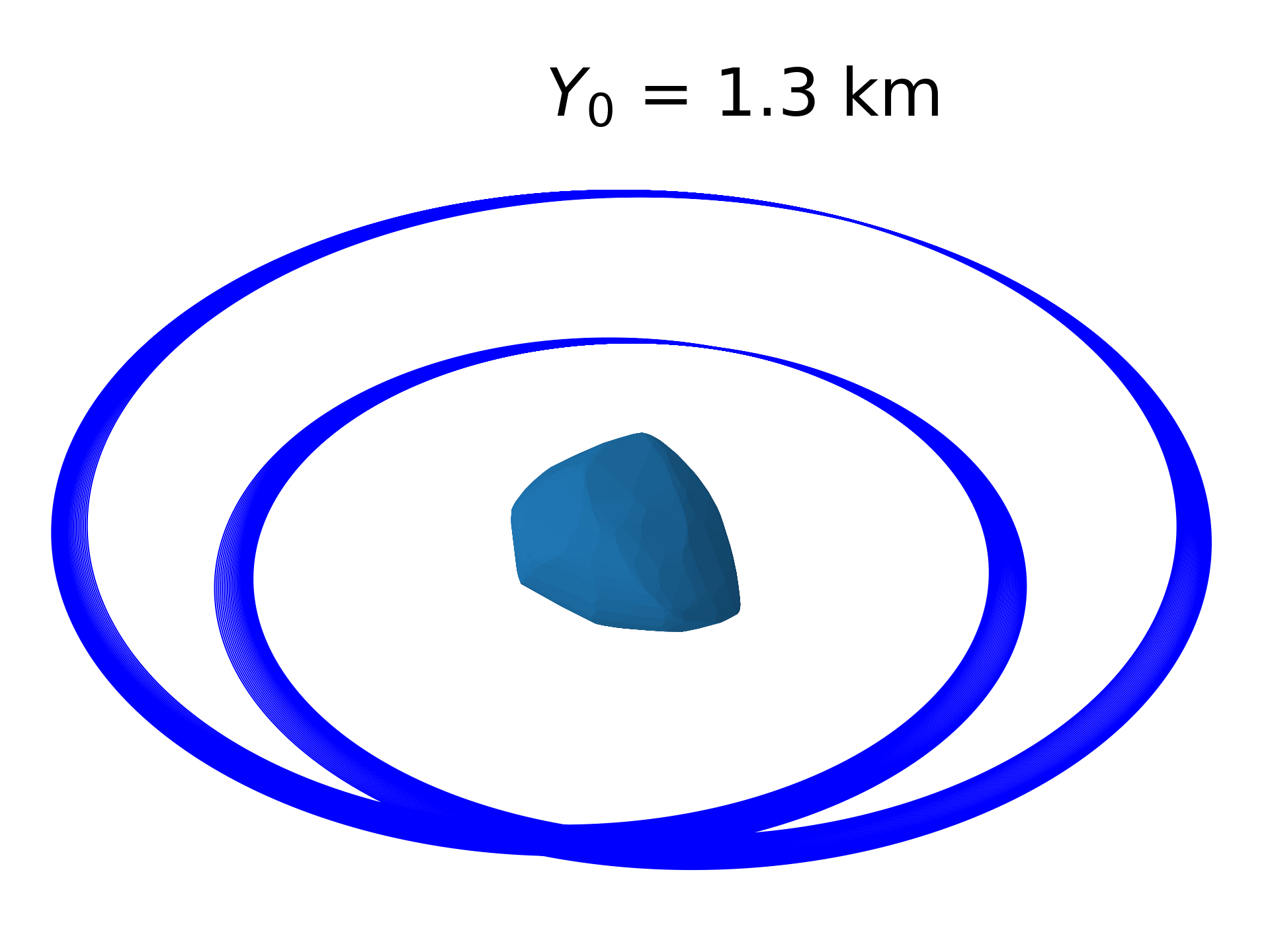}
         \includegraphics[width=0.40\linewidth]{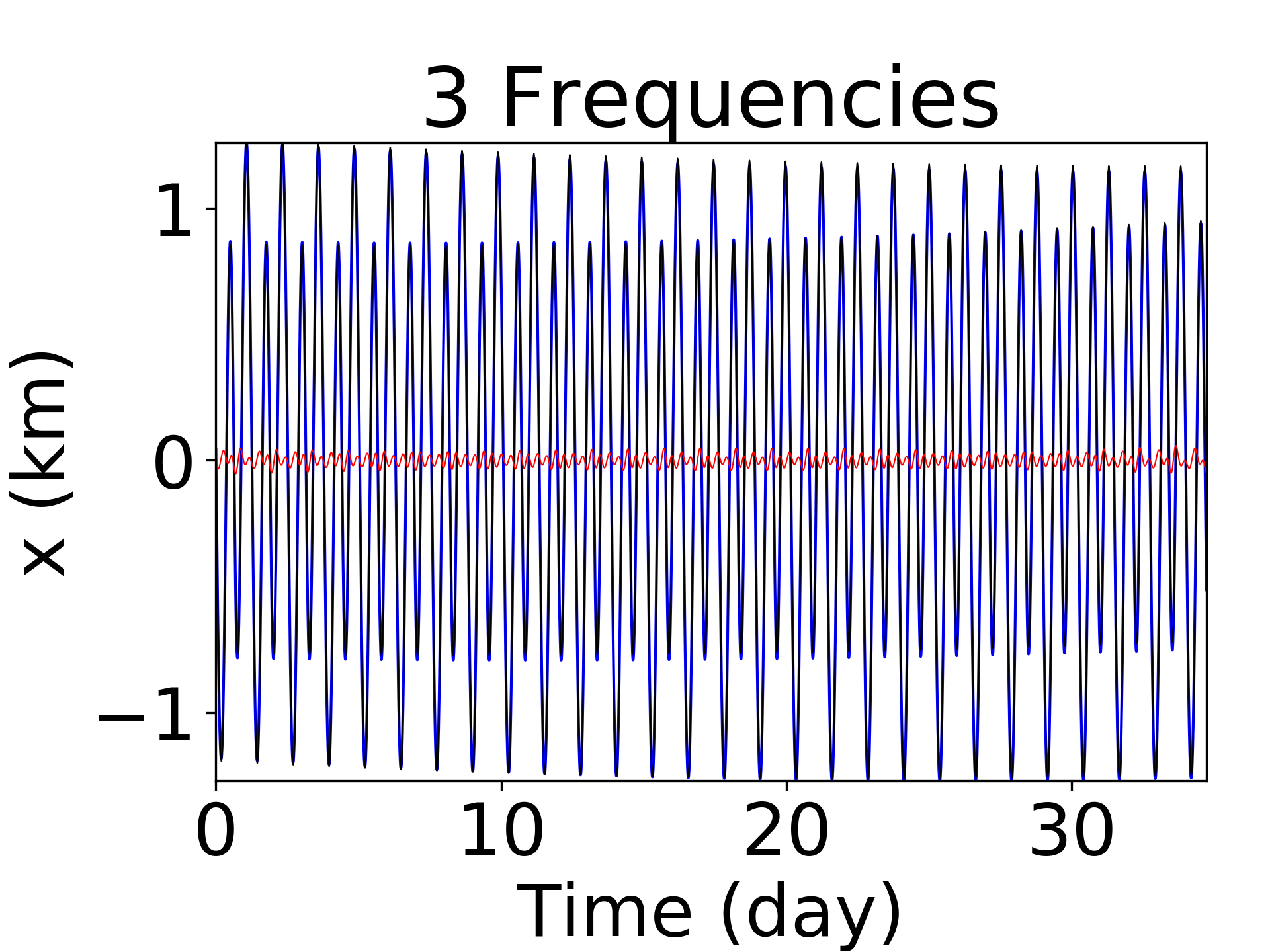}\\
         \includegraphics[width=0.50\linewidth]{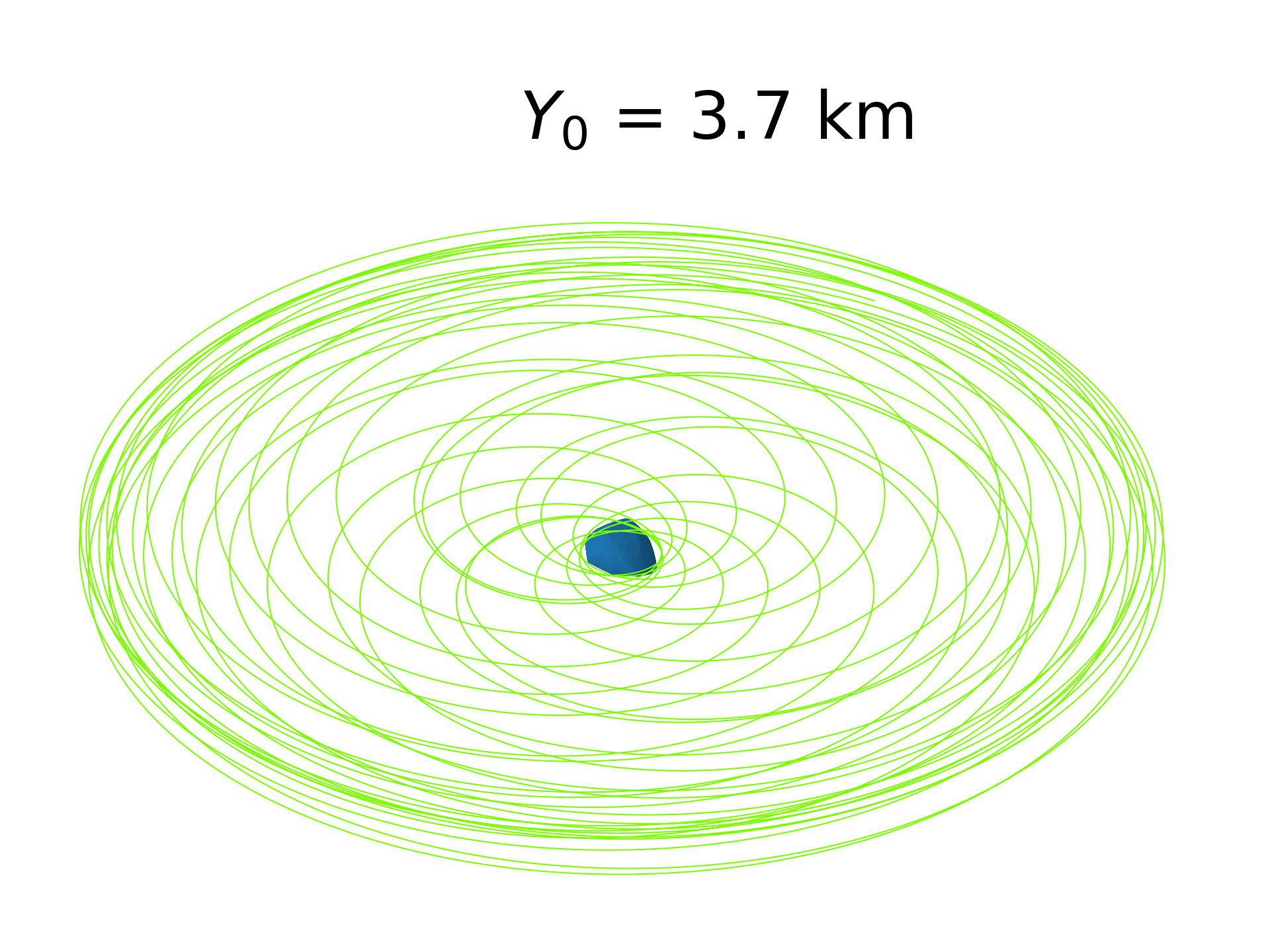}
         \includegraphics[width=0.40\linewidth]{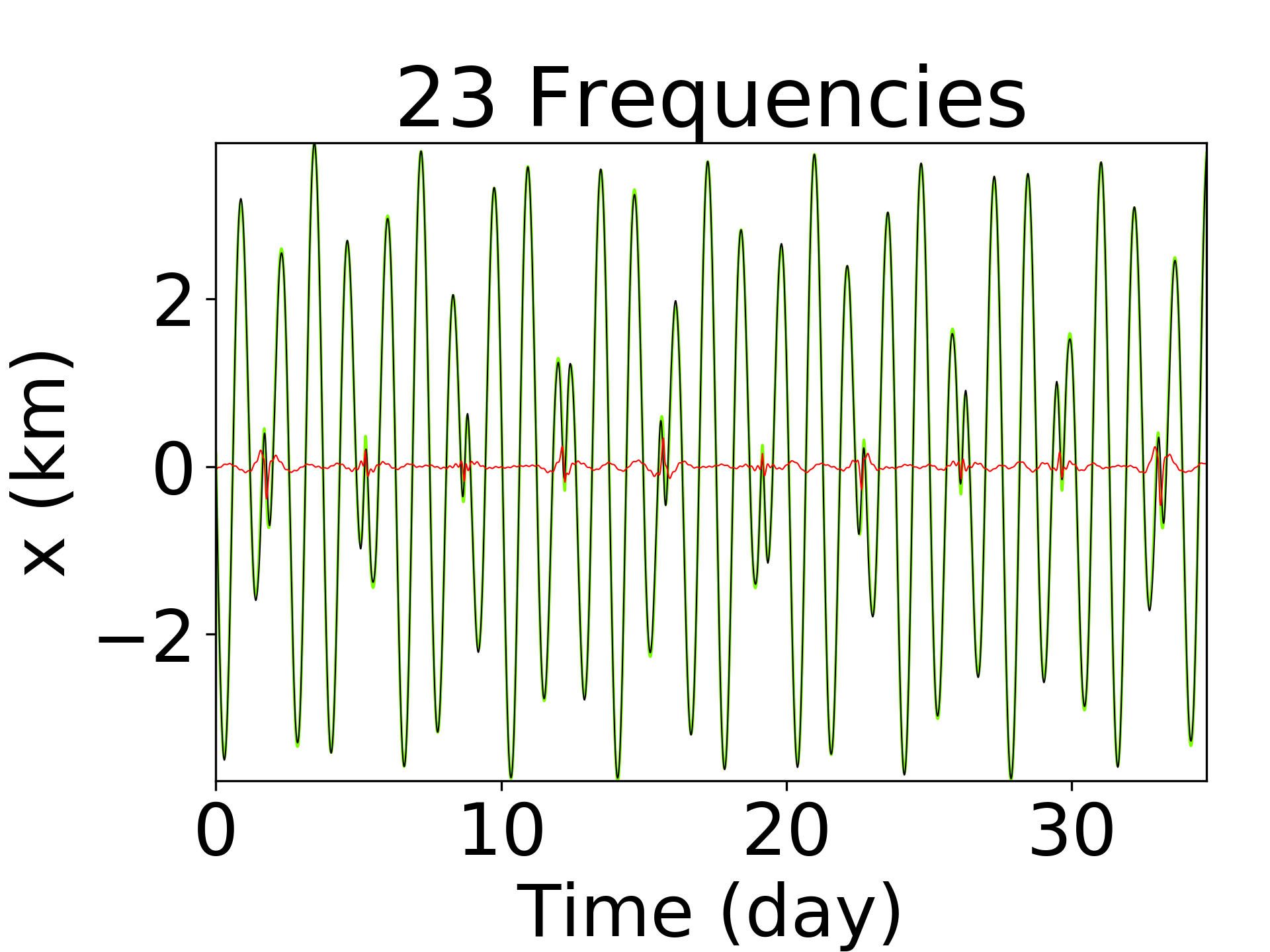}
         \caption{Intersection points of orbits around (99942) Apophis with the Surface of Section $x_{0} = z_{0} = \dot{x}_{0} = \dot{z}_{0} = 0$ and $\dot{y}_{0}$ was computed according to Eq. \ref{energies}. Here, we neglect any perturbation from the remaining bodies in the Solar System, including solar perturbations. } \label{fig07_poincare}
      \end{figure}

      An example of our results for $H=1.6\times 10^{-9}$ is presented in Fig. \ref{fig07_poincare}. This plot gives quick overview of the orbital structure. We can easily distinguish between regular and chaotic motion. The isolated points, for instance, represent chaotic orbits, while areas with no points represent areas that are not reachable by any orbit. When we have one point in the Surface of Section, we get a periodic orbit, while a quasi-periodic motion is depicted by a closed curve. In fact, the period of an orbit is not defined by the Surface of Section itself, because the number of dots depend on which section is chosen. In Fig. \ref{fig07_poincare}, we also present some orbits in the Apophis system. To characterize the periodicity of these orbits, we perform a frequency analysis of the x, y, and z-coordinate of each orbit and determine the leading frequencies. For that purpose, we first evaluated and removed the quadratic variation of the coordinate of the form $$\alpha + \beta t + \gamma t^{2}.$$ Then, we use the Fast Fourier Transform (FFT) to determine the leading frequencies. Our analysis is well adapted for dense polynomials using the software TRIP\footnote{\href{https://www.imcce.fr/Equipes/ASD/trip/trip.php}{https://www.imcce.fr/Equipes/ASD/trip/trip.php}}, developed at the IMCCE-Paris Observatory by \citet{gastineau_2011}. For the sake of clarity and to limit the computational time cost, we restrict our analysis to the first 30 days of integration time. We find complex periods and amplitudes representing our signal. Finally, we fit a nonlinear regression approach to model our signal using the Least-Square method, following an expression combining Fourier-type and Poisson-type components in the form:
      
      \begin{eqnarray*}\label{fit}
         x(t) = \sum_{i=1}^{N} & \bigg[ & A_{i}\sin(f_{i}t) + B_{i}\cos(f_{i}t) + \\\nonumber
                            &        & C_{i}t \sin(f_{i}t) + D_{i}t \cos(f_{i}t)\bigg].
      \end{eqnarray*}
      
      For the sake of simplicity and completeness, we present our results on the x-coordinate in the right-hand side of Fig. \ref{fig07_poincare}. We can observe a good quality of the fit resulting from the least-squares-analysis. The flat red curves represent the residuals obtained after subtracting a combination of the sinusoids (black curves) from the original signal. These sinusoids are listed in Table \ref{table06_sinusoids}. For more details about our frequency analysis, the interested reader can refer to \citet{aljbaae_2012} and \citet{aljbaae_2013}. Overall, our frequency analysis perfectly determines the orbits around our target. Our results consist of different mechanical energy gathered in the CPM-Asteroid database (Close Proximity Motion relative to an Asteroid). Moreover, we include in this database the influence of close approach with our planet, witch is the subject of the next section.

\input{TABLES/table06_sinusoids}

\begin{figure}[!htp]
         \centering
         \includegraphics[width=0.99\linewidth]{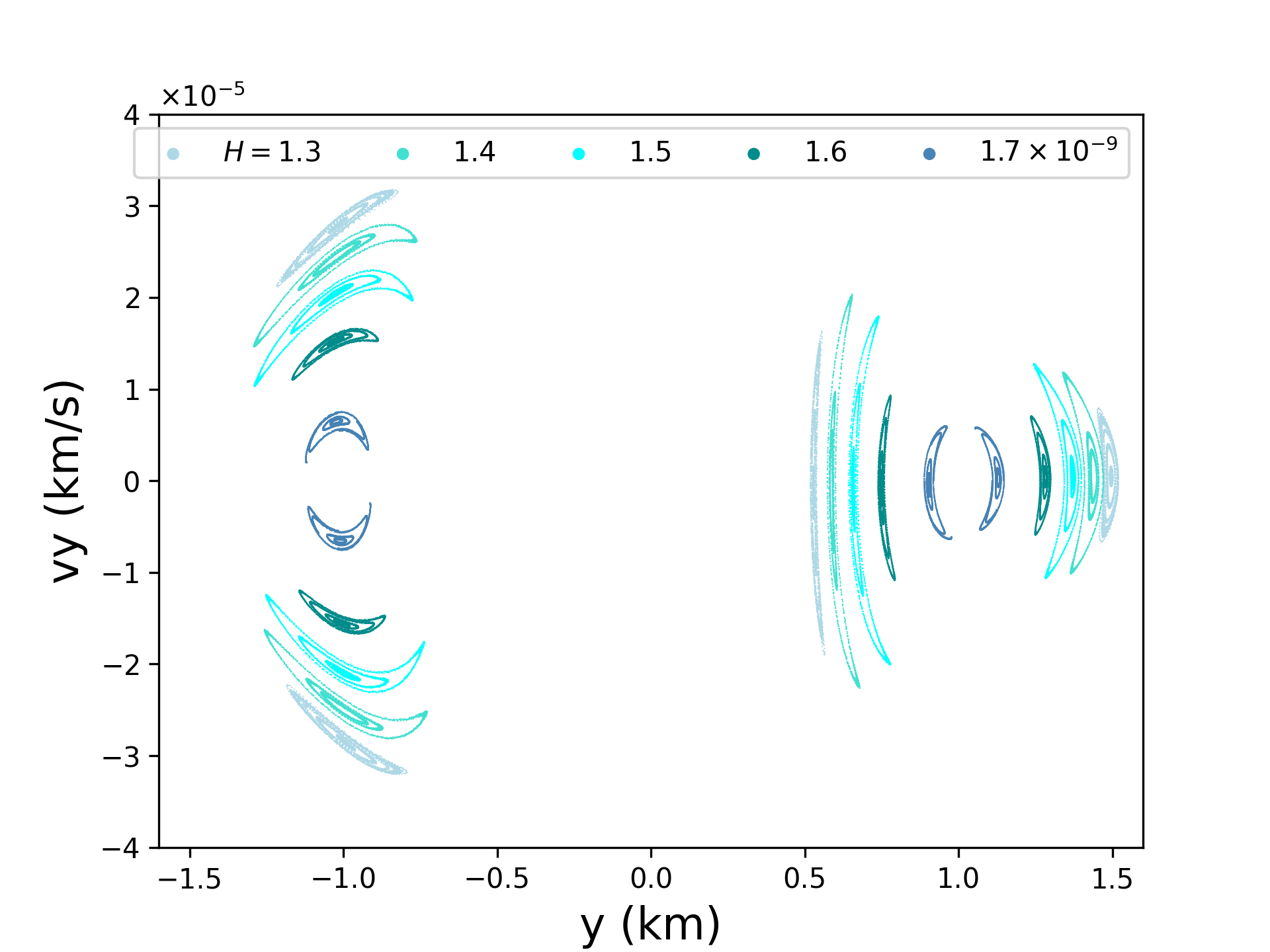}\\
         \caption{The evolution of the Dual quasi-period periodic island with H in the Surface of Section.} \label{fig08_dual_quasi_period}
      \end{figure}

      In the top panel of Fig. \ref{fig07_poincare}, we notice the existence of an island of a dual quasi-periodic response (blue closed curve). The evolution of this island in the Surface of Section is presented in Fig. \ref{fig08_dual_quasi_period}.
      
   \section{Close approach with Earth}\label{sec04_CE}
\begin{figure}[!htp]
         \centering
         \includegraphics[width=0.99\linewidth]{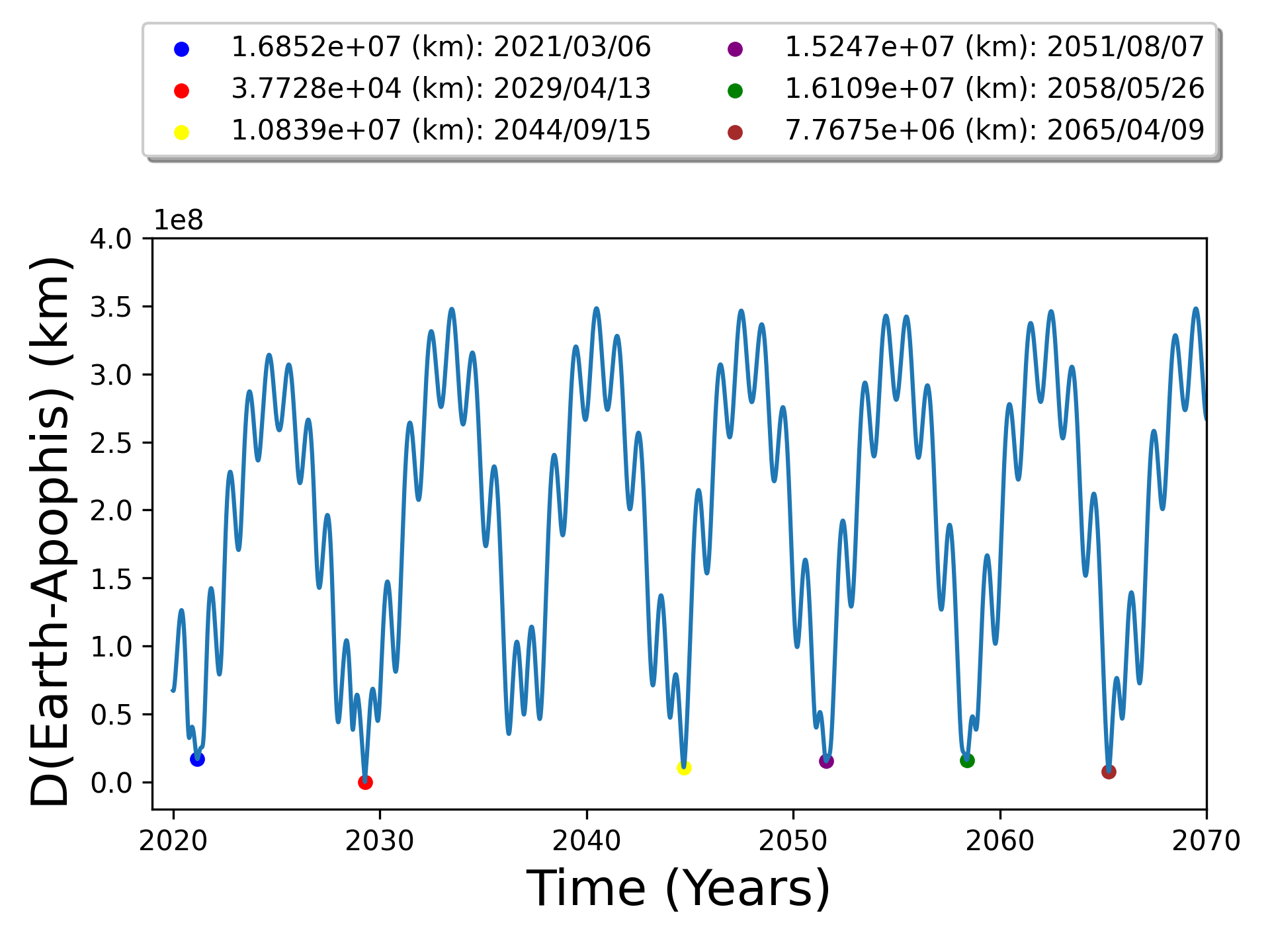}\\
         \caption{The close approach (Apophis-Eearth) provided by our numerical integration using Runge-Kutta methods with variable step size.} \label{fig09_CE_apophis}
      \end{figure}

      In this section, we study the dynamical system around (99942) Apophis during the close approach with our planet at $\sim$ 38,000 km on April 13$^{\text{th}}$ 2029 (Fig. \ref{fig09_CE_apophis}). For that purpose, we re-build the surfaces of section as presented in the previous section, with the difference that here we take into account the gravitational perturbations of the Sun, the 10 planet size bodies of our Solar System, including the Moon and Pluto, and the biggest 3 asteroids (Ceres, Pallas, and Vesta). \\   
   
\begin{figure}[!htp]
         \centering
         \includegraphics[width=0.99\linewidth]{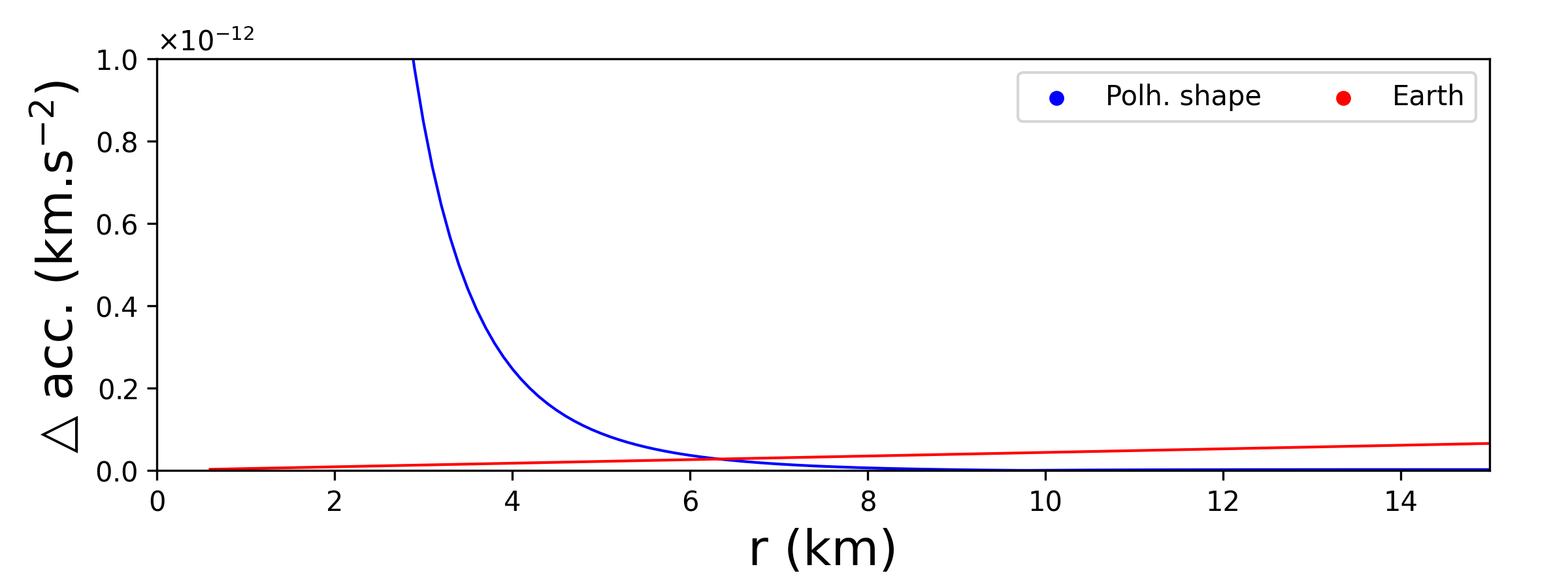}\\
         \caption{The perturbation on the acceleration of a spacecraft close to (99942) Apophis, due to the perturbation of the polyhedral shape of the central body (Blue) and due to the Earth (Red), 10 days before the close approach.} \label{fig10_perturbation_earth}
      \end{figure}

\begin{figure}[!htp]
         \centering
         \includegraphics[width=0.99\linewidth]{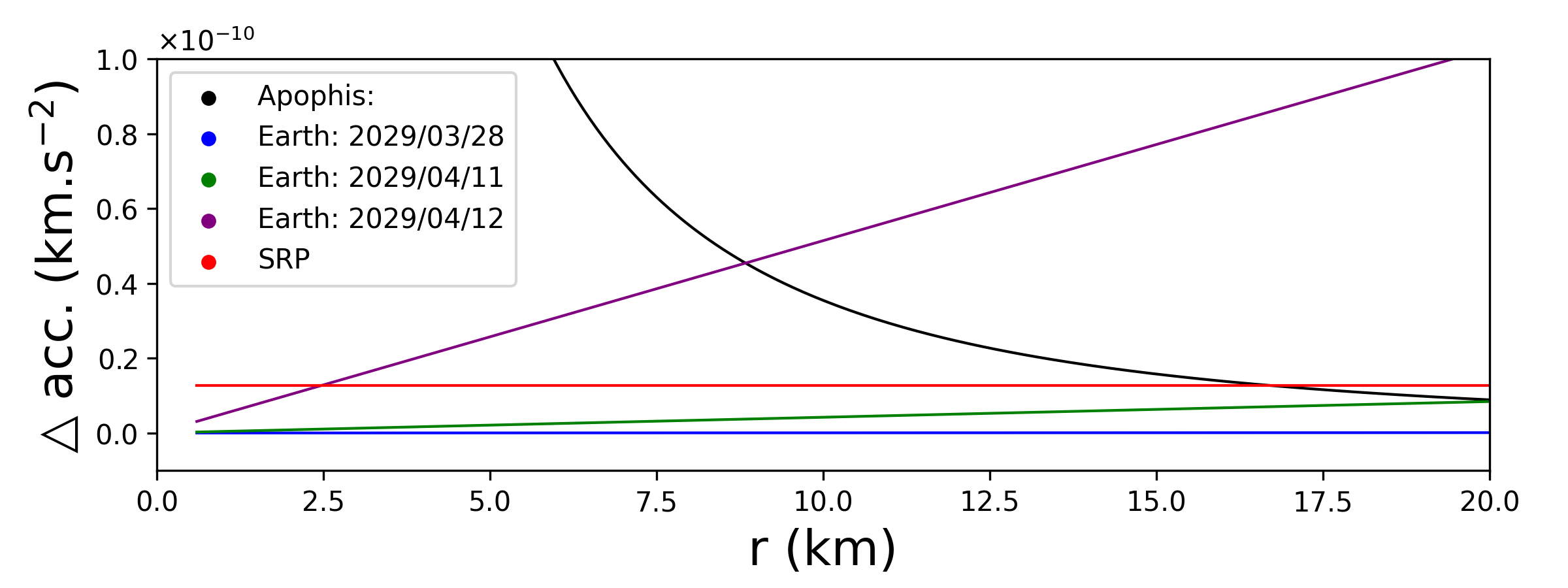}\\
         \includegraphics[width=0.99\linewidth]{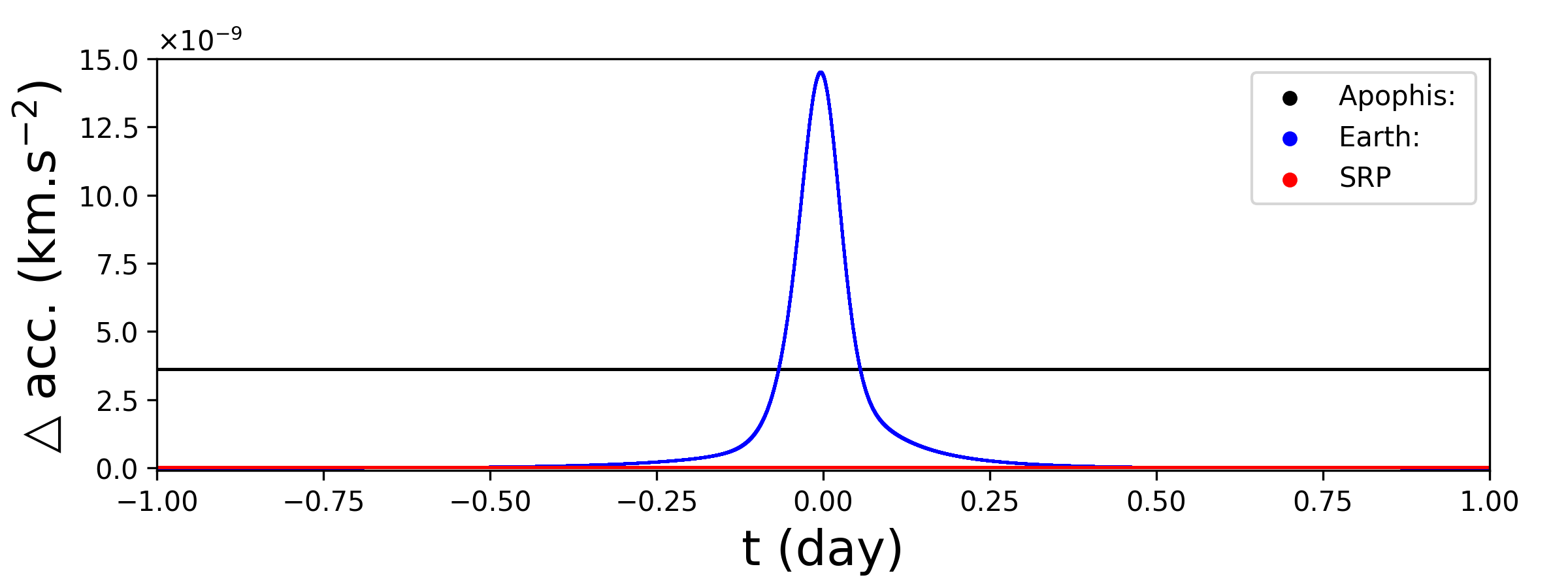}\\
         \caption{Top panel: the perturbation on the acceleration of a spacecraft close to (99942) Apophis: due to SRP (Red), due to the Earth at different epoch (blue, green, and purple). Bottom panel: the evolution of these perturbations overtime on the acceleration of a spacecraft fixed at a distance of 1 km from the surface of Apophis.} \label{fig11_perturbation_earth_srp}
      \end{figure}

      In fact, the Earth is by far the celestial body that most affects the dynamics around our target. In Fig. \ref{fig10_perturbation_earth}, we present the gravitational perturbation due to the polyhedral shape of Apophis (blue) and to the Earth (red) on the acceleration of a spacecraft close to the asteroid, 10 days before the minimum distance Apophis-Earth. We can see that the perturbation of our planet exceeds the perturbations of the shape beyond $\sim$6.4 km from the center of Apophis. This value is in accordance with the critical semimajor axis mentioned in \citet{sanchez_2017} and \citet{kinoshita_1991}, which is 5.719 km. However, this distance varies according to the position of our planet, as we can see in Fig. \ref{fig11_perturbation_earth_srp}. At the instant of the close approach, for instance, the critical semimajor axis becomes $\sim$0.6 km. That justifies our focus on a region that stretches only 10 km from the center of our target.
   
      The initial conditions (heliocentric positions and velocities) for all the bodies in Apophis system were provided by the JPL's HORIZONS ephemerides\footnote{\href{https://ssd.jpl.nasa.gov/?horizons}{https://ssd.jpl.nasa.gov/?horizons}} on March 1, 2029, at a distance of 0.156 au from our planet, 43 days before the closest distance Apophis-Earth. We use the spherical harmonics up to degree and order four to expand the gravitational potential of the Earth and Moon, as presented in \citet{sanchez_2014, sanchez_2017}. \\
      
      An important perturbation that arises from the Sun is the SRP. Besides the gravitational perturbations of the planets in our Solar System, above mentioned, we also considered the SRP in our study as described in \citet{beutler_2005}, where the radiation field due to the solar radiation is considered as parallel to the direction Sun-spacecraft. In this work, an OSIRIS-REx-like spacecraft is considered, presenting the following properties: a reflectance of 0.4, a mass of 1500 kg, and a cross-section of the spacecraft normal to the direction Sun-spacecraft of 25 m$^2$ with a mass-to-area ratio of 60 kg.m$^{-2}$, which yield a perturbation on the acceleration of the spacecraft that exceed the asteroid’s gravitational attraction at distances beyond $\sim$16.8 km close to the time of the close approach with Earth (1.00295 au from the Sun), as shown in Fig. \ref{fig11_perturbation_earth_srp} top panel, where we compare the perturbations on the acceleration of the spacecraft that is due to the SRP and due to our planet at different epochs in a region extending from 0.5 to 20 km from the surface of Apophis. In the bottom panel of this figure, we present the effect of these perturbations with respect to time, from one day before to one day after the CE on the acceleration of a spacecraft on a circular orbit at a distance of 1 km from the surface of our target. We can notice that the Earth perturbations quickly increases and becomes larger than the asteroid’s gravitational attraction itself. That leads to highly perturbed orbits in Apophis system, as we show later in this work. 
      
\begin{figure}[!htp]
         \centering
         \includegraphics[width=0.99\linewidth]{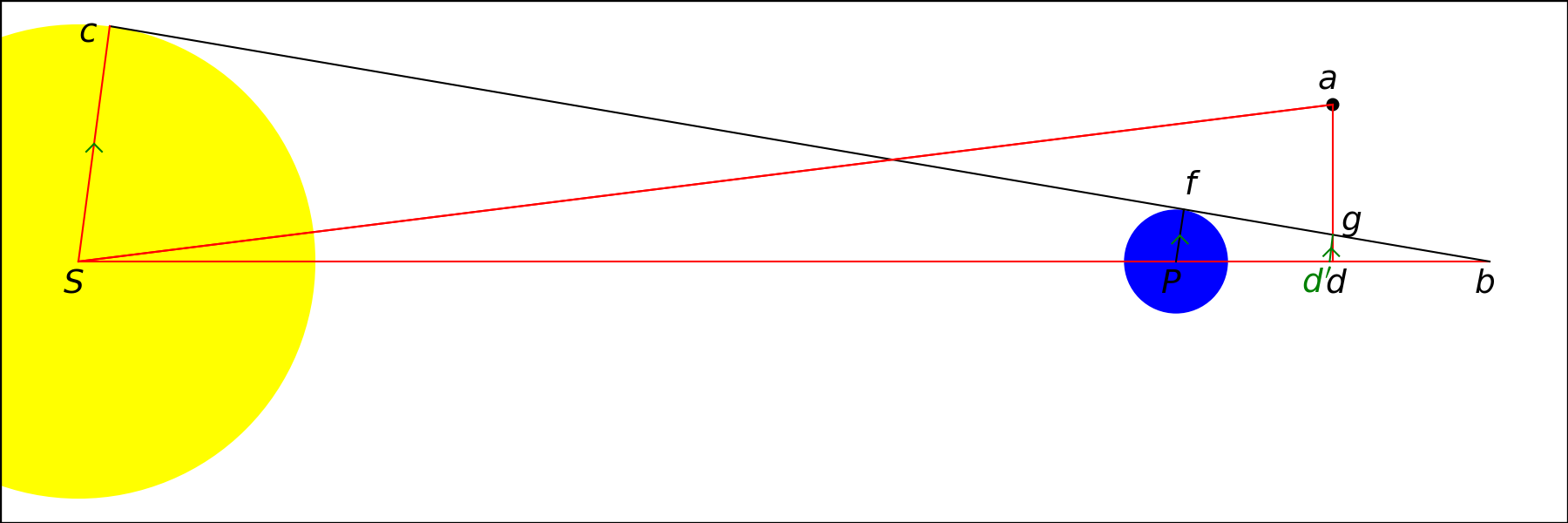}\\
         \includegraphics[width=0.99\linewidth]{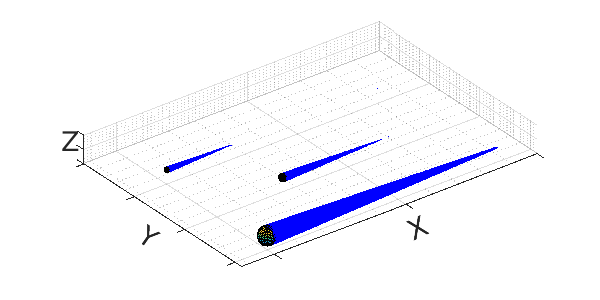}
         \caption{Non-scaled shadow geometry approach.} \label{fig12_Shadow}
      \end{figure}

      The shadowing of the sunlight by all the bodies in the system is also considered in our work. We assume that the Sun and the body in question are spherical with a radius of $R_{S}$ and $R_{P}$, respectively. The top panel of Fig. \ref{fig12_Shadow} illustrates the non-scaled shadow geometry for a body in the system. In fact, the sun is always far enough to consider the merger of $d$ and $d'$ and $\cos(\widehat{aSd})\simeq 1$. With this consideration in mind, the spacecraft enters the shadow when $|\vec{ad}| \leq |\vec{gd}|$ and the vectors $\vec{PS}$ and $\vec{Pa}$ are in opposite directions, where the point $a$ is the position of the spacecraft where one want to define if shadow occurs or not. To demonstrate the effectiveness of this algorithm, we present, in the bottom panel of Fig. \ref{fig12_Shadow}, the shadow of 3 spherical bodies in our simulation. However, it would be possible to consider the real shape of the body in the shadowing phenomenon by varying the value of $R_P$ at the point $f$, according to the polyhedral shape, using the CGAL library, which will affect significantly the execution time of our integration.\\
      
      Overall, the equations of motion that describe the motion of a test particle around (99942) Apophis during the close approach with the Earth are given by:
   
      \begin{eqnarray*}\label{motion1}
         \ddot{\text{r}}_{j}&=&-2\Omega \times \dot{\text{r}}_{j} -\Omega \times ( \Omega \times \text{r}_{j})+U_{\text{r}_{j}} + \mathcal{A}(\mathcal{P}) + \nonumber\\
                            & & \text{P}_{\text{E}} + \text{P}_{\text{M}} + \nu\mathcal{A}(\text{P}_{\text{R}})
      \end{eqnarray*}
      \textcolor{white}{.}\\
      where: $i, j=1,2, ..., 15$ refer to the concerned body (spacecraft, Sun, 10 planets, and the biggest 3 asteroids). $\text{r}$ is the position vector of the concerned body in the body-fixed frame, $\text{P}_{\text{E}}$ and $\text{P}_{\text{M}}$ are the accelerations due to the deformation of the Earth and of the moon, respectively. The vector $\mathcal{P}$ indicates the interaction between components $i$ and $j$ in the inertial frame. 
      \begin{eqnarray*}
         \mathcal{P} = \sum_{i=1,i\neq j}^{15}\mathcal{G}m_{i}\big(\frac{\varUpsilon_{i}-\varUpsilon_{j}}{|\varUpsilon_{i}-\varUpsilon_{j}|^{3}} - \frac{\varUpsilon_{i}}{|\varUpsilon_{i}|^{3}}\big)
      \end{eqnarray*}
      $\varUpsilon$ is the position vector in the inertial frame, $\mathcal{A}$ is an instantaneous rotation that takes the vector $\mathcal{P}$ from an inertial frame into a body-fixed frame. $\text{P}_{\text{R}}$ is the acceleration due to the direct radiation pressure applied only on the spacecraft, and $\nu$ is representing the shadowing phenomenon, taking the values 1 or 0, as defined earlier   in this work. 
      \begin{eqnarray*}
         \text{P}_{\text{R}} = (1 + \eta)~\text{au}^{2}~\frac{A}{m}\frac{S}{c} \frac{\text{r}_{\text{s}}-\text{r}_{\odot}}{|\text{r}_{\text{s}}-\text{r}_{\odot}|^{3}}
      \end{eqnarray*}
      \textcolor{white}{.}\\
      where: $\eta$ is the reflectance properties of the spacecraft surface. au is the Astronomical Unit, $A$ is the cross section of the spacecraft normal to its direction to the Sun. $m$ is the mass of the spacecraft. $S$ is the solar constant and $c$ is the speed of light in vacuum. The value of $\frac{S}{c}$ is $4.56316 \times 10^{-6}$ N/m$^{2}$ \citep{beutler_2005}. $\text{r}_{\text{s}}$ and $\text{r}_{\odot}$ are the coordinate vector of the spacecraft and the Sun, respectively. As we already mentioned, we considered the case of a spacecraft with low area-to-mass ratio ($\sim$ 0.017). \\
   
\begin{figure}[!htp]
         \includegraphics[width=0.98\linewidth]{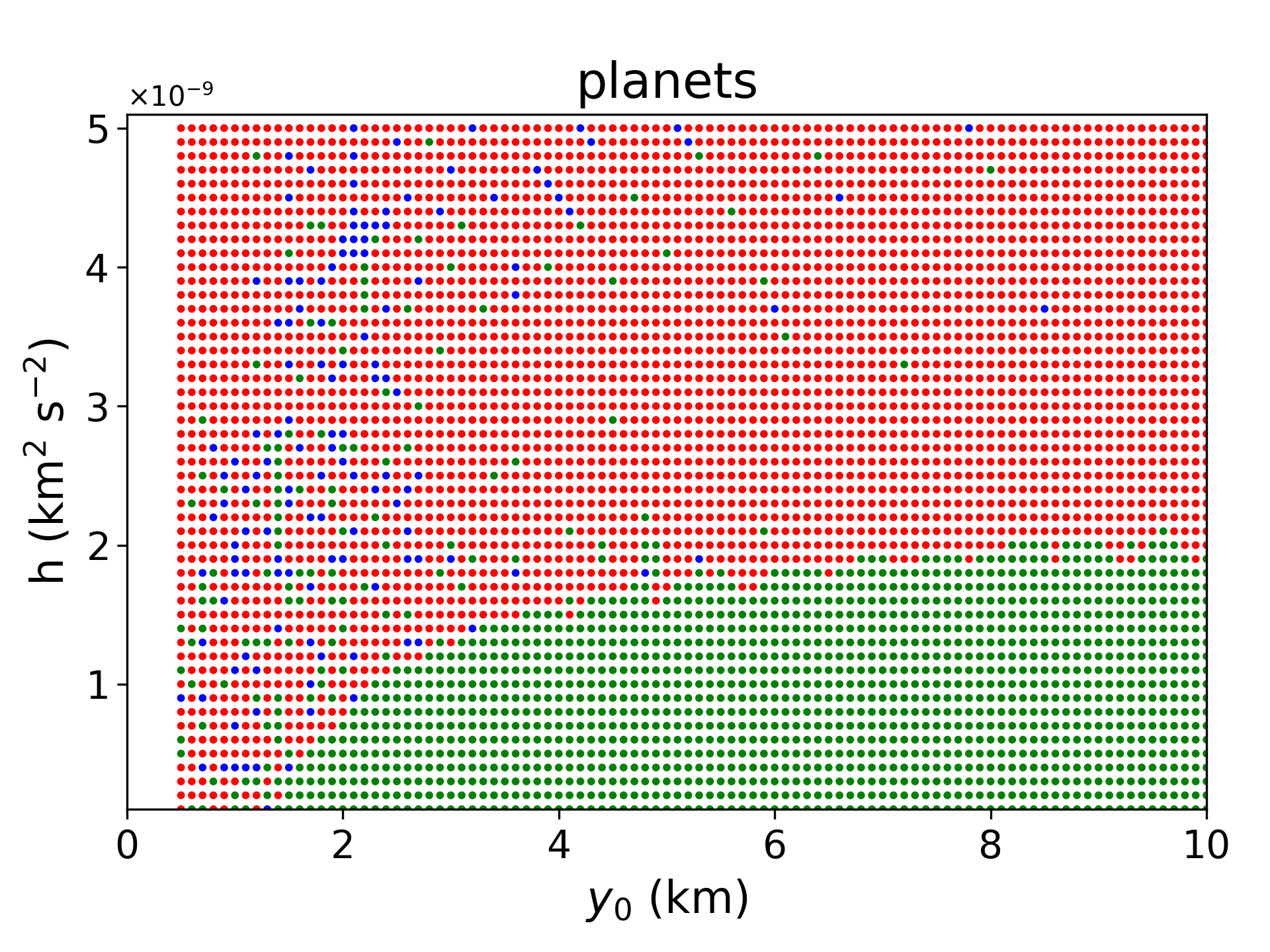}\\
         \includegraphics[width=0.98\linewidth]{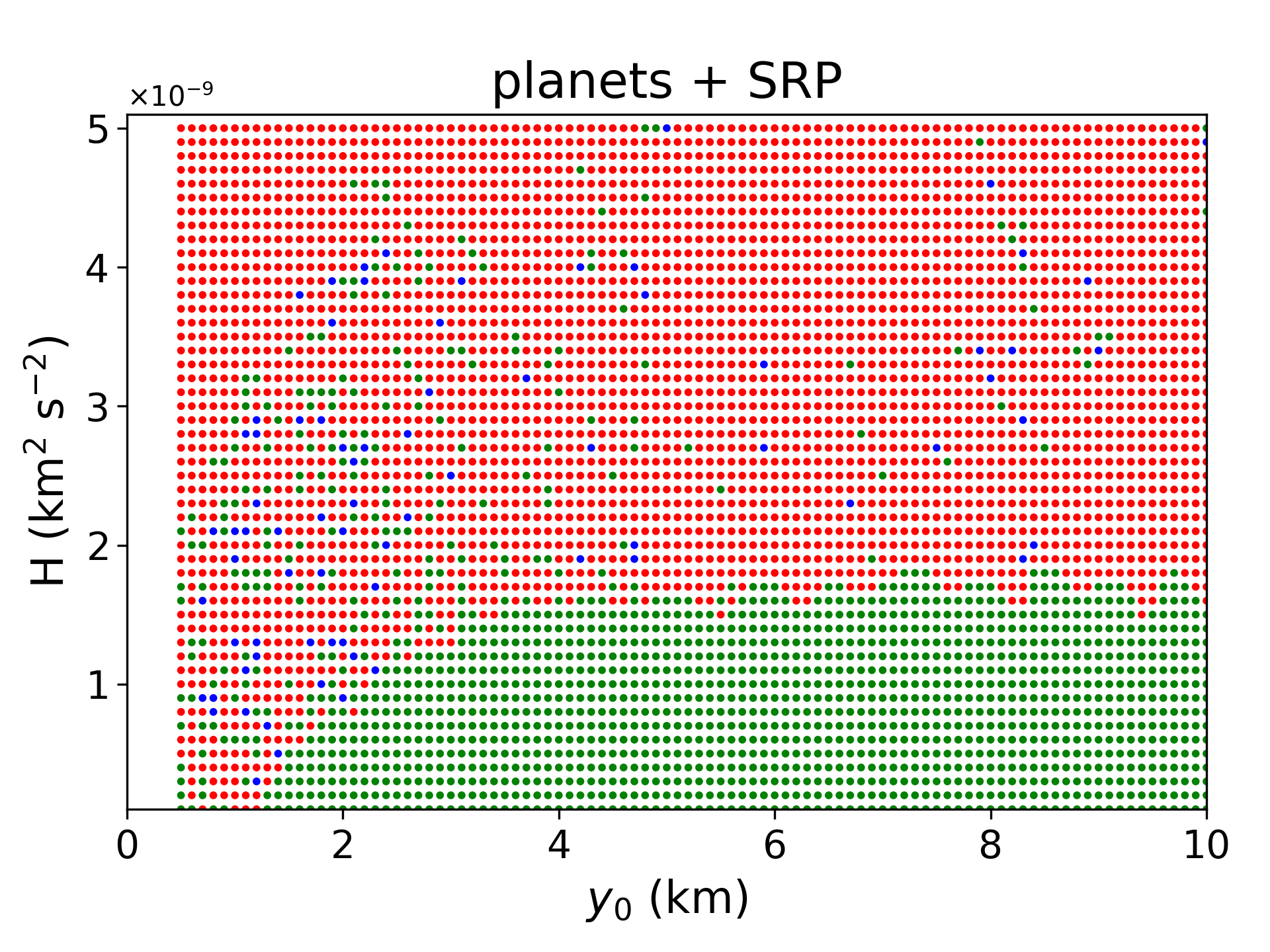}\\
         \caption{Type of orbits around the asteroid (99942) Apophis. Considering or not the perturbations of the planets in our Solar System and the SRP. The colour symbols are the same as in Fig. \ref{fig06_type_orbits}.} \label{fig13_type_orbits}
      \end{figure}

      In Fig. \ref{fig13_type_orbits}, we present the type of all the orbits generated with our complete model, considering the planets in our Solar System without the SRP (top panel) and with the SRP (bottom panel). Comparing with Fig. \ref{fig06_type_orbits} we can see that the planets in our Solar System destroyed the most of the orbits around Apophis, making the spacecraft collide or escape from the system, as we will see in more details later on in this section. We only found some bounded orbits very close to the central body. However, the SRP destabilized about 50\% of them, changing their distribution in the ($dy_0$, h) plane. Such a drastic effect of the SRP was already seen in \citet{chanut_2017} studying the dynamics around the asteroid Bennu, and also in \citet{sanchez_2019} studying the Less-Disturbed Orbital Regions Around the Near-Earth Asteroid 2001 SN$_{263}$.      

\begin{figure}[!htp]
         \centering
         \includegraphics[width=0.48\linewidth]{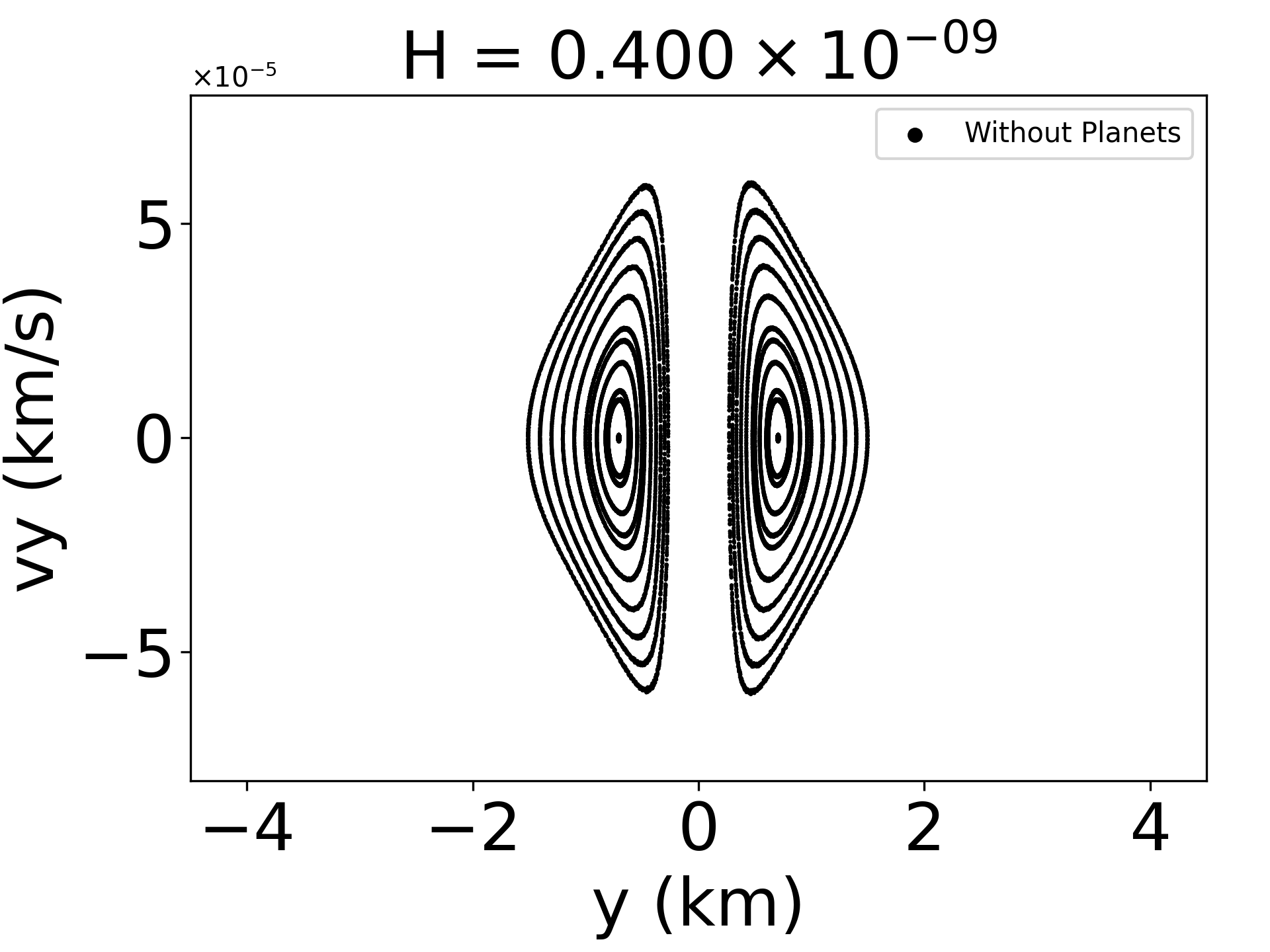}
         \includegraphics[width=0.48\linewidth]{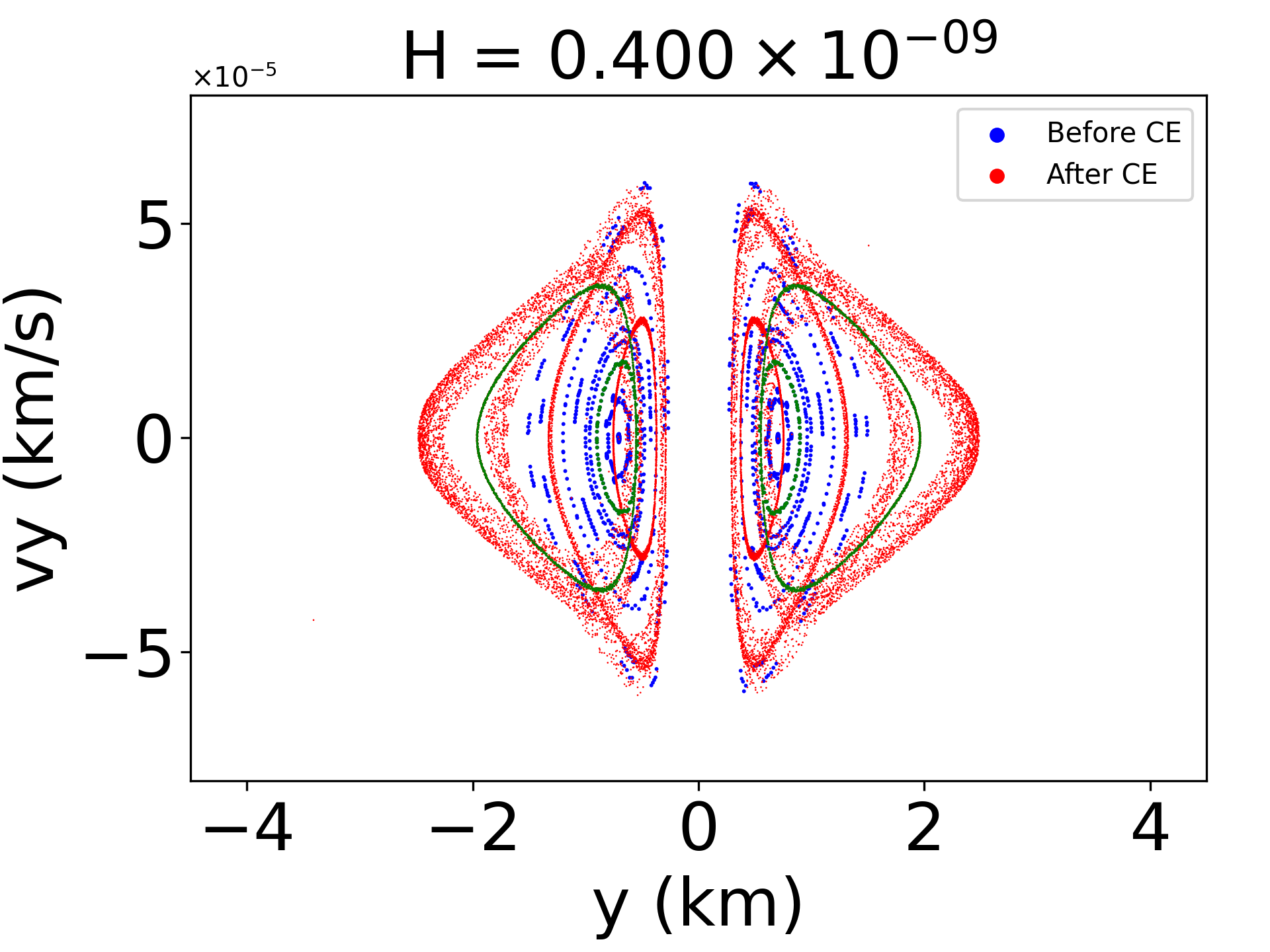}\\
         \includegraphics[width=0.48\linewidth]{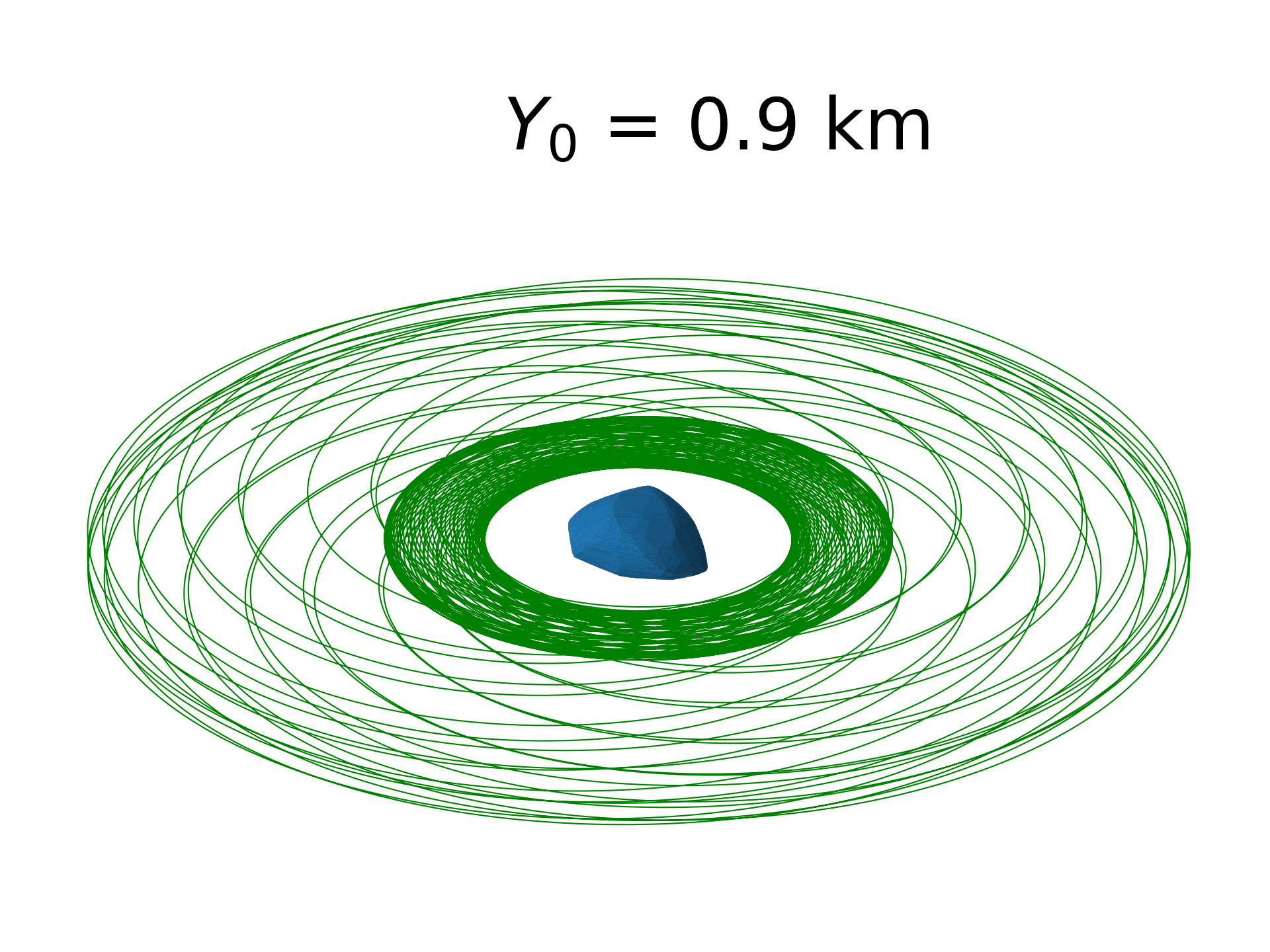}
         \includegraphics[width=0.48\linewidth]{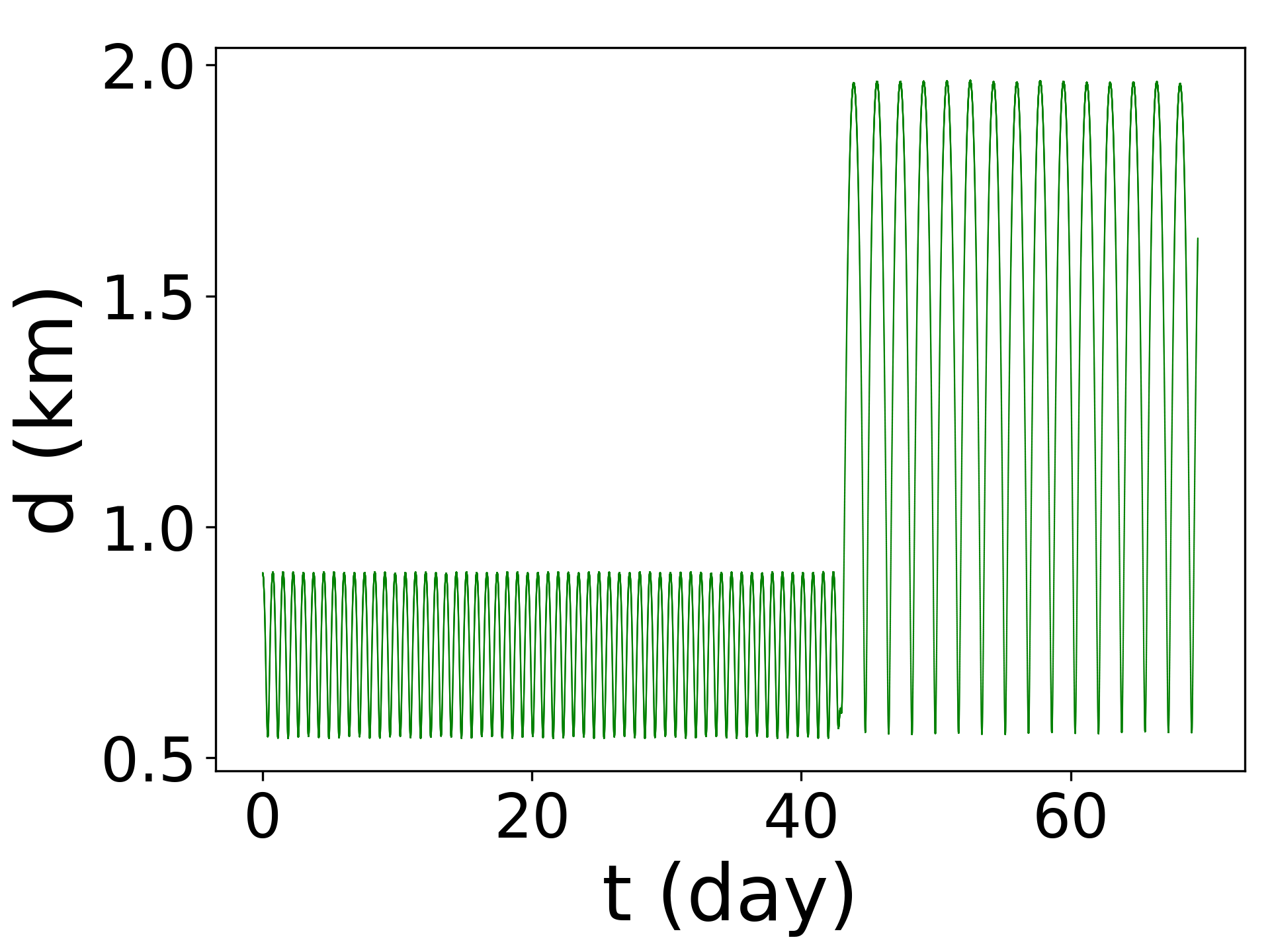}
         \caption{Intersection points of orbits around (99942) Apophis, starting from March 1, 2029. Here, we considered the perturbation from the remaining bodies in the Solar System and neglected the SRP.} \label{fig14_poincare_ce}
      \end{figure}

\begin{figure}[!htp]
         \centering
         \includegraphics[width=0.48\linewidth]{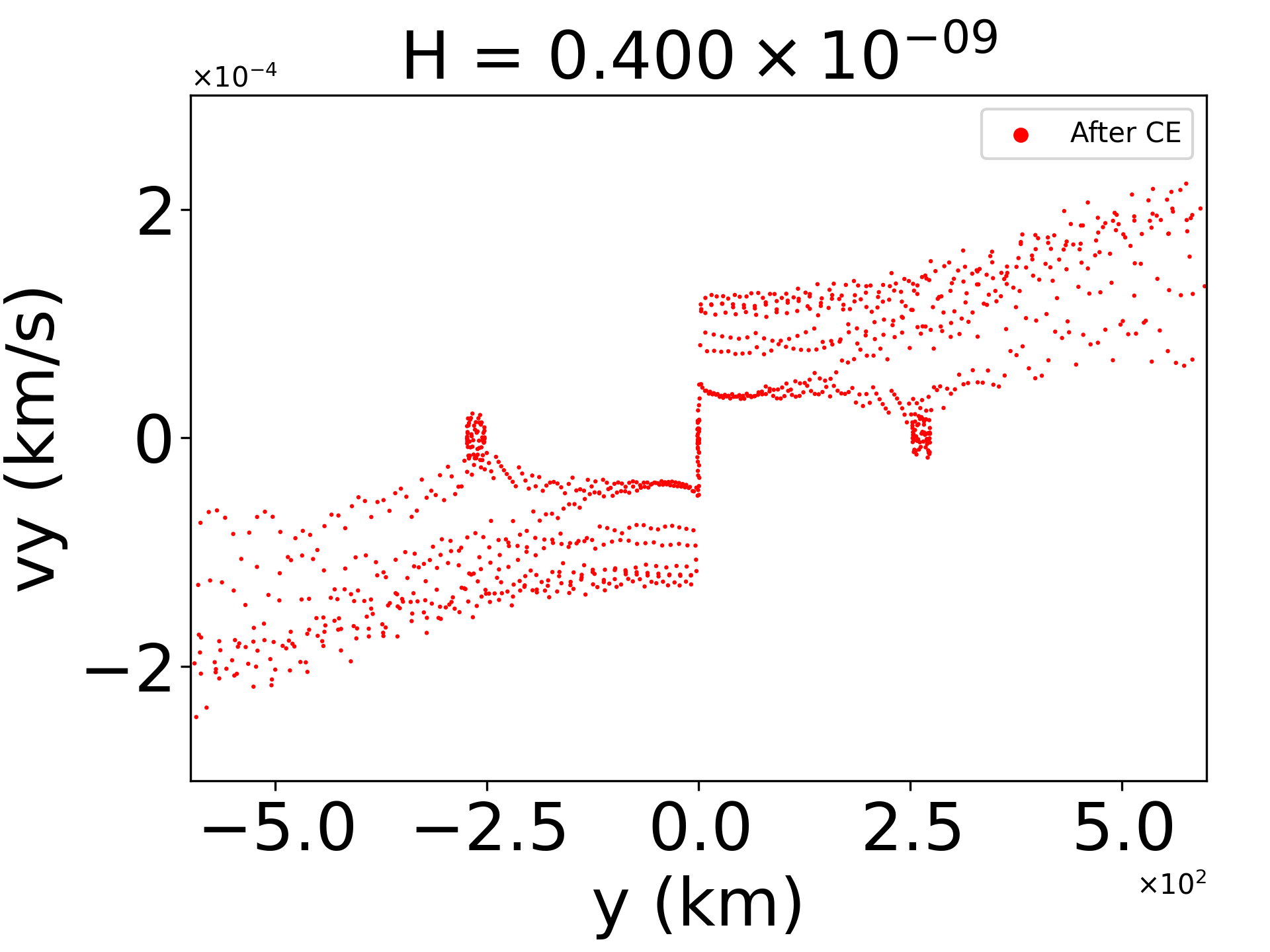}
         \includegraphics[width=0.48\linewidth]{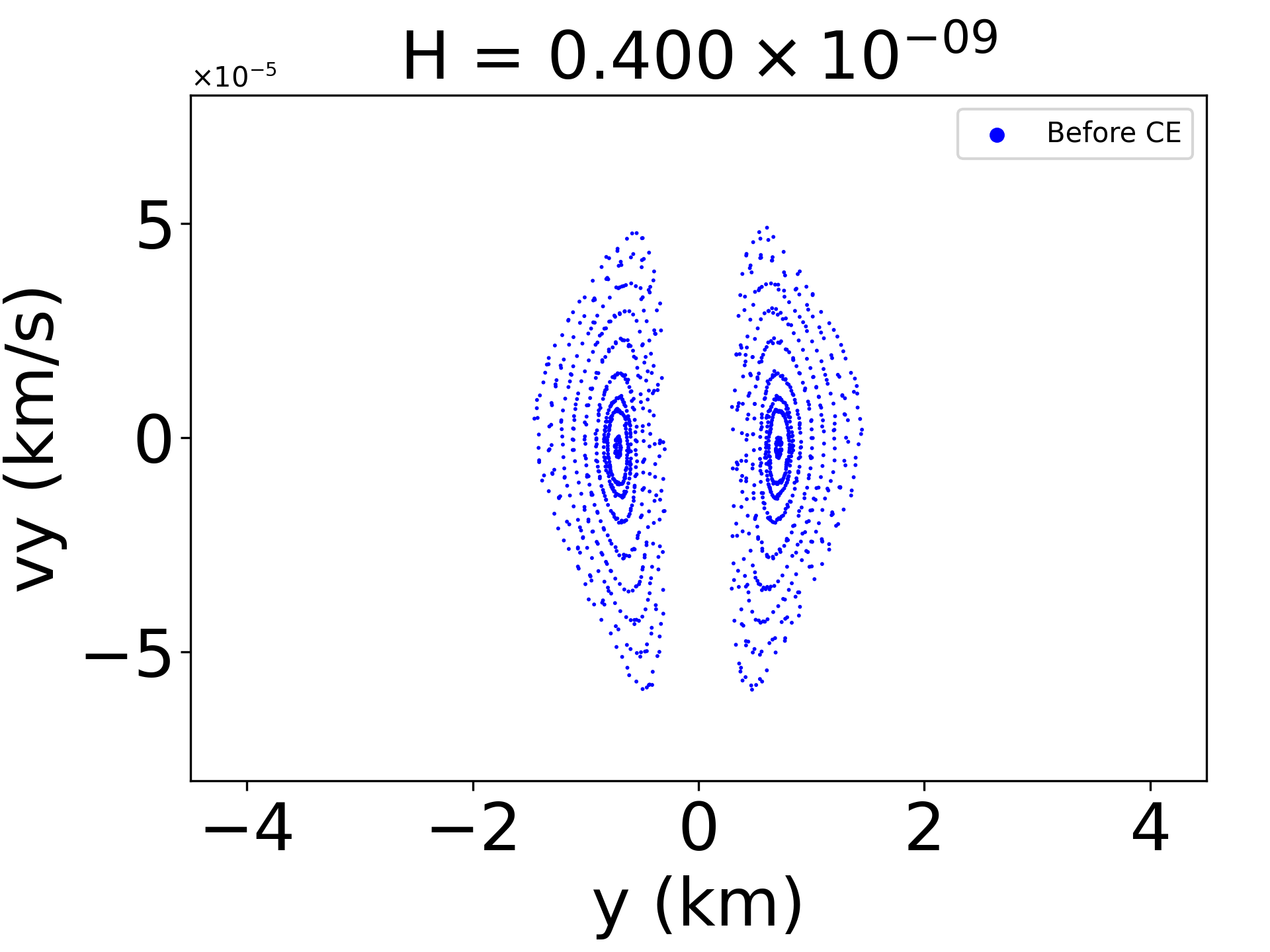}\\
         \includegraphics[width=0.48\linewidth]{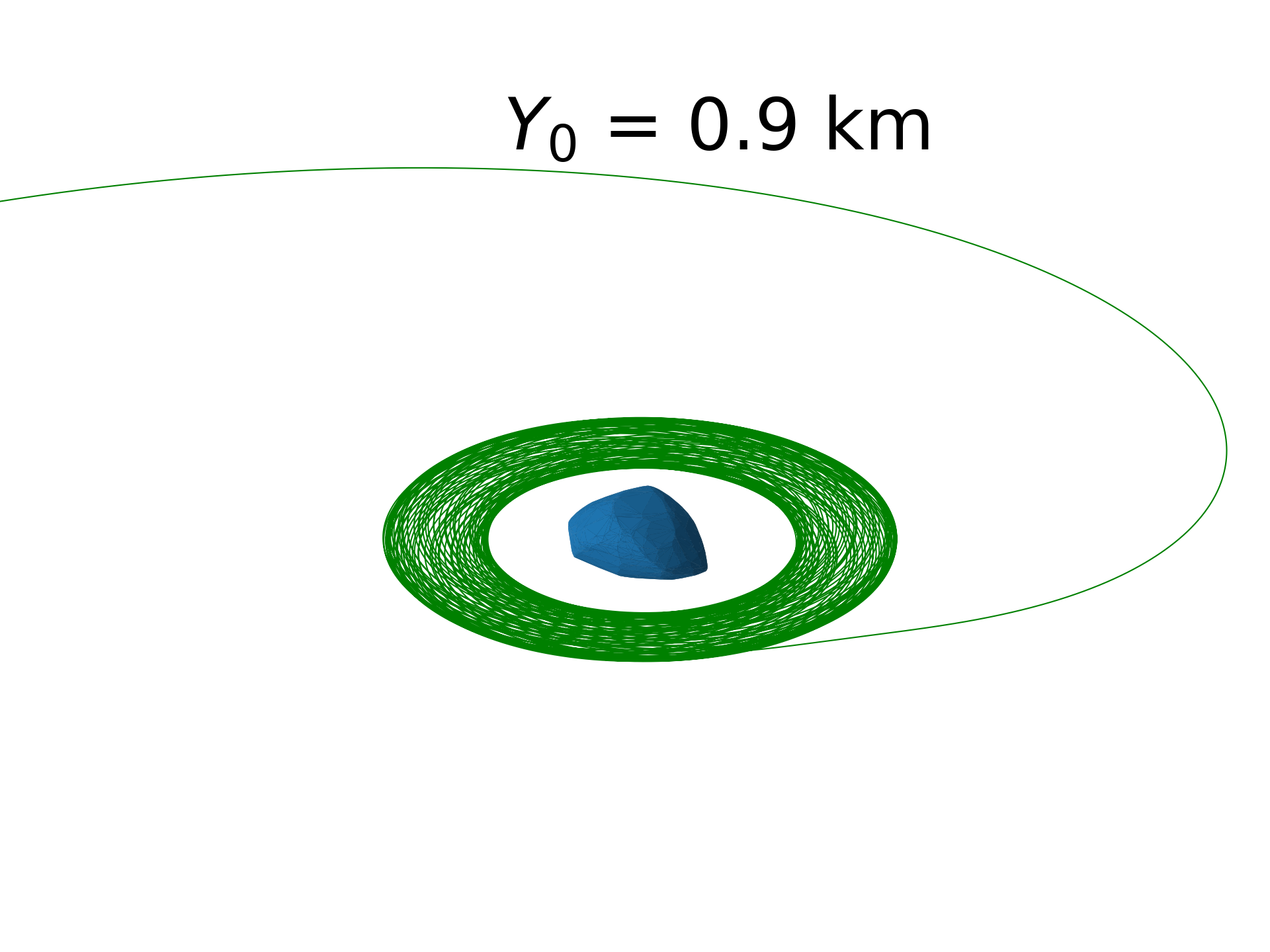}
         \includegraphics[width=0.48\linewidth]{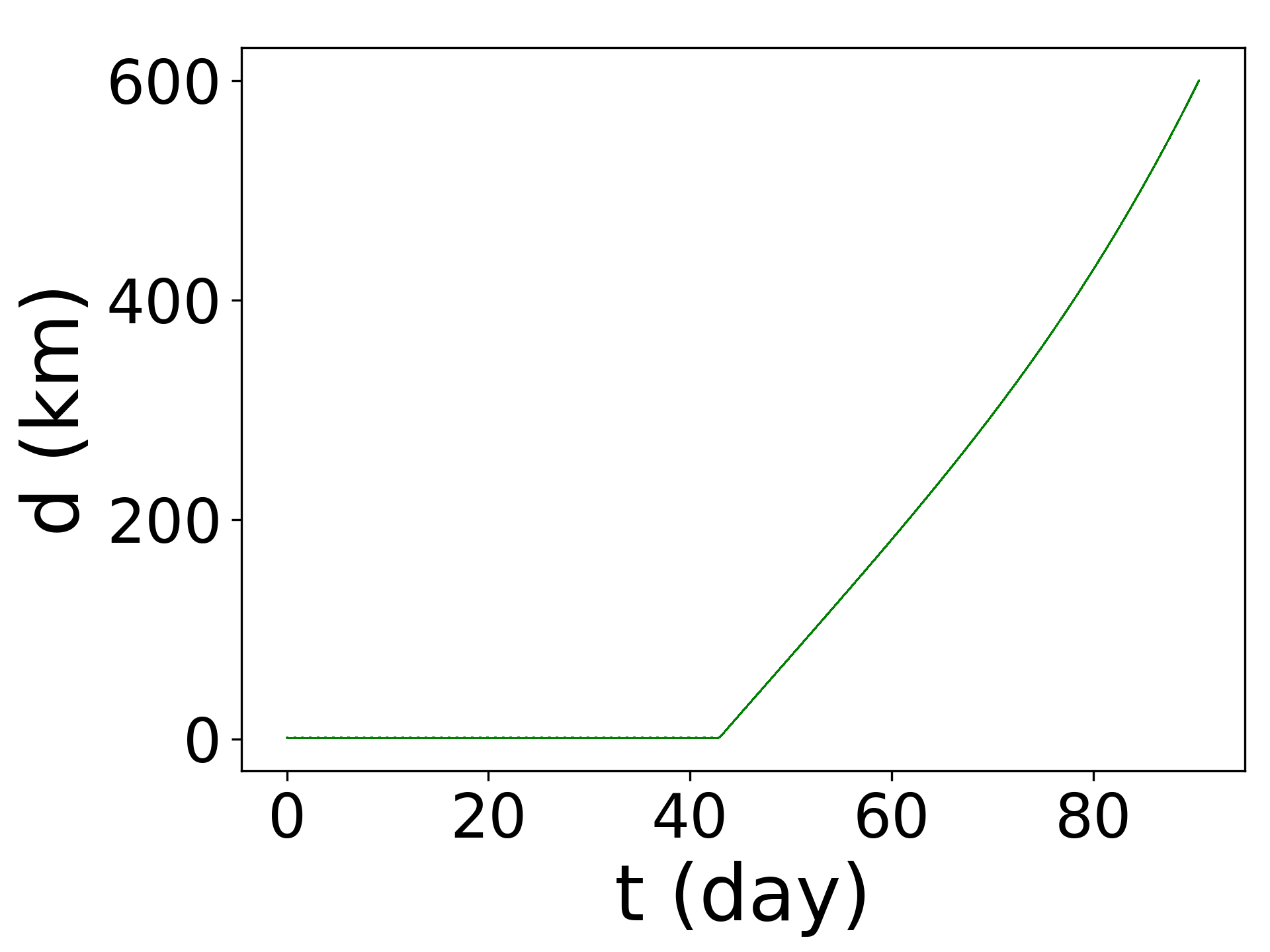}
         \caption{Intersection points of orbits around (99942) Apophis, starting from March 1, 2029. Here, we considered the perturbation from the remaining bodies in the Solar System and the SRP.} \label{fig15_poincare_ce_srp}
      \end{figure}

      In order to better understand the dynamics around our target, we present, in Fig. \ref{fig14_poincare_ce}, the surfaces of the section for $H=0.4\times 10^{-9}$, starting from March 1, 2029 considering (top right panel) or not (top left panel) the perturbations of the planets in the Solar System and neglecting the SRP. Comparing these two panels, we can notice that a new configuration appears after 43 days of our integration, that correspond to the instant of the close approach with our planet. This point will be seen clearly by following the evolution of the distance between our test particle and the central body, as in the bottom panels in Fig. \ref{fig14_poincare_ce}, where we show an example of bounded orbits around Apophis. For the seek of clarity, we presented the first 70 days of the orbit. However, including the SRP in our model changes completely the structure of the surfaces of the section, giving the tendency for highly chaotic orbits. In Fig. \ref{fig15_poincare_ce_srp} we present our results for $h=0.4\times 10^{-9}$. Again, we notice that the distance between our test particle and the central body significantly changed just after the close approach with our planet. We should now turn our attention to identify the less perturbed region around Apophis.
   
   \section{Search for less perturbed regions around (99942) Apophis} \label{sec05_stability_research}
\begin{figure}[!htp]
         \centering
         \includegraphics[width=1\linewidth]{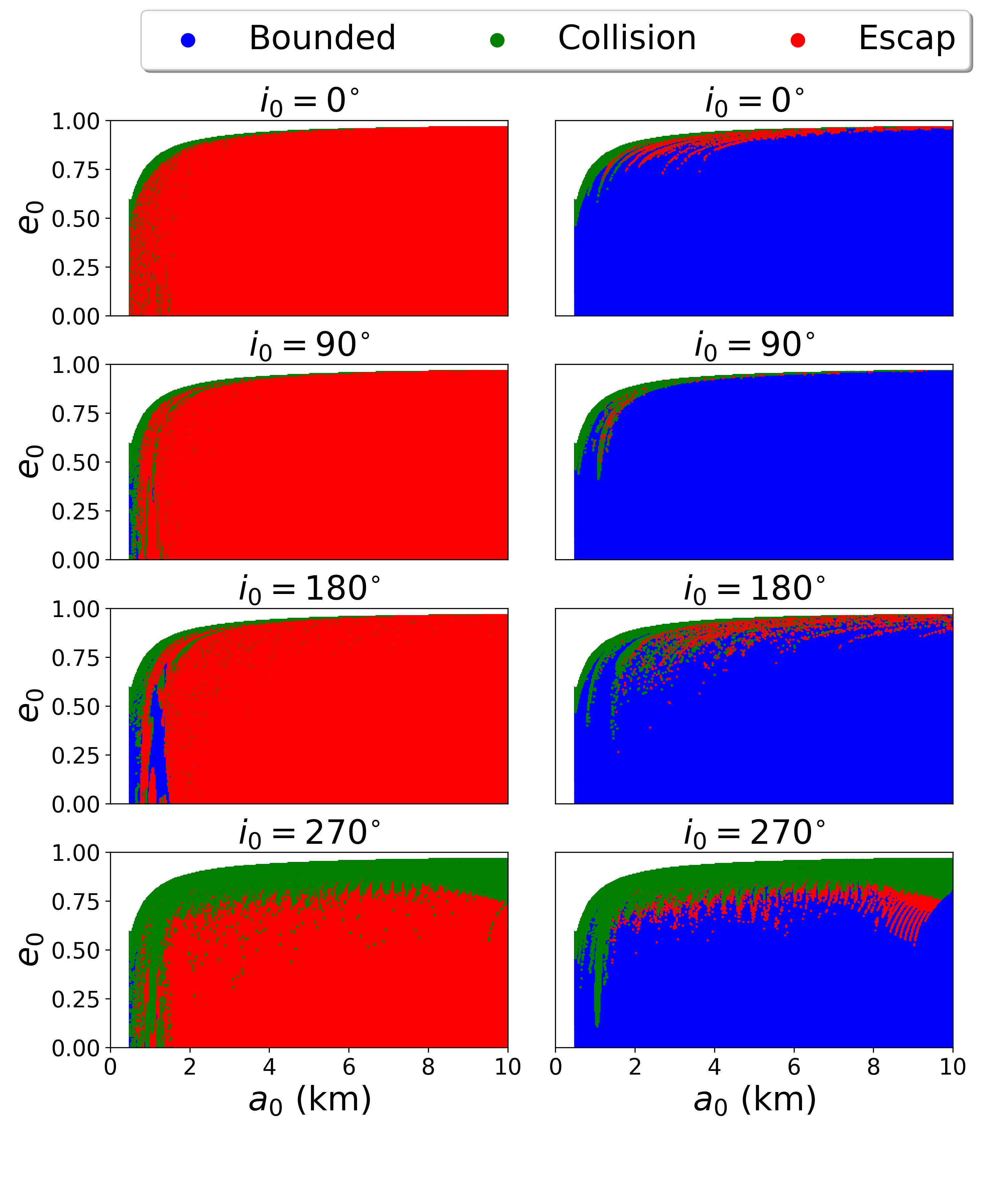}
         \caption{Type of orbits around the asteroid (99942) Apophis for 60 days (left column) and 40 days (right column) starting from March 1, 2029. }\label{fig16_type_orbits}
      \end{figure}

      In order to identify suitable regions to place a spacecraft around (99942) Apophis on March 1, 2029, we made a numerical analysis of orbits in a region with a semimajor axis between 0.5 and 10 km from the center of (99942) Apophis with an interval of 25 m. We vary the initial eccentricities from 0 to 1 with a step size of 0.005, and tested 4 different inclinations ($0^{\circ}$, $90^{\circ}$, $180^{\circ}$, $270^{\circ}$). The argument of the perigee ($\omega$), the longitude of ascending node ($\Omega$), and the mean anomaly of the small probes are initially $0^{\circ}$. Again, the vast majority of the tested orbits collide or escape from the system at the time of the close approach with Earth. In Fig. \ref{fig16_type_orbits} we present the type of all the tested orbits integrated for 40 days (right column) and 60 days (left column). Considering that left and right panels correspond to the same initial conditions evolved by the same equations but for a different final time, it means that most of the orbits escape after 40 days of integration due to the close approach with the Earth.\\
   
\begin{figure}[!htp]
        \centering
        \includegraphics[width=0.48\linewidth]{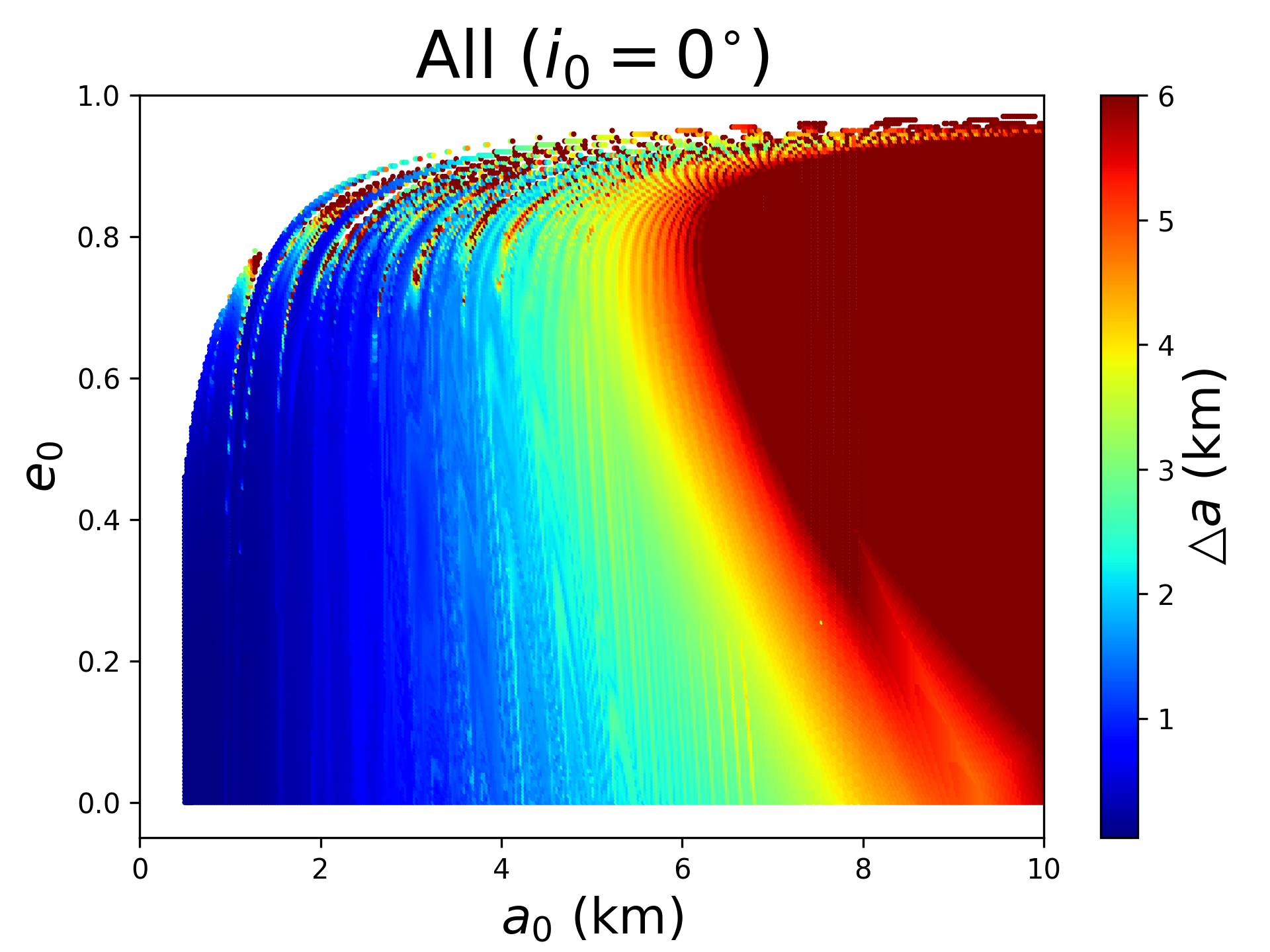}
        \includegraphics[width=0.48\linewidth]{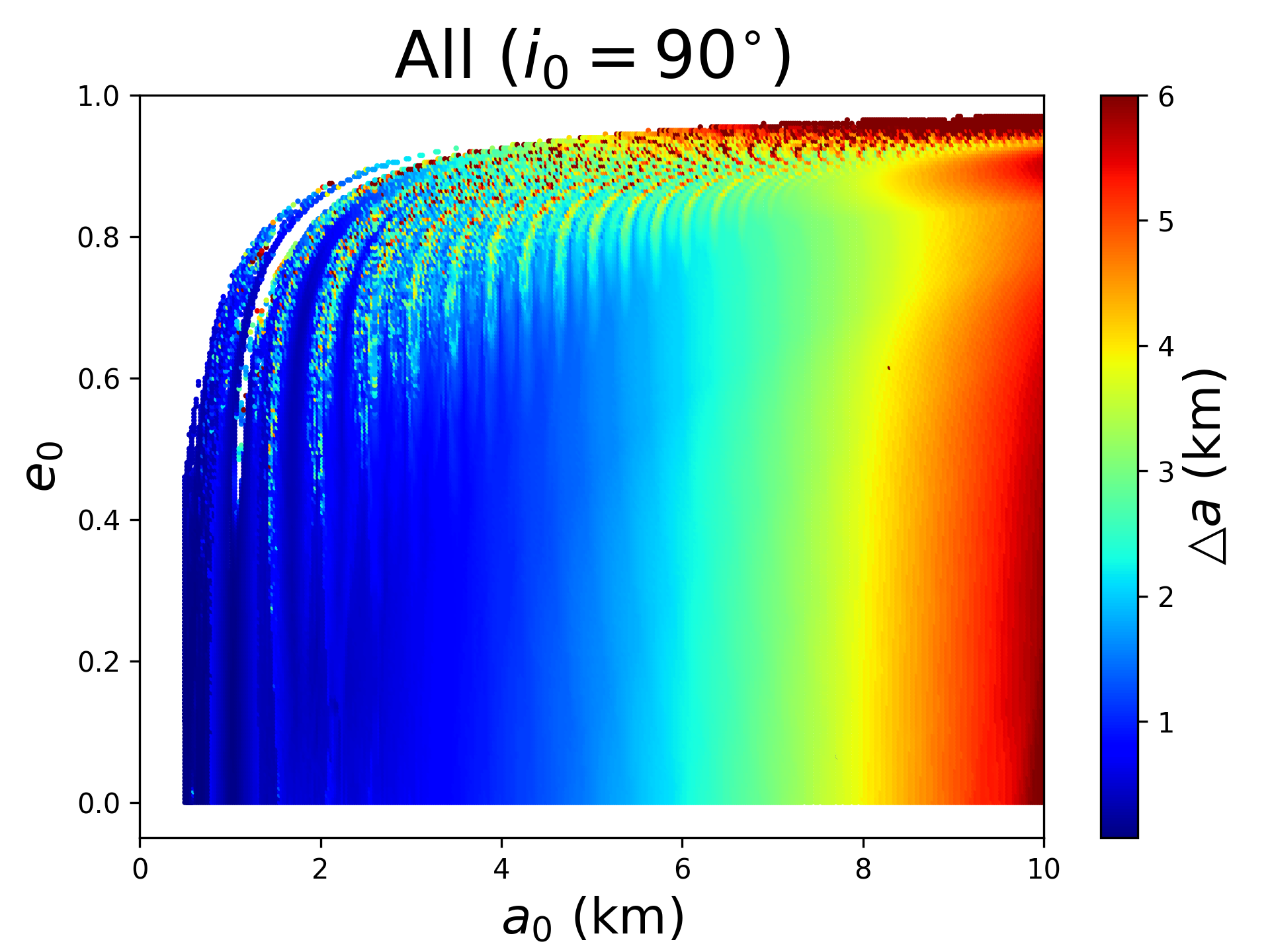}\\
        \includegraphics[width=0.48\linewidth]{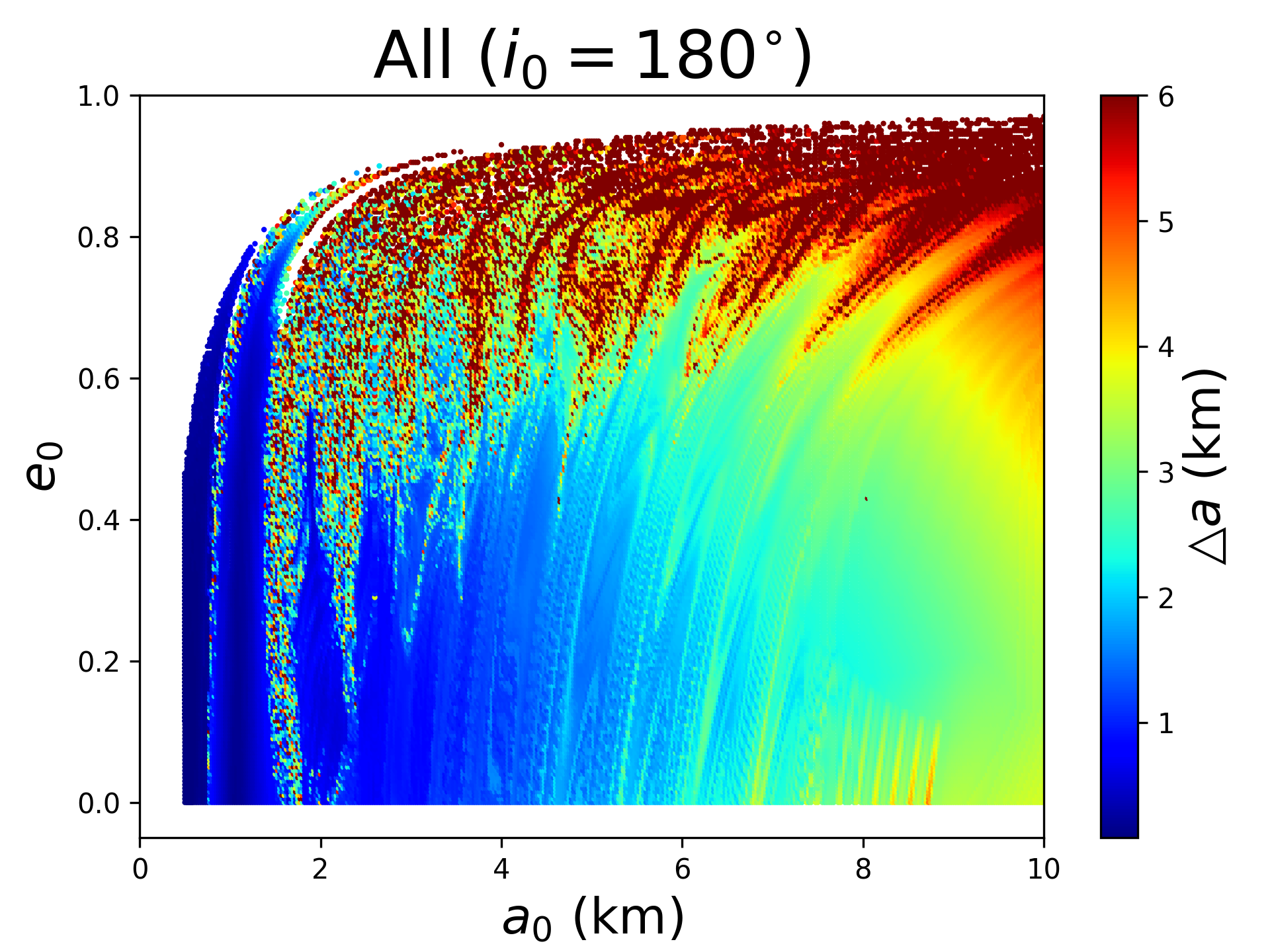}
        \includegraphics[width=0.48\linewidth]{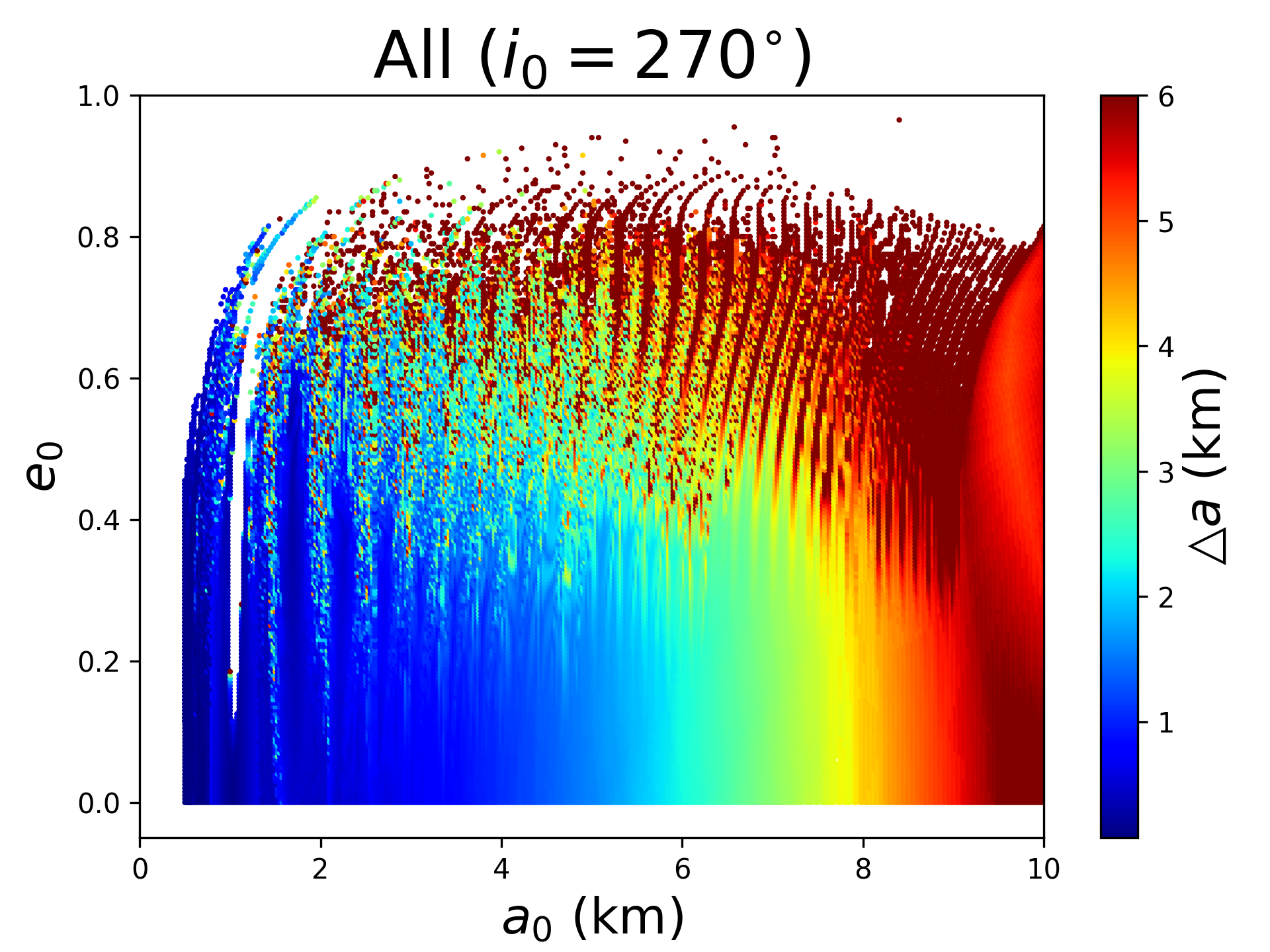}\\
        \caption{The variation maps of the semimajor axis coming from the ensemble perturbations on real system of Apophis.}\label{fig17_0_all_map_delta_a}
     \end{figure}

\begin{figure}[!htp]
        \centering
        \includegraphics[width=0.99\linewidth]{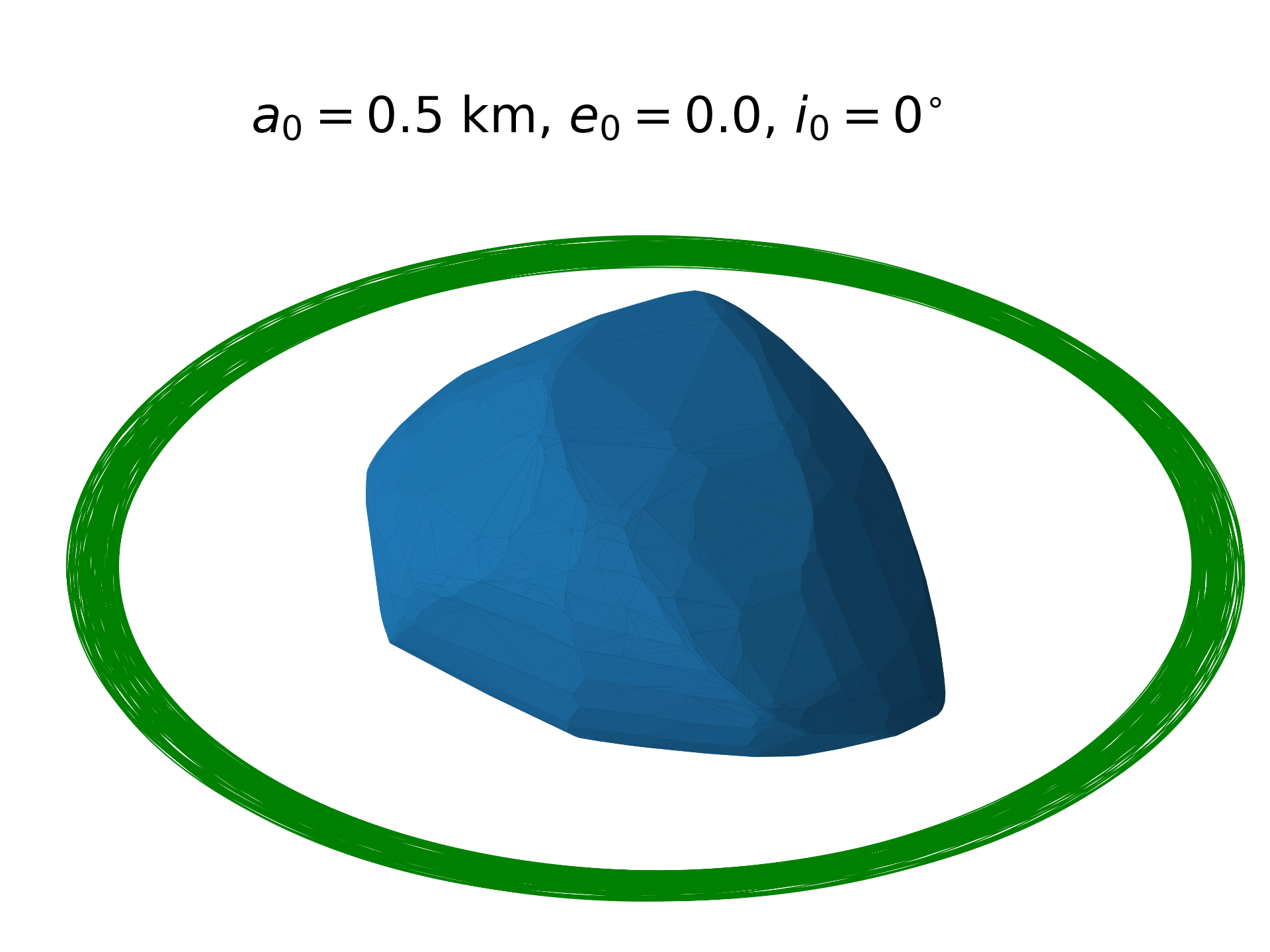}
        \caption{An example of the less perturbed orbits close of Apophis over 40 days.}\label{fig18_example_orbits}
    \end{figure}

      Considering our results for 40 days of integration, we use the variation of the semimajor axis ($\bigtriangleup a$) as a criterion to identify the less perturbed regions in the system. Our results are presentd in Fig. \ref{fig17_0_all_map_delta_a}. The smooth parts of the map with small values of $\bigtriangleup a$ could be a possible option to insert a spacecraft into natural orbits around Apophis before the close approach with our planet. The minimum value founded of this variation is 0.05 km and a variation of the corresponding eccentricity is ($\bigtriangleup e$) of 0.128, which is still a non-negligible variation presented in Fig. \ref{fig18_example_orbits}. However, an interesting part of the region around Apophis is heavily perturbed, that appears in the map beyond 4 km from the center of Apophis.   
   
   \section{Orbital control around (99942) Apophis}\label{sec06_rbital_control}
      As we already saw in this paper, the most problematic behaviors of the dynamics around our target come from the close approach with our planet. In order to solve the stabilization problem for the system of equation \ref{motion1}, we applied in this section the robust path following control law as presented in \citet{negri_2020a, negri_2020b}, The advantage of a path following control to a reference tracking control is that in the first situation, only the geometry of the orbit is controlled, with no a priori time parameterization. A detailed discussion of the applicability and practical considerations of the path following the law derived in \citet{negri_2020a} for asteroid missions is done in \citet{negri_2020b}. The acceleration correction is calculated in the radial-transverse-normal coordinates (RTN), where the versors of the spacecraft are defined as follows:
   
      \begin{eqnarray*}\label{versors}
          \hat{r} = \frac{\vec{r}}{r}, ~~~~~~~~~~~~ \hat{h} = \frac{\vec{h}}{h}, ~~~~~~~~~~~~ \hat{\theta} = \hat{h} \times \hat{r} 
      \end{eqnarray*}
      where, $\vec{h} = \vec{r} \times \dot{\vec{r}}$ is the angular momentum of the spacecraft, $\vec{r}$ and $\dot{\vec{r}}$ are the position and velocity vectors, in the frame where the orbit will be controlled. We defined the eccentricity vector ($\vec{e}$) and the sliding surface ($\vec{s} $) as 
      \begin{eqnarray*}
          \vec{e} &=& \frac{1}{\mu} \bigg(\dot{\vec{r}}\times \vec{h} - \mu \hat{r}\bigg)\\
          \vec{s} &=& \begin{bmatrix}
                       (\vec{e}-\vec{e}_{d}).(\lambda_{R}\hat{r} + \hat{\theta})\\
                       h-h_{d}\\
                       \hat{h}_{d}.(\lambda_{N}\hat{r} + \hat{\theta})
                     \end{bmatrix}
      \end{eqnarray*}
      where, $\lambda_{R}$ and $\lambda_{N}$ are design parameters. They will determine the asymptotic convergence to the sliding surface. In our application, they are fixed to a value of 0.002. $\vec{e}_{d}$ and $\hat{h}_{d}$ are the desired eccentricity vector and angular momentum versor, respectively. They are given by: 
      \begin{eqnarray*}
         \vec{e}_{d} &=& \begin{bmatrix}
                          \cos(\Omega_{d})\cos(\omega_{d}) - \sin(\Omega_{d})\sin(\omega_{d})\cos(i_{d})\\
                          \sin(\Omega_{d})\cos(\omega_{d}) - \cos(\Omega_{d})\sin(\omega_{d})\cos(i_{d})\\
                          \sin(\omega_{d})\sin(i_{d})
                       \end{bmatrix}        \\
         \hat{h}_{d} &=& \begin{bmatrix}
                          \sin(i_{d})\sin(\Omega_{d})\\
                        - \sin(i_{d})\cos(\Omega_{d})\\
                          \cos(i_{d})
                       \end{bmatrix}
      \end{eqnarray*}   
      where, $i_{d}, \Omega_{d}, $ and $\omega_{d}$ are the desired inclination, longitude of the ascending node and argument of periapsis, respectively.
      The acceleration corrections in the RTN coordinates can be written as:
      \begin{eqnarray*}
         \vec{u}_{RTN} = (U_{R}, U_{T}, U_{N}) = - F^{-1}\big(G + K \text{sat}(\vec{s}, \vec{\Phi})\big)
      \end{eqnarray*}
      where, 
      \begin{eqnarray*}
         F &=& \frac{1}{h\mu}\begin{bmatrix}
                             -h^{2}  & (2\lambda_{R}h - (\dot{\vec{r}}.\hat{r})r)h & -\mu r \vec{e}_{d}.\hat{h}\\
                                0    & \mu r h                                     & 0                         \\
                                0    &           0                                 & \mu r  \hat{h}_{d}.\hat{h}
                           \end{bmatrix}\\
         G &=& \frac{h}{r^{2}}\begin{bmatrix}
                               (\vec{e}-\vec{e}_{d}).(\lambda_{R}\hat{\theta} - \hat{r})\\
                               0                                             \\
                               \hat{h}_{d}. (\lambda_{N} \hat{\theta} - \hat{r})
                            \end{bmatrix}\\
          K &=& \begin{bmatrix}
                    0.001 & 0.0 & 0.0\\
                    0.000 & 0.1 & 0.0\\
                    0.000 & 0.0 & 0.1
                 \end{bmatrix}, \Phi = \begin{bmatrix}
                                           0.005\\
                                           0.500\\
                                           0.050
                                       \end{bmatrix}
       \end{eqnarray*}
       $\text{sat}(\vec{s}, \vec{\Phi})$ is the saturation function, proposed to avoid the discontinuous control input.
       \begin{eqnarray*}
          \text{sat}(\alpha, \beta) =  \begin{cases}
                                           +1 \qquad \alpha > \beta \\
                                           +\frac{\alpha}{\beta} \quad -\beta \leq \alpha \leq \beta \\
                                           -1 \qquad \alpha < -\beta
                                       \end{cases}
       \end{eqnarray*}   
    
       Finally, the acceleration corrections in our frame of reference is given by
       \begin{eqnarray*}
         \vec{u} = U_{R}\hat{r} + U_{T}\hat{\theta} + U_{N}\hat{h}
       \end{eqnarray*}
    
       For more detailed and deeper analysis on all the theoretical considerations considering all practical aspects, we refer the reader to \citet{negri_2020a, negri_2020b}. 
    
\begin{figure}[!htp]
          \centering
          \includegraphics[width=0.48\linewidth]{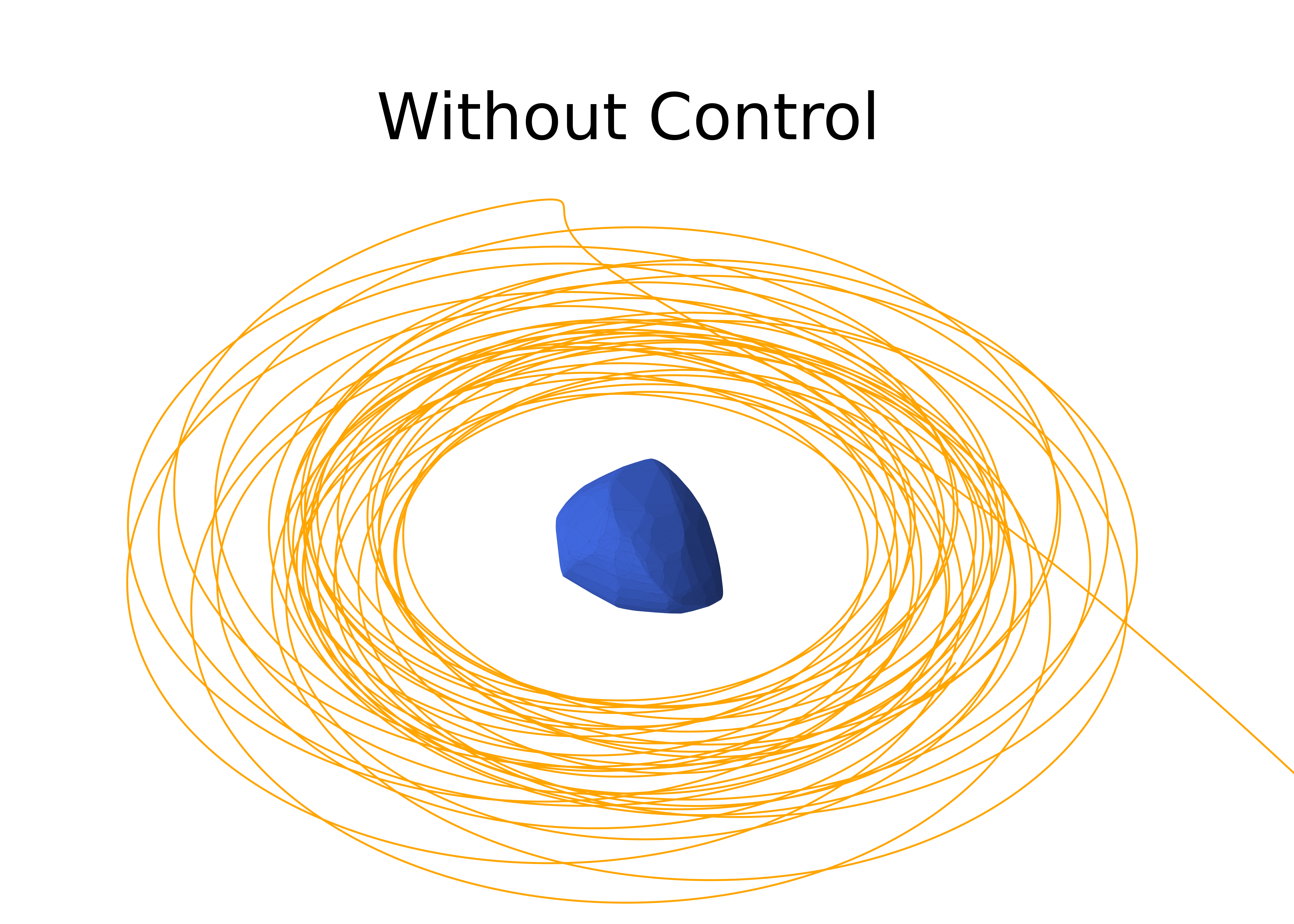}
          \includegraphics[width=0.48\linewidth]{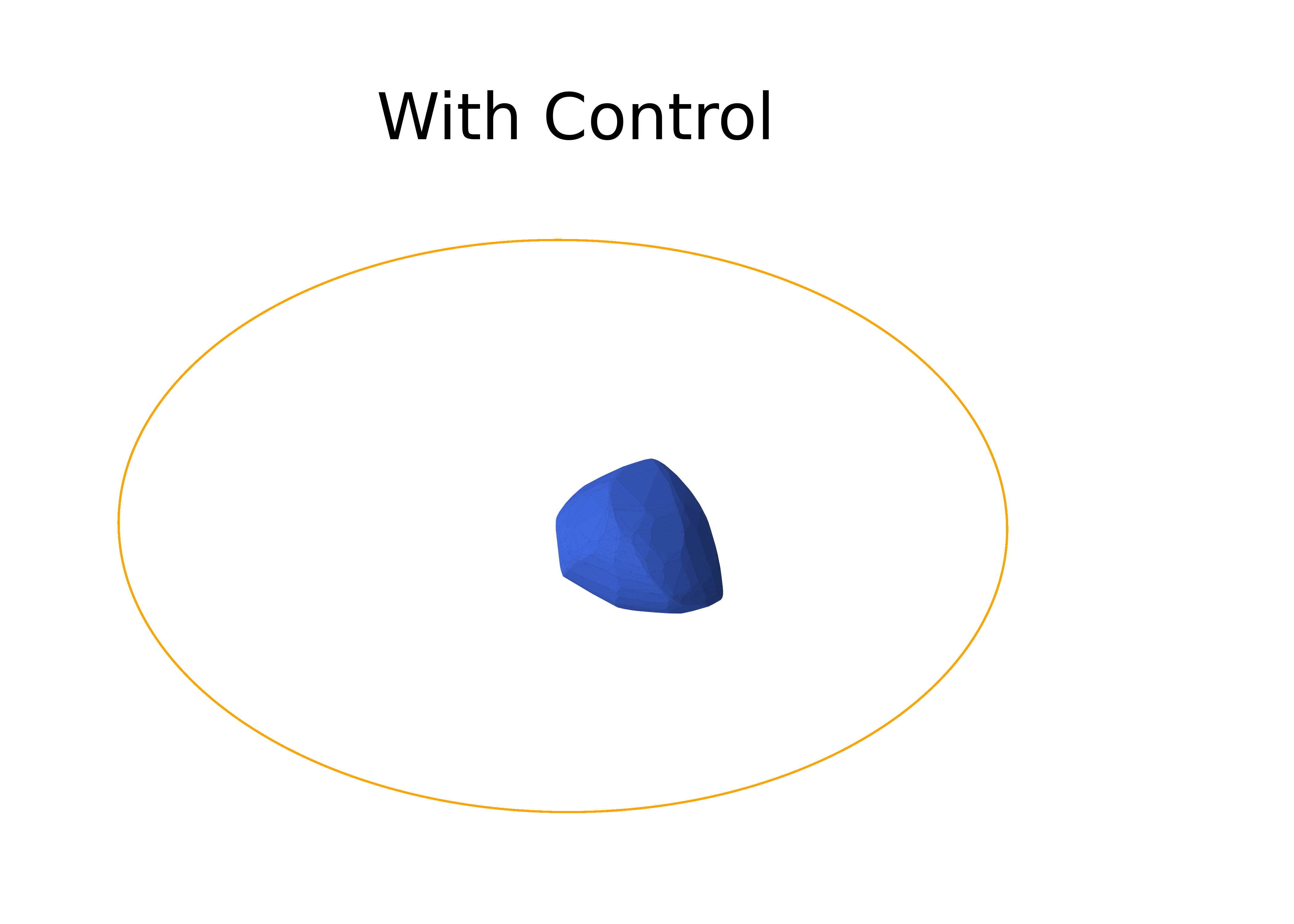}\\
          \includegraphics[width=0.99\linewidth]{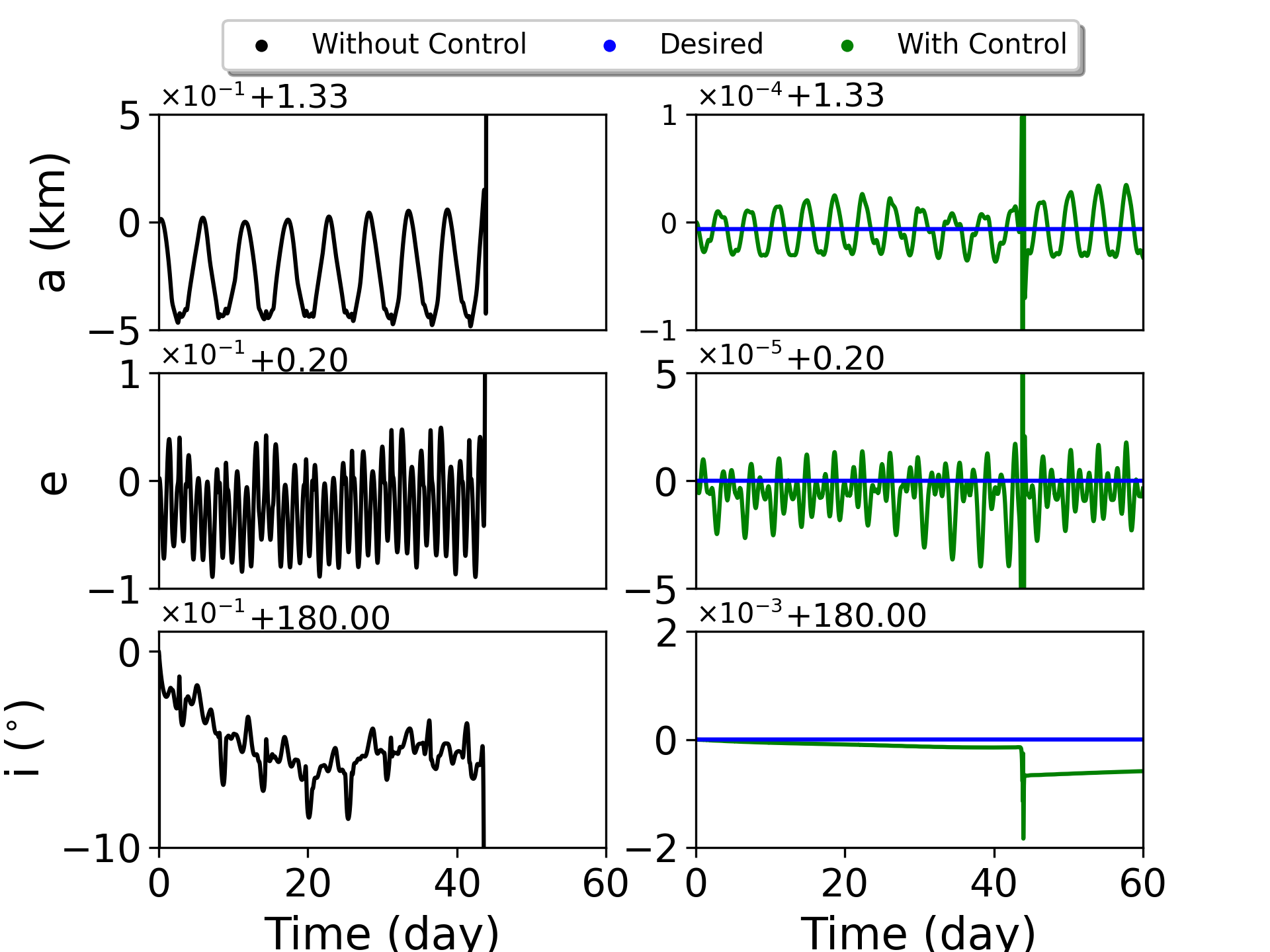}\\
          \caption{Controlled orbit close to (99942) Apophis, in the inertial frame. $a_0=0.5$ km, $e_0=0.2$, $i_0=180^{\circ}$, and other orbital parameters are fixed to 0.}\label{fig19_controlled_orbit_srp}
      \end{figure}

\begin{figure}[!htp]
          \centering
          \includegraphics[width=0.99\linewidth]{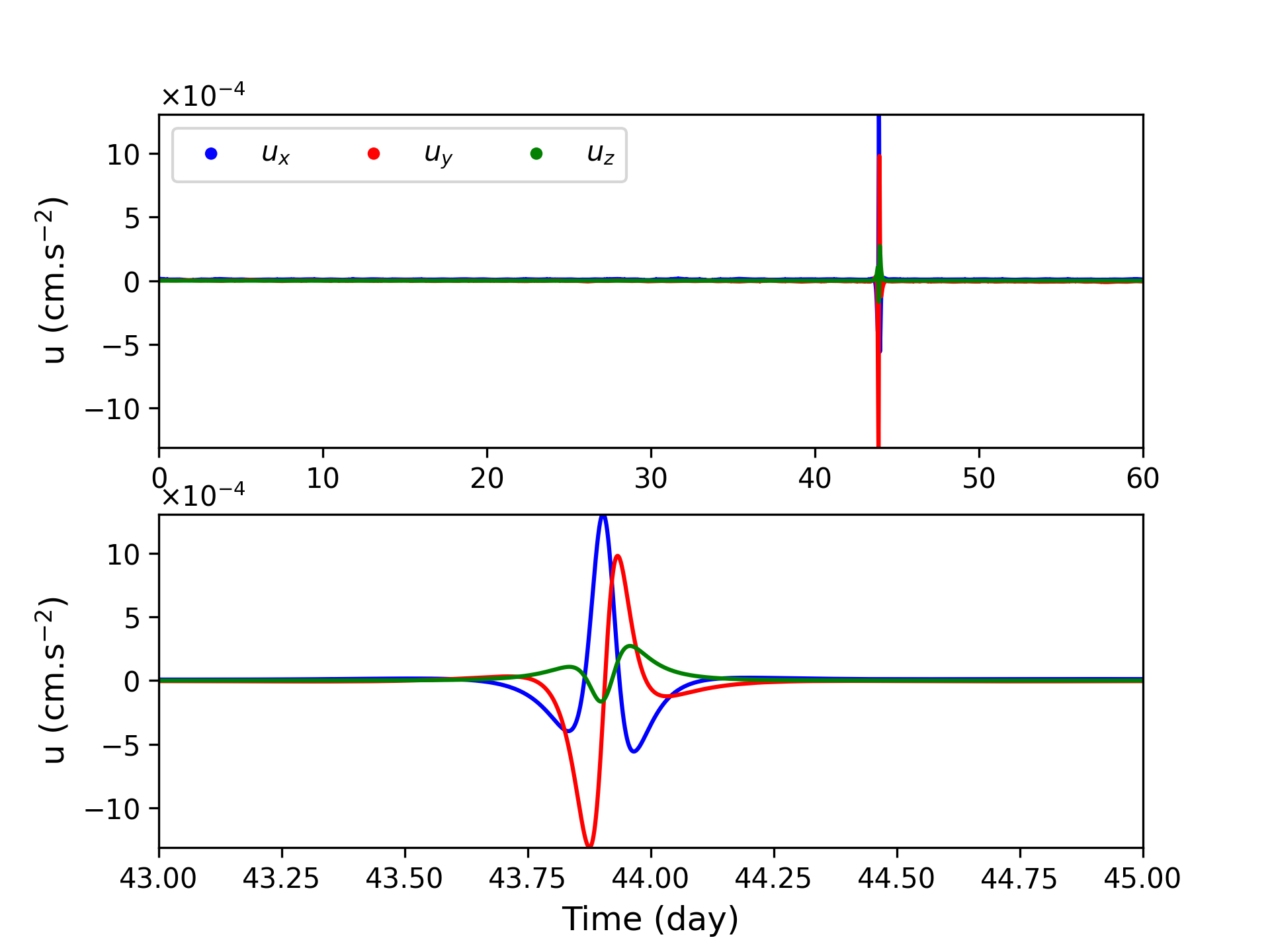}
          \caption{The control components of the orbit shown in Fig. \ref{fig19_controlled_orbit_srp}}\label{fig20_control_components}
      \end{figure}

       To demonstrate the effectiveness of our control, we considered all the perturbations (as mentioned in Sec. \ref{sec04_CE} above) on a spacecraft in an orbit with an initial semimajor axis of 1.325 km, initial eccentricity of 0.2 and initial inclination of 180$^{\circ}$ while the other orbital elements are fixed to 0. In Fig. \ref{fig19_controlled_orbit_srp} we present the orbit with (right-hand side) and without (left-hand side) control. The spacecraft without control will escape Apophis in 43 days, just after the close approach with our planet. However, our control, with a desired orbital parameters are the same as the initial ones, successfully stabilizing this orbit with a total $\bigtriangleup V$ of 0.495 m/s for 60 days of operation, which is a very law value. One can notice satisfactory small deviations from the reference orbit with a single peak in the controlled orbital elements, which corresponds to the moment of the close approach with Earth, where the components of the control input become larger and thus require more energy, as shown in Fig \ref{fig20_control_components}.
    
   \section{CPM-Asteroid database}\label{sec07_CPM-Asteroid}
      As a result of all our computations, we constructed the CPM-Asteroid (Close Proximity Motion relative to an Asteroid) database, which contains at its present extent the surfaces of section in the potential of (99942) Apophis in the body-fixed frame of reference, distributing our initial conditions in the x- and y-axis, which are related to symmetry. We considered 50 values of $H$, varying from $0.1$ to $5.0\times10^{-9}$. In parallel, we deliver for each orbit the corresponding tables giving the Fourier and Poisson series for the x-, y-, and z-coordinate. We used a web application framework developed with Shiny in R, and made our results available in the GitHub repository under an MIT public license. CPM-Asteroid is available in \href{https://github.com/safwanaljbaae/CPM-Asteroid}{https://github.com/safwanaljbaae/CPM-Asteroid}. Any other data presented in this paper can be obtained directly from the corresponding author upon reasonable request.  
   \section*{Conclusions}
      In this work, we have carried out a detailed study of the dynamics around the asteroid (99942) Apophis, one of the most interesting Near-Earth Asteroids due to its Earth close approach on April 13th.,2029. We tried to provide a preliminary realistic analysis of the orbit dynamics about the asteroid. Inspired by previous works on modeling the gravitational potential of nonspherical bodies, we calculated the mass of each tetrahedron of the (99942) Apophis shape and assigned it to a point mass in its center, representing our asteroid as a sum of 2024 points that correspond to the number of faces in the shape. That allows us to considerably reduce the computation processing time using other methods. As a preliminary step to propose an Apophis mission, we neglected, in this work, the effect of tumbling and the influence of the Earth's tides on the spin state. This interaction is still completely unknown and out the scope of this work. We obtained the physical properties and analyzed the equilibria near our target considering only the effects of a uniformly rotating 2024 points gravity field. The surfaces of section are calculated in the potential of (99942) Apophis in the body-fixed frame to show the behaviors of large scale orbits considering or not the perturbations of the planets in our Solar System and the SRP, which can considerably affect the dynamics around our target. The close approach with our planet imposes a fast and relatively strong perturbation making the vast majority of the tested orbits collide or escape from the system. An OSIRIS-REx-like spacecraft is considered for numerical analysis of orbital dynamics associated with (99942) Apophis, considering the full perturbations on the system. We employed the Runge-Kutta 7/8 variable step-size algorithm covering a period of 60 days, starting from March 1, 2029. The initial state vector of the particles is calculated using the classical orbital parameters ($a$, $e$, $i$, $\varpi$, $w$, and $f$). The variations of the semimajor axis are used to identify the less perturbed region in the system. We can state that the region with initial semimajor axis smaller than 4 km and initial eccentricity smaller than 0.4 affected by relatively small perturbations before the close approach with our planet. However, there is no stable regions around our target during the close approach. We applied the sliding mode control theory in order to solve the stabilization problem for the system. With a total $\bigtriangleup V$ of 0.495 m/s for 60 days of operation, we successfully stabilized an orbit with an initial semimajor axis of 0.5 km. Finally, we argue that our computations in this work could be refined in the future by taking into account the changes of the spin axis and rate of (99942) Apophis during the 2029 close encounter with our planet. Nevertheless, we estimate that our work provids a resonable approach of dynamical analysis of future spacecraft missions related to the target. It will be even very difficult, from a ballistic point of view, to launch a probe close to (99942) Apophis, but the idea deserves some interest.
    \section{acknowledgements}
       The authors would like to thank the Coordination for the Improvement of Higher Education Personnel (CAPES), which supported this work via the grant 88887.374148/2019-00, and  the São Paulo State Science Foundation (FAPESP, grant 2017/20794-2). We are grateful to Dr. {\bf Wael Al Zoughbi}, MD, PHD from the University of Weill Cornell Medicine for the discussions that motivated us to create the database CPM-Asteroid mentiond in this work. 

   \section*{Conflict of interest}
      The authors declare that they have no conflict of interest.


\bibliographystyle{spbasic}      
\bibliography{mybib.bib}   
\end{document}

%% file: TABLES/table1_coefficient.tex
\begin{table}
\begin{minipage}[!htp]{1\linewidth}
\caption{The volume of (99942) Apophis with its correction coefficients relative to mass (2$^{\text{nd}}$ column) and density (5$^{\text{th}}$ column). The corrected diameter is displayed in the sixieme column.}\label{table1_coefficient}
\resizebox{1.0\textwidth}{!}{
\rowcolors{1}{grey80}{}
\begin{tabular}{ccccccc}
\hline
\multicolumn{4}{c}{Compatibility of Vol. with}        &  Diameter  & Diameter    &  shape  \\
\multicolumn{2}{c}{Mass} & \multicolumn{2}{c}{Density}& (Original) & (Corrected) &    \\
Compatible & Coefficient & Compatible &   Coefficient &    km      &     km      &    \\
\hline
NO      & $0.285\mp 0.158$       &  NO           &   $0.285_{-0.0057}^{+0.0062}$        &   1.358    &  0.387  & \citet{pravec_2014}\\
NO      & $1.152\mp 0.638$       &  NO           &   $1.152_{-0.0232}^{+0.0252}$        &   0.336    &  0.387  & \citet{brozovic_2018}\\

\hline
\end{tabular}}
\end{minipage}
\end{table}

%% file: TABLES/table02_harmonics.tex
\begin{table}
\begin{minipage}[htp]{1\linewidth}
\caption{Apophis Gravity Field Coefficients up to order 4, using 3996 triangular plates of the nonconvex shape model provided by Brozovi{\'c} et al. (2018). This coefficients are computed with the respect to a constant density of 1.75 $g.cm^{-3}$, a total mass of 5.310 $ \times 10^{10} $  kg (derived from the polyhedron volume) and a reference distance of 0.193 km.} \label{table02_harmonics}
\begin{center}
\resizebox{0.8\textwidth}{!}{
\rowcolors{1}{grey80}{}

   \begin{tabular}{ccrr}
   \hline
   order & degree & $ C_{nm}$ & $ S_{nm}$ \\
   \hline
   
  0   &   0   &   1.0000000000    &   \multicolumn{1}{c}{-} \\ 
  1   &   0   &   0.0000000000    &   \multicolumn{1}{c}{-} \\ 
  1   &   1   &   0.0000000000    &   0.0000000000 \\ 
  2   &   0   &   -0.0783221323    &   \multicolumn{1}{c}{-} \\ 
  2   &   1   &   0.0000000000    &   0.0000000000 \\ 
  2   &   2   &   0.0263789204    &   -0.0000000000 \\ 
  3   &   0   &   0.0448720700    &   \multicolumn{1}{c}{-} \\ 
  3   &   1   &   0.0071441362    &   -0.0022479382 \\ 
  3   &   2   &   -0.0086800988    &   -0.0031712753 \\ 
  3   &   3   &   -0.0045154950    &   0.0001852096 \\ 
  4   &   0   &   0.0063633515    &   \multicolumn{1}{c}{-} \\ 
  4   &   1   &   -0.0036166896    &   -0.0009014188 \\ 
  4   &   2   &   -0.0004122196    &   0.0002392422 \\ 
  4   &   3   &   0.0001533535    &   0.0002872799 \\ 
  4   &   4   &   0.0000804961    &   0.0000311549 \\ 
\hline
\end{tabular}
}
\end{center}
\end{minipage}
\end{table}

%% file: TABLES/table03_eq_poins.tex
\begin{table}
\begin{minipage}[htp]{1\linewidth}
   \caption{Locations of the 4 equilibrium points of (99942) Apophis, with the relative errors (\%) with respect to the classical polyhedron method (Tsoulis and Petrovic, 2001) and the value of the Jacobi constant C (using the shape model with 1014 vertices and 2024 faces).} \label{table03_eq_poins}
\resizebox{1.0\textwidth}{!}{
\rowcolors{1}{grey80}{}
   \begin{tabular}{crrrcl}
   \hline
    & \multicolumn{1}{c}{x (km)} & \multicolumn{1}{c}{y (km)} & \multicolumn{1}{c}{z (km)} &   Errors (\%)  & \multicolumn{1}{c}{$C(km^{2} s^{-2}$)}\\
   \hline
   \multicolumn{5}{c}{\textcolor{blue}{Tsoulis and Petrovic}} \\ \hline 
$E_{1}$ &               1.031511633 &              -0.015403571 &              -0.001193548 &      &      -0.521305965 $ \times 10^{-8} $  \\ 
$E_{2}$ &              -0.047314147 &               1.020569106 &               0.000543576 &      &     -0.517908205 $ \times 10^{-8} $  \\ 
$E_{3}$ &              -1.032399019 &              -0.008968726 &              -0.001245442 &      &      -0.521512451 $ \times 10^{-8} $  \\ 
$E_{4}$ &              -0.044140503 &              -1.020703353 &               0.000489828 &      &      -0.517905804 $ \times 10^{-8} $  \\ 
\hline
   \multicolumn{6}{c}{\textcolor{blue}{Chanut et al. (2015a); Aljbaae et al. (2017)}} \\ \hline 
$E_{1}$ &               1.031356216 &              -0.015946043 &              -0.001144834 &   0.01427    &   -0.521254519 $ \times 10^{-8} $  \\ 
$E_{2}$ &              -0.045642642 &               1.020709220 &               0.000513721 &   0.00625    &    -0.517929162 $ \times 10^{-8} $  \\ 
$E_{3}$ &              -1.031250000 &              -0.044075355 &              -0.001210880 &   0.02388    &    -0.521445551 $ \times 10^{-8} $  \\ 
$E_{4}$ &              -0.043612570 &              -1.020784265 &               0.000466467 &   0.00569    &   -0.517925320 $ \times 10^{-8} $  \\ 

\hline
   \multicolumn{6}{c}{This work} \\ \hline 
$E_{1}$ &               1.031020908 &              -0.014074537 &              -0.000975681 &   0.04942    &   -0.521130609 $ \times 10^{-8} $  \\ 
$E_{2}$ &              -0.041280470 &               1.021029506 &               0.000441225 &   0.01941    &   -0.517974994 $ \times 10^{-8} $  \\ 
$E_{3}$ &              -1.031741294 &              -0.007887703 &              -0.001003436 &   0.06458    &   -0.521297563 $ \times 10^{-8} $  \\ 
$E_{4}$ &              -0.039861524 &              -1.021073280 &               0.000403140 &   0.01895    &   -0.517971558 $ \times 10^{-8} $  \\ 
\hline
\end{tabular}}
\end{minipage}
\end{table}

%% file: TABLES/table04_eigenvalues.tex
\begin{table}
\begin{minipage}[htp]{1\linewidth}
\caption{Eigenvalues of the coefficient matrix of the 4 external equilibrium points.} \label{table04_eigenvalues}
\begin{center}
\resizebox{0.88\textwidth}{!}{
\rowcolors{1}{grey80}{}
   \begin{tabular}{crrrr}
   \hline
   
   eigenvalues & \multicolumn{1}{c}{$E_{1}$} & \multicolumn{1}{c}{$E_{2}$} & \multicolumn{1}{c}{$E_{3}$} & \multicolumn{1}{c}{$E_{4}$}\\ \hline
   
    & \multicolumn{1}{c}{$ \times 10^{-4}$} & \multicolumn{1}{c}{$ 10^{\times -4}$} & \multicolumn{1}{c}{$ 10^{\times -4}$} & \multicolumn{1}{c}{$ \times 10^{-4}$}  \\
    \hline 
    \multicolumn{5}{c}{\textcolor{blue}{Tsoulis and Petrović (2001)}} \\

$\lambda_{1}$ &          +0.5826 $i$ &          +0.5752 $i$ &          +0.5862 $i$ &          +0.5755 $i$                       \\ 
$\lambda_{2}$ &          -0.5826 $i$ &          -0.5752 $i$ &          -0.5862 $i$ &          -0.5755 $i$                       \\ 
$\lambda_{3}$ &          +0.5805 $i$ &          +0.5547 $i$ &          +0.5813 $i$ &          +0.5535 $i$                       \\ 
$\lambda_{4}$ &          -0.5805 $i$ &          -0.5547 $i$ &          -0.5813 $i$ &          -0.5535 $i$                       \\ 
$\lambda_{5}$ &             -0.1314  &          +0.1438 $i$ &             -0.1497  &          +0.1473 $i$                       \\ 
$\lambda_{6}$ &             +0.1314  &          -0.1438 $i$ &             +0.1497  &          -0.1473 $i$                       \\
\hline
\multicolumn{5}{c}{\textcolor{blue}{Chanut et al. (2015a); Aljbaae et al. (2017)}} \\
$\lambda_{1}$ &          +0.5805 $i$ &          +0.5749 $i$ &          +0.5812 $i$ &          +0.5749 $i$                       \\ 
$\lambda_{2}$ &          -0.5805 $i$ &          -0.5749 $i$ &          -0.5812 $i$ &          -0.5749 $i$                       \\ 
$\lambda_{3}$ &          +0.4709 $i$ &          +0.3662 $i$ &          +0.4744 $i$ &          +0.3671 $i$                       \\ 
$\lambda_{4}$ &          -0.4709 $i$ &          -0.3662 $i$ &          -0.4744 $i$ &          -0.3671 $i$                       \\ 
$\lambda_{5}$ &             -0.1727  &          +0.2571 $i$ &             -0.1822  &          +0.2559 $i$                       \\ 
$\lambda_{6}$ &             +0.1727  &          -0.2571 $i$ &             +0.1822  &          -0.2559 $i$                       \\ 
\hline
\multicolumn{5}{c}{This work} \\
$\lambda_{1}$ &          +0.5821 $i$ &          +0.5752 $i$ &          +0.5851 $i$ &          +0.5754 $i$                       \\ 
$\lambda_{2}$ &          -0.5821 $i$ &          -0.5752 $i$ &          -0.5851 $i$ &          -0.5754 $i$                       \\ 
$\lambda_{3}$ &          +0.5802 $i$ &          +0.5561 $i$ &          +0.5807 $i$ &          +0.5551 $i$                       \\ 
$\lambda_{4}$ &          -0.5802 $i$ &          -0.5561 $i$ &          -0.5807 $i$ &          -0.5551 $i$                       \\ 
$\lambda_{5}$ &             -0.1272  &          +0.1384 $i$ &             -0.1426  &          +0.1416 $i$                       \\ 
$\lambda_{6}$ &             +0.1272  &          -0.1384 $i$ &             +0.1426  &          -0.1416 $i$                       \\ 
   \hline
\end{tabular}
}
\end{center}
\end{minipage}
\end{table}

%% file: TABLES/table05_processing_time.tex
%
%


\begin{table}
\begin{minipage}[htp]{1\linewidth}
\caption{CPU execution time to caculate the potential around (99942) Apophis, using a Pentium 3.60GHz computers.} \label{table05_processing_time}
\begin{center}
\resizebox{0.8\textwidth}{!}{
\rowcolors{1}{grey80}{}
   \begin{tabular}{lccccc}
   \hline
                       &  Tsoulis &  Mascon 8 & This work     \\
   \hline    
   1,002,000 points    &  65m27.337s     &         8m9.347s     & 0m18.337s     \\
   60 days orbit       &  463m24.751s     &         38m39.966s   & 7m21.495s     \\
   \rowcolor{grey80}$\bigtriangleup r$ (m)  &      90.30           &      90.05     &   89.50   \\
   \hline
\end{tabular}
}

\end{center}
\end{minipage}
\end{table}

%% file: TABLES/table06_sinusoids.tex
\begin{table}
\begin{minipage}[htp]{1\linewidth}
   \caption{Coefficients of Fourier and Poisson series for the x-coordinate of some orbits.} \label{table06_sinusoids}
\begin{center}
\resizebox{0.7\textwidth}{!}{
\rowcolors{1}{grey80}{}
   \begin{tabular}{rrrrr}
   \hline
   \multicolumn{5}{c}{$Y_{0} = 0.5$ km}   \\
   \multicolumn{5}{c}{$\alpha$: -0.175134$\times 10^{-1}$ ~~~~~~~ $\beta$: 0.371340$\times 10^{ -2}$ ~~~~~~~ $\gamma$:-0.132787$\times 10^{ -3}$}\\
       Period (days)& \multicolumn{1}{c}{$A_{i}\times 10^{ -3}$} &  \multicolumn{1}{c}{$B_{i}\times 10^{ -7}$} & \multicolumn{1}{c}{$C_{i}\times 10^{ -7}$} & \multicolumn{1}{c}{$D_{i}\times 10^{ -7}$}     \\
       5.005097     &      1060.429636  &    -41727.356973  &      -708.137062  &         1.147953  \\
       8.879365     &      -987.100624  &     35536.946087  &      -109.801098  &        -0.973850  \\
      12.753623     &      -268.809728  &      7370.387359  &        83.234385  &         0.375540  \\
      16.627881     &      -108.634779  &      2038.475917  &        67.728599  &         0.356383  \\
       1.130352     &       -63.980540  &      2874.906050  &       205.567497  &        -2.799160  \\
      20.502139     &       -51.875076  &       233.489164  &        54.129184  &         0.436433  \\
      24.376396     &       -27.178968  &        -0.322973  &        43.711115  &         0.184518  \\
      28.250654     &       -15.102884  &      -227.821751  &        29.804747  &         0.195668  \\
       0.198953     &         0.621542  &    -18981.383234  &      -691.424041  &         4.343197  \\
       2.744033     &        12.580168  &       116.637427  &        78.171624  &        -0.603787  \\
      32.124911     &        -8.742252  &      -395.521096  &        20.368620  &         0.222374  \\
      35.999166     &        -5.218701  &      -679.446427  &        18.204319  &         0.348981  \\
      39.873423     &        -3.170148  &      -932.064314  &         0.838382  &         0.367507  \\
       6.614609     &         2.731951  &      2781.713423  &      -259.831928  &        -0.129600  \\        
        \hline
        \multicolumn{5}{c}{$Y_{0} = 1.0$ km}                 \\
        \multicolumn{5}{c}{$\alpha$: 0.267806$\times 10^{-1}$ ~~~~~~~ $\beta$: -0.331719E$\times 10^{ -2}$ ~~~~~~~ $\gamma$: 0.853069$\times 10^{ -4}$}\\
       Period (days)& \multicolumn{1}{c}{$A_{i}\times 10^{0}$} &  \multicolumn{1}{c}{$B_{i}\times 10^{ -2}$} & \multicolumn{1}{c}{$C_{i}\times 10^{ -6}$} & \multicolumn{1}{c}{$D_{i}\times 10^{ -4}$}     \\
      10.261697     &        -0.982367  &         0.134383  &        -0.973593  &        -0.565243  \\
        \hline
        \multicolumn{5}{c}{$Y_{0} = 1.3$ km}                 \\
        \multicolumn{5}{c}{$\alpha$: 0.434399$\times 10^{-1}$ ~~~~~~~ $\beta$: -0.480825E$\times 10^{ -2}$ ~~~~~~~ $\gamma$: 0.110765$\times 10^{ -3}$}\\
       Period (days)& \multicolumn{1}{c}{$A_{i}\times 10^{0}$} &  \multicolumn{1}{c}{$B_{i}\times 10^{ -3}$} & \multicolumn{1}{c}{$C_{i}\times 10^{ -6}$} & \multicolumn{1}{c}{$D_{i}\times 10^{ -5}$}     \\
       9.975383     &        -0.980238  &        21.335815  &         0.662484  &       -18.411893  \\
       4.999458     &        -0.424993  &         0.464195  &       268.456684  &         0.980552  \\
      14.951223     &         0.134310  &        -6.183156  &       -71.496261  &         6.445293  \\
        \hline
        \multicolumn{5}{c}{$Y_{0} = 3.7$ km}                 \\
        \multicolumn{5}{c}{$\alpha$: -0.134317$\times 10^{+1}$ ~~~~~~~ $\beta$: 0.251937E$\times 10^{ 0}$ ~~~~~~~ $\gamma$: -0.657195$\times 10^{ -2}$}\\
       Period (days)& \multicolumn{1}{c}{$A_{i}\times 10^{-2}$} &  \multicolumn{1}{c}{$B_{i}\times 10^{ -3}$} & \multicolumn{1}{c}{$C_{i}\times 10^{ -5}$} & \multicolumn{1}{c}{$D_{i}\times 10^{ -5}$}     \\
       5.016080     &      -262.495691  &        13.636583  &        35.273369  &       -21.921920  \\
       6.820045     &      -111.897686  &        36.362375  &        21.796639  &        22.360328  \\
       8.624110     &        39.974526  &       -23.612523  &        36.174524  &       -15.594241  \\
       3.212095     &       -33.123935  &        -6.247984  &        -0.322134  &        -8.746280  \\
      10.428166     &       -21.850022  &        18.579728  &        21.580907  &         6.209745  \\
      12.232221     &        13.306911  &       -15.523684  &         4.718020  &        -4.800397  \\
      14.036276     &        -9.339702  &        12.601490  &        14.215852  &         7.664860  \\
       1.408633     &         9.103327  &         5.895655  &         2.617009  &         2.211337  \\
      15.840326     &         6.559770  &       -11.331941  &        -1.487833  &        -3.552058  \\
      17.644372     &        -5.078517  &         9.401595  &        11.830137  &         6.625115  \\
      19.448412     &         3.800195  &        -8.697650  &        -4.270305  &        -2.767244  \\
       0.439118     &        -4.761676  &       -59.022149  &       124.021966  &       271.103923  \\
       0.185284     &       -94.915949  &      1548.827683  &      7129.717648  &     -3068.896937  \\
      21.252446     &        -3.094827  &         7.330216  &        10.334042  &         5.300542  \\
       4.012193     &         2.332160  &       -13.063991  &        45.039407  &        17.509439  \\
       2.198787     &        -2.582965  &         8.099549  &        -4.761521  &       -31.158042  \\
      23.056472     &         2.401123  &        -6.811102  &        -5.787199  &        -2.369398  \\
      24.860492     &        -2.008734  &         5.899593  &         9.013578  &         3.774695  \\
       5.417179     &         2.416741  &         2.600684  &       -31.006796  &        20.533877  \\
       5.841547     &         0.882406  &         0.655588  &        25.089849  &       -96.524539  \\
      26.664625     &         1.603889  &        -5.257909  &        -6.937826  &        -3.018056  \\
       7.784807     &        -1.230749  &        14.766599  &        20.567543  &      -190.389295  \\
      28.468681     &        -1.336557  &         5.267708  &         6.749202  &         0.250524  \\
      \hline    

\end{tabular}
}
\end{center}
\end{minipage}
\end{table}